\documentclass[11pt]{report}
\usepackage{graphicx,epsf,bm,bbm,color,amsmath,ulem,amssymb,psfrag,fullpage,subfigure}

\usepackage[colorlinks=true]{hyperref}
\usepackage{psfrag}
\usepackage{titlepic}

	\newcommand{\beq}{\begin{equation}}
	\newcommand{\be}{\begin{equation}}
	\newcommand{\beqn}{\begin{eqnarray}}
	\newcommand{\eeq}{\end{equation}}
	\newcommand{\ee}{\end{equation}}
	\newcommand{\eeqn}{\end{eqnarray}}

\newcommand{\bem}{\begin{pmatrix}}
\newcommand{\eem}{\end{pmatrix}}

\newcommand{\f}{\frac}

\newcommand{\vk}{\vec{k}}
\newcommand{\vp}{\vec{p}}
\newcommand{\vq}{\vec{q}}
\newcommand{\vK}{\vec{K}}
\newcommand{\np}{n_{>}}
\newcommand{\nm}{n_{<}}
\newcommand{\la}{\lambda}
\newcommand{\lap}{\lambda'}

\def \a{\vec a}
\def \A{\vec A}
\def \G{\vec G}
\def \D{\vec D}
\def \R{\vec R}

\def \vep{\varepsilon}
\def \ep{\epsilon}

\DeclareMathAlphabet{\mathpzc}{OT1}{pzc}{m}{it}

\begin{document}

\title{{\it \small Habilitation \`a diriger des recherches}\\
Dirac fermions in graphene and analogues:\\
magnetic field and topological properties}
\author{Jean-No\"el Fuchs\\ 
{\small Laboratoire de Physique th\'eorique de la Mati\`ere Condens\'ee (UPMC, Paris)} \\
{\small and Laboratoire de Physique des Solides (Universi\'e Paris-Sud, Orsay)}}
\date{\today}
\titlepic{\includegraphics[width=4cm]{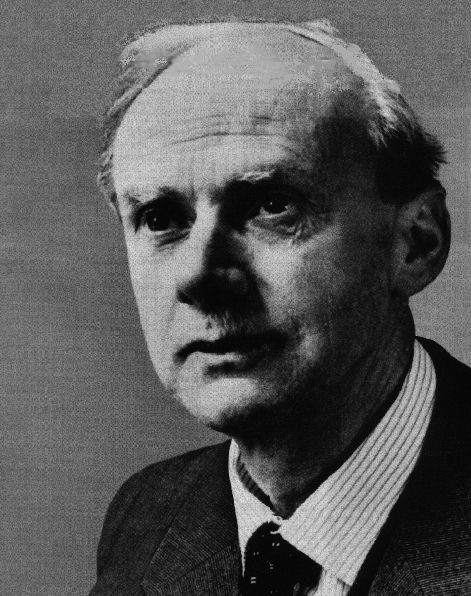}\includegraphics[width=10cm]{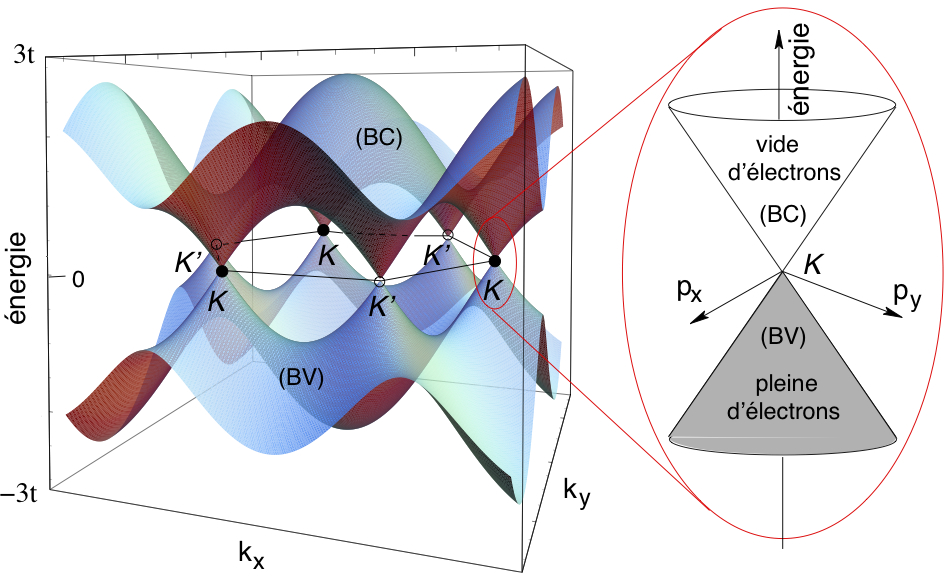}\\Paul Dirac and his fermions in graphene}
\maketitle

	
\tableofcontents
	
	\chapter*{Foreword}
	
	This document is my habilitation thesis (habilitation \`a diriger des recherches, in french). It summarizes my research activity since october 2004, which corresponds to my recruitment as a ma\^itre de conf\'erences (assistant professor) in Orsay. Before that I had been working on spin waves in cold atomic gases as a PhD student in Paris and later on interacting one-dimensional ultra-cold atomic gases (BCS-BEC crossover and Casimir effect) as a post-doc in Innsbruck (Austria). In the following, I will exclude anything related to these two last subjects and will concentrate on the unifying theme of the rest of my work: graphene and more generally two-dimensional condensed matter systems featuring Dirac fermions as quasi-particles, focusing either on the presence of a magnetic field or on topological properties. My interest for graphene and Dirac fermions started in march 2006 through the influence of Mark Goerbig. It was not long before a gang of four (not quite as famous as the original one) was constituted with the addition of Fr\'ed\'eric Pi\'echon and Gilles Montambaux. In what follows, it is mainly work with these collaborators that I am presenting.
	
	The defense took place on May 31$^{st}$ 2013 in Orsay. The composition of the jury was: Denis Basko (referee), H\'el\`ene Bouchiat (jury member), David Carpentier (referee), Allan H. MacDonald (referee), Francesco Mauri (president of the jury) and Fr\'ed\'eric Pi\'echon (jury member). 
	
	
	
	The outline of this habilitation thesis is the following: I start with a short introduction to the field of two-dimensional Dirac fermions in condensed matter, then I summarize my research activity on this subject in two chapters -- 1) magnetic field and 2) topological properties -- and finally I present some projects and perspectives for future work.

	\chapter{Introduction to Dirac fermions}
	
	{\it  Dirac to Feynman: ``I have an equation; do you have one too?''}
	
	
	\section{The Dirac Hamiltonian}
	Paul Dirac invented his equation as a relativistic generalization of the Schr\"odinger equation in order to describe the quantum mechanical motion of an electron \cite{Diracbook}. It was originally devised to apply to massive electrons moving in three-dimensional (3D) space and later gave birth to quantum electrodynamics. Dirac's construction involves matrices (called $\alpha_j$ with $j=0,1,..,d$ where $d=3$ is the space dimension) that satisfy the Clifford algebra $\{\alpha_i,\alpha_j\}=2\delta_{i,j}$ and which are needed to write the electron's Hamiltonian
\be
H_D=c \sum_{j=1}^d p_j \alpha_j + m_0 c^2 \alpha_0	
\label{eq:dirac}
\ee
where $m_0$ is the electron rest mass, $c$ is the velocity of light and $(p_1,...)=(p_x=-i\hbar \partial_x,...)$ are momentum operators. The corresponding dispersion relation is $\vep(\vp)=\pm \sqrt{m_0^2c^4+c^2\vec{p}^2}$. The number and the size of Dirac matrices $\alpha_j$ depends on the spatial dimension $d$. For example, for $d=3$, four $4\times 4$ matrices are needed, while for $d=2$ (resp. $d=1$), three (resp. two) $2\times 2$ matrices are enough to satisfy the Clifford algebra. Hence, Dirac's construction can be applied to situations others than that of 3D massive electrons, such as other spatial dimensions or other types of particles -- e.g. massless ($m_0=0$, as proposed by H. Weyl) or uncharged (as first suggested by E. Majorana), both once suspected to apply to neutrinos. 
	
	In condensed matter physics, Dirac fermions (which is the name given to particles obeying the Dirac equation) emerge as effective low-energy quasiparticles in some band structures. The most famous example is certainly graphene \cite{CastroNeto2009} -- a two-dimensional sheet of carbon atoms arranged in a honeycomb lattice -- but it is far from being the only. As shown below, graphene hosts 2D massless Dirac fermions that come in four flavors due to spin and valley degeneracy.
	
	In the following, we will mostly concentrate on graphene, but we first make a parenthesis to mention other condensed matter systems featuring Dirac fermions in order to show that it is not such an exceptional situation. A single layer of boron nitride (BN) hosts 2D massive Dirac fermions (this is also the case of the recently studied MoS$_2$), see e.g. \cite{Semenoff}. Organic salts such as $\alpha$-BEDT-TTF$_2$I$_3$ under high pressure feature quasi-2D massless Dirac fermions in the so-called zero-gap state \cite{Katayama2006}. In $d$-wave superconductors, the superconducting gap closes in a few nodes in the reciprocal space, around which quasi-particles are massless Dirac fermions (the so called nodal quasi-particles) \cite{Sachdev}. In the recently discovered family of topological insulators \cite{HasanKane}, Dirac fermions also play a role: surface states of 3D strong topological insulators are described by 2D massless Dirac fermions that come in a single flavor (hence there nickname of $1/4$-graphene). In the 2D HgTe/CdTe quantum wells, when the thickness of the well is fine-tuned to a critical value, the system is at the frontier between a topological insulator (quantum spin Hall phase) and a trivial insulator. The critical state is gapless and described by a 2D massless Dirac equation with only ``spin'' but no valley degeneracy (equivalent to $1/2$-graphene). There are also a lot of toy-models that host Dirac fermions (e.g. 2D tight-binding models such as the brick-wall lattice, the Kagom\'e lattice, the dice lattice, the square lattice at half a quantum of flux per plaquette, etc.) \cite{AsanoHotta}, some of which can be simulated in artificial matter such as cold atoms in an optical lattice \cite{Tarruell2012}, microwaves in a lattice of dielectric resonators \cite{Mortessagne2013}, photonic crystals \cite{Segev2007}, semiconductor superlatices \cite{Gilbertini} or molecular graphene \cite{Manoharan}. Up to this point, we only mentioned 2D versions of Dirac fermions. However, 3D realizations also exist in condensed matter: it has long be known that semi-metals like bismuth or graphite host massive and anisotropic 3D Dirac fermions. More recently, 3D massless Dirac fermions ($4\times 4$ Dirac matrices) \cite{KaneMele2012} and 3D Weyl fermions (also massless, but with $2\times 2$ Pauli matrices) were proposed to exist in some crystals such as pyrochlore iridates \cite{Vish} and in toy-models \cite{Delplace2012}.
	
\section{Graphene: 2D massless Dirac fermions}
		\begin{figure}[htb]
		\begin{center}
		\subfigure[]{\includegraphics[height=4.2cm]{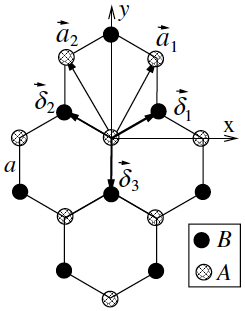}}
		\subfigure[]{\includegraphics[height=4.2cm]{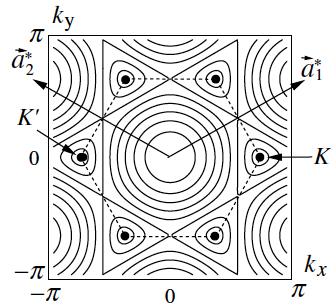}}
		\subfigure[]{\includegraphics[height=5cm]{graphenebandstructure3.jpg}}
		\caption{\label{fig:graphenebandstructure}(a) The honeycomb lattice: the two Bravais lattice vectors, the two atoms $A$ and $B$ in the unit cell and the three nearest neighbor vectors (of length $a$) are indicated. (b) Reciprocal space $(k_x,k_y)$ [in units of $1/a$]. The first Brillouin zone is indicated as a dashed hexagon. Also shown are the two reciprocal lattice vectors, the $K$ and $K'$ points and the iso-energy curves of the tight-binding model. From Bena et al. \cite{Bena2009}. (c) Tight-binding band structure of graphene: energy $\vep$ as a function of momentum ($k_x,k_y$), where $-t$ is the nearest neighbor hopping amplitude. There are two inequivalent contact points at $K$ and $K'$ between the conduction (positive energy) and valence (negative energy) bands. In undoped graphene, the ``Fermi surface'' consists of two points at $K$ and $K'$. A zoom shows the linear dispersion of massless Dirac fermions near the $K$ point. From Fuchs et al. \cite{ImagePhy}.}
		\end{center}
		\end{figure}
	We now turn back to graphene and see how massless Dirac fermions emerge there. The honeycomb lattice of graphene is shown in fig.~\ref{fig:graphenebandstructure}(a). Each carbon atom is linked to three neighboring atoms, through $\sigma$ bonds, leaving a single electron per atom for transport (known as a $\pi$ electron). The honeycomb lattice consists of a triangular Bravais lattice with a basis of two atoms (usually called $A$ and $B$) per unit cell. The distance between two atoms is $a\approx 0.14$~nm. The reciprocal Bravais lattice is also triangular and the first Brillouin zone (BZ) is hexagonal, see fig.~\ref{fig:graphenebandstructure}(b). Remarkable points in recirpocal space are the BZ center (called $\Gamma$), its six corners (only two of which are inequivalent modulo a reciprocal lattice vector and called $K$ and $K'$ or collectively the $K$ points) and the six points on the edge of the BZ at mid-distance between $K$ and $K'$ (only three of which are inequivalent and called $M_1$, $M_2$ and $M_3$ or collectively the $M$ points). The simplest description of conduction electrons in graphene is provided by the nearest-neighbor tight-binding model of Wallace \cite{Wallace1947}. Each carbon atom contributes a $2p_z$ orbital perpendicular to the graphene plane and a single conduction electron. The Hilbert space is constructed from these orbitals, which are assumed to form an orthonormal basis. Although the overlap between nearest neighbor atomic orbitals is neglected, a finite hopping amplitude $-t$ is assumed with $t\sim 3$~eV. For each wavevector $\vk=(k_x,k_y)$ in BZ, the Hamiltonian in sublattice subspace ($A,B$) reads
	\be
	H(\vk)=\left(\begin{array}{cc}0&f(\vk)^*\\f(\vk)&0\end{array}\right) \textrm{ with } f(\vk)=-t\sum_{j=1}^3 e^{i\vk \cdot \vec{\delta}_j}
	\label{eq:hw}
	\ee
where $\vec{\delta}_j$ are vectors that connect nearest neighbors, see fig.~\ref{fig:graphenebandstructure}(a). On-site energies (diagonal terms $AA$ and $BB$ in eq.~(\ref{eq:hw})) have been taken as the zero of energy. Hopping is only from $A$ to $B$ and vice-versa (off-diagonal terms $AB$ and $BA$ in eq.~(\ref{eq:hw})), which makes the lattice bipartite. In other words, the model has a chiral (or sublattice) symmetry\footnote{In $(A,B)$ subspace, it is the $\sigma_z$ matrix that performs a chiral or sublattice transformation. Chiral symmetry, in this context, means that $\{\sigma_z,H\}=0$, which implies particle-hole symmetry of the spectrum.}. As a consequence the energy spectrum $\vep_{\alpha}(\vk)=\alpha |f(\vk)|$ has particle-hole symmetry, where $\alpha=\pm$ is the band index ($\pm$ is for the positive/negative energy band). It is plotted in fig.~\ref{fig:graphenebandstructure}(c). The gap between the two bands closes at the two points $K$ and $K'$. Indeed $f(\pm \vec{K})=0$ with $\vK=(\frac{4\pi}{3\sqrt{3}a},0)$ the position of the $K$ point in fig.~\ref{fig:graphenebandstructure}(b), whereas $-\vK$ is that of the $K'$ point. We introduce the valley index $\xi=\pm$ to identify the $K$ ($\xi=+$, $\xi \vK=\vK$) and the $K'$ ($\xi=-$, $\xi \vK=-\vK$) points. There is a theorem -- known as fermion doubling and similar in spirit to the Nielsen and Ninomiya theorem of lattice gauge theories \cite{NN1983} -- that guarantees the appearance of Dirac points in pairs in 2D lattice models with a certain symmetry (see below). 
With a single electron per carbon atom, the negative energy band is completely filled at zero temperature and the positive energy band is empty. The Fermi energy is therefore $\vep_F=0$ and the ``Fermi surface'', which would naturally be a line in 2D, actually consists of two isolated points (more akin to a 1D ``Fermi surface'') at $K$ and $K'$. The filled/empty band is therefore the valence/conduction band.

At long wavelength $qa\ll 1$, in the vicinity of the two Fermi points, we linearize the Hamiltonian (\ref{eq:hw}) $H(\vk=\xi \vK +\vq)\approx \vq \cdot \vec{\nabla}_{\vk}H|_{\xi \vK}\equiv H_\xi (\vq)$ to obtain
	\be
	H_\xi (\vq)= \frac{3ta}{2}\left(\begin{array}{cc}0&\xi q_x-iq_y\\\xi q_x+iq_y&0\end{array}\right)=\hbar v_F (q_x \xi \sigma_x+ q_y \sigma_y)
	\ee
where $v_F \equiv  \frac{3ta}{2\hbar}$ is the Fermi velocity (which is here a constant independent of the carrier density and $v_F\approx \frac{c}{300}\approx 10^6$~m/s) and $\sigma_x,\sigma_y$ are the two first Pauli matrices. This is a 2D Dirac Hamiltonian  \cite{DM1984,Semenoff} with $v_F$ playing the role of an effective velocity of light, zero rest mass $m_0=0$ and the Dirac matrices are the three $2\times 2$ Pauli matrices $(\alpha_0,\alpha_1,\alpha_2)=(\sigma_z,\xi \sigma_x,\sigma_y)$. The energy spectrum (close to $K$ and $K'$) now consists of two Dirac cones $\vep_{\alpha}(\vq)=\alpha \hbar v_F q$, each similar to the dispersion of ultra-ralativistic particles with the replacement $c\to v_F$, see the zoom in fig.~\ref{fig:graphenebandstructure}(c). If the true spin is also included (which is not automatic for the 2D Dirac equation, in contrast to Dirac's original construction in 3D), there are four flavors of massless Dirac fermions due to spin and valley degeneracy, as the spectrum does not depend on $\xi$. It is usual to introduce several types of spin 1/2 when discussing graphene's low energy theory. The most fundamental is the sublattice pseudo-spin $\boldsymbol{\sigma}=(\sigma_x,\sigma_y,\sigma_z)$ in the $A,B$ subspace. Then, there is the valley isospin $\boldsymbol{\tau}=(\tau_x,\tau_y,\tau_z)$ in the $K,K'$ subspace (related to fermion doubling) and finally the true spin $\boldsymbol{s}=(s_x,s_y,s_z)$ in the $\uparrow,\downarrow$ subspace. All the matrices involved are Pauli matrices in different subspaces. Upon requantizing the momentum $\hbar \vq \to \vp\equiv -i\hbar \vec{\nabla}$, the Hamiltonian (for a single valley $\xi$ and for a single spin projection $s$) reads
\be
H_\xi= v_F (p_x \xi \sigma_x+ p_y \sigma_y) \textrm{ or } H=v_F (p_x \tau_z\sigma_x+ p_y \sigma_y)
\label{eq:dg}
\ee
as $\xi$ is an eigenvalue of $\tau_z$. 
This Hamiltonian will be the starting point of many investigations in the following chapters.

\section{Boron nitride: 2D massive Dirac fermions}
\begin{figure}[htb]
		\begin{center}
		\subfigure[]{\includegraphics[height=5cm]{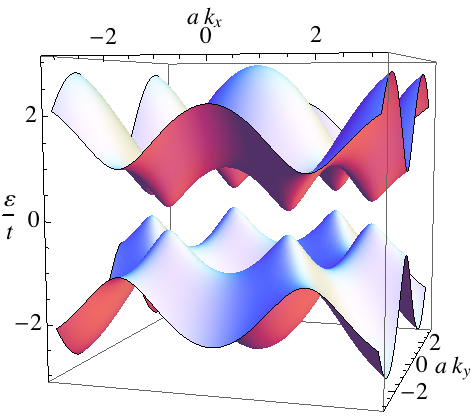}}
		\subfigure[]{\includegraphics[height=4cm]{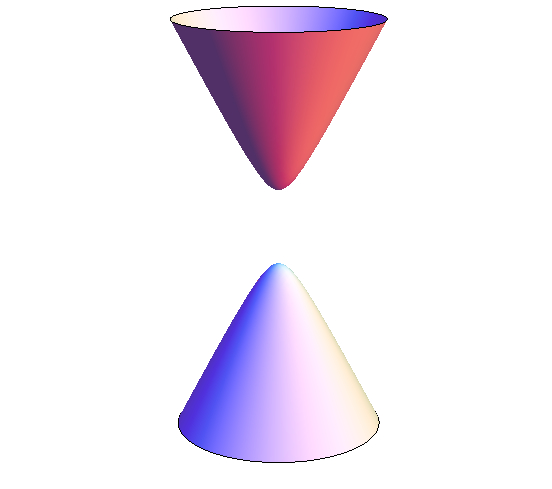}}
		\caption{\label{fig:bnbandstructure}(a) Band structure of boron nitride (or ``gapped graphene'') in the simplest tight-binding model: energy $\vep$ as a function of the wavevector $(k_x,k_y)$ in the first Brillouin zone. The gap $2\Delta$ was taken as $0.45 t$, where $t$ is the hopping amplitude. (b) Low energy massive Dirac cone (zoom near one of the $K$ points).}
		\end{center}
		\end{figure}

A monolayer of hexagonal boron nitride is very similar to graphene except that the two atoms $A$ and $B$ are now boron and nitrogen instead of two identical carbon atoms \cite{Semenoff}. The on-site energies are therefore no more equal and, in the Hamiltonian (\ref{eq:hw}), diagonal terms $\vep_A$ and $\vep_B\neq \vep_A$ should be added. Choosing the zero of energy, we may take $\vep_A=-\vep_B=\Delta>0$. The spectrum of the tight-binding model becomes $\vep_{\alpha}(\vk)=\alpha \sqrt{\Delta^2+|f(\vk)|^2}$, see fig.~\ref{fig:bnbandstructure}(a), where $2\Delta$ is the gap. The corresponding low energy Hamiltonian (\ref{eq:dg}) reads
\be
H_\xi= v_F p_x \xi \sigma_x+ v_F p_y \sigma_y+ m v_F^2 \sigma_z \textrm{ with } mv_F^2\equiv \Delta
\ee
with the spectrum $\vep_{\alpha}(\vp)=\alpha \sqrt{(mv_F^2)^2+(v_Fp)^2}$, see fig.~\ref{fig:bnbandstructure}(b). In this case, the inversion symmetry $A\leftrightarrow B$ present in graphene is lost. As a result, a $\sigma_z$ term is allowed and the Dirac fermions become massive, with rest mass $m\equiv \Delta/v_F^2$. The sublattice $\sigma_z$ symmetry is also lost but not the particle-hole symmetry of the spectrum.

\medskip

Many properties of graphene's Dirac equation would be worth discussing here -- such as symmetries (chiral/sublattice, inversion, charge conjugaison, time-reversal, etc.), helicity/chirality, Klein tunneling, absence of backscattering, zitterbewegung, etc. --  but this would take us too far, so that we refer the interested reader to the review \cite{CastroNeto2009} or the textbook \cite{KatsnelsonBook}, for example. Some of these properties will be introduced shortly when needed in the following chapters. However, we cannot resist briefly discussing the existence of accidental contact (Dirac) points and there appearance in pairs (fermion doubling). The following section lies somewhat aside of the main stream and can be skipped in a first reading.

\section{Existence of Dirac points and fermion doubling}
In this section, we consider a 2D lattice model with two bands. We wish to discuss under which circumstances contact points between the two bands exist and, when it is the case, the fact that they appear in pairs of opposite chirality (fermion doubling) \cite{AsanoHotta,Kogan2013,Hatsugai2011}.

First, consider the existence of contact points. A single contact point involves two states and is sometimes referred to as a degeneracy point. The Hamiltonian can be written as $H(\vk)=\vep_0(\vk)\sigma_0+\mathbf{R}(\vk)\cdot \boldsymbol{\sigma}$, where $\mathbf{R}=(R_1,R_2,R_3)$ is a real 3-component vector, $\boldsymbol{\sigma}=(\sigma_x,\sigma_y,\sigma_z)=(\sigma_1,\sigma_2,\sigma_3)$ are the three Pauli matrices and $\vk=(k_x,k_y)$ is a two-dimensional vector in the first Brillouin zone. The dispersion relation is $\vep(\vk)=\vep_0(\vk)\pm|\mathbf{R}(\vk)|$. We want to find $\vk=(k_x,k_y)$ such that $\mathbf{R}(\vk)=(R_1(\vk),R_2(\vk),R_3(\vk))=0$. This is overspecified, and therefore highly improbable, because we only have two parameters and three equations to satisfy. So, if we want to find a contact, we need a condition so that one of the $R_j$ (let say $R_3$) vanishes. Such a condition is usually provided by a discrete symmetry such as space-time inversion or chirality. If one has space-time inversion symmetry (the product of time-reversal and spatial inversion transformations), then $H(\vk)=\sigma_1 H(\vk)^* \sigma_1$, as a consequence of which $R_3=0$ for all $\vk$. The role of space-time inversion can also be played by the chiral (sublattice) symmetry\footnote{Here we have in mind the simplest tight-binding model of graphene in which the sublattice symmetry $\sigma_3$ is an example of a chiral symmetry and is a consequence of the honeycomb lattice being bipartite. In general, a chiral symmetry is said to exist if there is a unitary operator $S$ that squares to the identity and which anti-commutes with the Hamiltonian $\{S,H(\vk)\}=0$ \cite{Hatsugai2011}.}, $H(\vk)=-\sigma_3 H(\vk)\sigma_3$, which implies $R_3(\vk)=0$ and $\vep_0(\vk)=0$ for all $\vk$. In both cases, the condition of a contact point becomes $R_j(\vk)=0$ with $j=1$ and $2$, which is no more overspecified, so that the contact becomes feasible. This is known as an accidental contact as its position in the BZ is not imposed by a point-group symmetry. An example of a accidental contact is that happening in deformed graphene or in the organic salts $\alpha$-(BEDT-TTF)$_2$I$_3$ (in both examples, space-time inversion is present). An example of an essential contact is that in undeformed graphene, which has point group $C_{6v}$. In the case of an essential degeneracy, the contact happens at a high symmetry point of the BZ. In summary, a contact point is feasible in a 2D lattice model with 2 bands if there is either a chiral symmetry or space-time inversion symmetry. 

Now, let us see why contact points appear in pairs (fermion doubling). We first suppose that a contact point exists at $\vk_D$ between two bands in a 2D lattice model, which corresponds to $\mathbf{R}(\vk_D)=0$. Then, if time reversal symmetry is present, $H(\vk)=H(-\vk)^*$, such that $(R_1(-\vk),R_2(-\vk),R_3(-\vk))=(R_1(\vk),-R_2(\vk),R_3(\vk))$. Therefore $\mathbf{R}(\vk_D)=0$ implies that $\mathbf{R}(-\vk_D)=0$. Unless $\vk_D\equiv -\vk_D$ modulo a reciprocal lattice vector (this is the position of the so-called time-reversal invariant points), there is a pair of contact points at $\vk_D$ and $-\vk_D$. In the second part of this thesis, we will see that time-reversal invariant points are precisely where the merging of Dirac points occurs. Note also that the role of time reversal symmetry may be played by another discrete symmetry such as space inversion (parity), $H(\vk)=\sigma_1 H(-\vk) \sigma_1$, which also implies that the pair is at $\vk_D$ and $-\vk_D$. So for the moment, the theorem goes as: in a 2D lattice model with either time-reversal or inversion symmetry, contact points come in $(\vk_D,-\vk_D)$ pairs unless they are located at time-reversal invariant points. 

Actually, Hatsugai \cite{Hatsugai2011} has shown that the existence of a chiral symmetry (see the preceding footnote) implies not only the feasibility of contact points but also the fact that they appear in pairs -- without requiring an additional symmetry such as time-reversal. However, in such a case, the pair of contact points is not necessarily at $\vk_D$ and $-\vk_D$. In addition, he showed that the contact points within a pair have opposite chiralities. Therefore, in a 2D lattice model with a chiral symmetry, contact points appear in pairs and have opposite chiralities (this is the 2D version of the Nielsen-Ninomiya theorem \cite{NN1983}). Graphene is quite a special case as it has many symmetries among which time-reversal, inversion and sublattice (chiral) symmetries. Boron nitride has time-reversal symmetry but neither inversion nor sublattice symmetry. Boron nitride has no contact points; however, it does feature a pair of massive Dirac fermions at low energy.  


\bigskip

After this general introduction, we now present our work in two chapters. The first deals with orbital properties of Dirac fermions in a perpendicular magnetic field, such as Landau levels, quantum Hall effect, magneto-optics, magneto-plasmons, magneto-phonon resonance and magneto-transport. The second is concerned with topological properties of Dirac fermions, such as their winding number and the merging transition of Dirac points.

	\chapter{Massless Dirac fermions in a strong magnetic field}
	
	{\it  The wavefunction of the zero-energy Landau level in one valley resides on only one sublattice (parity anomaly).}
	
	\vspace{0.5cm}
	
In this chapter, we consider the electronic properties of graphene and alike in a strong perpendicular magnetic field. The main focus is on orbital properties -- and not so much on spin properties (such as the Zeeman effect) -- of massless Dirac fermions in the presence of a magnetic field. A large portion of what is presented here was done in collaboration with Mark Goerbig and can also be found in his habilitation thesis, which was published in \cite{MarkRMP}. In the following, we first present the energy spectrum of massless Dirac fermions in a magnetic field (relativistic Landau levels), studying in particular deviations from the ideal linear dispersion relation (e.g. trigonal warping, tilting of the cones, etc.). Then we turn to the relativistic (integer) quantum Hall effect and discuss in particular a Peierls instability that partially lifts the valley degeneracy. Next, interactions between electrons are included in order to investigate particle-hole excitations such as magneto-plasmons. We also study interactions between electrons and optical phonons leading to a magneto-phonon resonance in graphene. Eventually, we consider how magneto-transport measurements in disordered samples can reveal the nature of conductivity limiting impurities in graphene. 
	
	\section{Landau levels}
	\subsection{``Relativistic'' Landau levels}
We start from the effective Hamiltonian for graphene (see the introduction chapter):
	\be
	H_\xi=v_F (\xi \sigma_x p_x +  \sigma_y p_y)
	\ee
Instead of working with a bispinor with components $\left(\begin{array}{c}A\\B\end{array}\right)$ in both valley, it is more convenient to switch the order of components in the $K'$ valley, $\left(\begin{array}{c}B\\A\end{array}\right)$, so that $(\sigma_x,\sigma_y,\sigma_z)\to (\sigma_x,-\sigma_y,-\sigma_z)$ when $\xi=-$ and the Hamiltonian reads:
	\be
	H_\xi=\xi v_F (\sigma_x p_x +  \sigma_y p_y)
	\ee
	\begin{figure}[htb]
	\begin{center}
	\subfigure[]{\includegraphics[height=5cm]{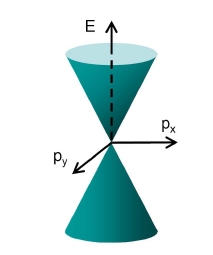}}
	\subfigure[]{\includegraphics[height=4cm]{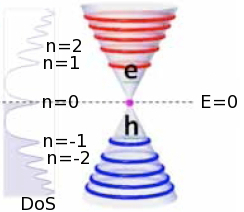}}
	\caption{\label{fig:cone}(a) Dispersion relation $\vep=\pm v_F p$ of massless Dirac fermions in zero magnetic field. (b) Relativistic Landau levels $\vep_n=\pm v_F\sqrt{2\hbar eBn}$ and the corresponding (disorder broadened) density of states. From Luican et al. \cite{Luican2011}.}
	\end{center}
	\end{figure}

	In the presence of a magnetic field, the Peierls substitution gives $\vec{p}=(p_x,p_y) \to \vec{\Pi}=\vec{p}+e\vec{A}$ where $-e<0$ is the electron charge and $\vec{A}$ is the vector potential such that $\vec{\nabla}\times \vec{A}=B \vec{e}_z$ is the magnetic field. The Hamiltonian becomes:
	\be
		H_\xi=\xi v_F (\sigma_x \Pi_x + \sigma_y \Pi_y) = \xi v_F [\sigma_x (p_x+eA_x) + \sigma_y (p_y+eA_y)]
	\ee
	The canonical momentum $\vec{p}$ is conjugate to the position operator $[r_j,p_k]=i\hbar \delta_{j,k}$ but depends on the gauge, whereas $\vec{\Pi}$ is the gauge-invariant mechanical momentum, which is not conjugate to the position operator. In the presence of a non-zero magnetic field, the two components of the mechanical momentum do not commute anymore $[\Pi_x,\Pi_y]=-ie\hbar B$. In analogy with the one-dimensional harmonic oscillator we introduce the creation $a^\dagger$ and annihilation $a$ operators as linear combinations of $\Pi_x$ and $\Pi_y$ such that $[a,a^\dagger]=1$: here $\sqrt{2e\hbar B}a/a^\dagger \equiv \Pi_x\mp i\Pi_y$. Now the Hamiltonian reads
		\be
			H_{\xi}=\xi v_F\sqrt{2e\hbar B}\left(\begin{array}{cc}0&a \\a^\dagger&0 \end{array} \right)\, ,
		\ee
which is easily diagonalized by taking its square:
	\be
				H_\xi^2=v_F^2 2e\hbar B\left(\begin{array}{cc}a^\dagger a +1 &0 \\0&a^\dagger a \end{array} \right)			\ee
	We call $|n\rangle$ the eigenvectors of the number operator $a^\dagger a$, $a^\dagger a  |n\rangle = n |n\rangle$ with $n \in \mathbb{N}$, so that the energy eigenvalues for both valleys are
	\be
	\vep_{\alpha,n}=\alpha v_F\sqrt{2e\hbar Bn}
	\ee
where $\alpha=\pm$ is the band index and refers to the conduction/valence band, see fig.~\ref{fig:cone}(b). This result was first obtained by McClure \cite{McClure}. Each Landau level (LL) has a macroscopic degeneracy $2N_\phi = 2 \frac{B \mathcal{A}}{h/e}=2\frac{\phi}{\phi_0}$ where $\phi=B\mathcal{A}$ is the total flux threading the system of area $\mathcal{A}$, $\phi_0\equiv \frac{h}{e}$ is the flux quantum and the factor of two is due to valley degeneracy. The corresponding eigenvectors are
	\be
	\frac{1}{\sqrt{2}}\left(\begin{array}{c}|n-1\rangle \\ \xi \alpha |n\rangle\end{array} \right) \textrm{ if } n\geq 1 \textrm{ and } \left(\begin{array}{c}0 \\ |n=0\rangle\end{array} \right) \textrm{ if } n=0
	\ee
which may also be written as:
	\be
	\frac{1}{\sqrt{2}}\left(\begin{array}{c}(1-\delta_{n,0})|n-1\rangle \\ \sqrt{1+\delta_{n,0}}\xi \alpha |n\rangle\end{array} \right)
	\ee
The most remarkable feature is the existence of a zero-energy Landau level ($n=0$). This is related to the parity anomaly of the 2D Dirac equation, see e.g. \cite{Semenoff}. Its energy is independent of the magnetic field but its existence is due to the magnetic field, as testified by its $2N_\phi$ degeneracy\footnote{In zero magnetic field, there are exactly four states at zero energy in an infinite system, corresponding to the two Dirac points and not taking the spin degeneracy into account. In other words, each contact (Dirac) point corresponds to {\it two} zero energy states. Fermion doubling implies that there are actually four.}. This is not due to the linear spectrum, but to the Berry phase of $\pi$ carried by each Dirac point, as shown below. Note also that in this $n=0$ Landau level (LL), and only in this one, the state belonging to one valley resides on only one of the sublattices. Schematically $K\sim B$ and $K'\sim A$ when $n=0$. This is the parity anomaly, as inversion symmetry ($A\leftrightarrow B$) appears to be broken in this LL. It will play an important role later when discussing magnetic field induced instabilities of the $n=0$ LL. 

Semiclassical quantization is worth discussing briefly here as it deepens our understanding of LLs in graphene. The semi-classical quantization condition of Onsager and Lifshitz \cite{Onsager,LK} states that among closed classical cyclotron orbits, those that survive in the quantum realm are such that $S(\vep)l_B^2=2\pi(n+\gamma)$, where $n$ is an integer, $0\leq \gamma<1$ is a yet undetermined phase shift, $l_B\equiv \sqrt{\frac{\hbar}{eB}}$ is the magnetic length and $S(\vep)=\int_{\vep(\vk)\leq \vep} dk_x dk_y$ is the surface enclosed by the cyclotron orbit in reciprocal space. This surface is simply related to the density of states $\rho(\vep)$ (per spin, per valley and per unit area) by $S(\vep)=(2\pi)^2\int^\vep d\vep\, \rho(\vep)=\pi (\frac{\vep}{\hbar v_F})^2$. Therefore, the semi-classical quantization condition gives the Landau levels $\vep_{\alpha,n}\approx \alpha v_F\sqrt{2\hbar eB(n+\gamma)}$, where $\alpha=\pm$ is the band index. The phase shift $\gamma=\frac{1}{2}-\frac{\Gamma}{2\pi}$ is due to a Maslov index of $2$ (giving the usual $\frac{1}{2}$ factor for the LLs of the parabolic two-dimensional electron gas) and a Berry phase $\Gamma$ for a cyclotron orbit \cite{MikitikSharlai,Fuchs2010}. This point will be discussed in detail in the sections \ref{bp} and \ref{sq} on Berry phases. In the case of Dirac cones $\Gamma=\pi$ and therefore $\gamma=0$. In the end, the semiclassical LLs are $\vep_{\alpha,n} \approx \alpha v_F\sqrt{2\hbar eBn}$ when $n\gg 1$, which is the validity condition of the semiclassical approximation. Here this approximation recovers the exact result.
	
	To summarize, we compare the relativistic LLs just obtained $\vep_{\alpha,n}=\alpha v_F\sqrt{2e\hbar Bn}$ with the usual LLs $\vep_n=\frac{\hbar eB}{m}(n+\frac{1}{2})$ obtained from the massive Schr\"odinger Hamiltonian $H=\frac{\vec{\Pi}^2}{2m}$: (i) both signs versus only positive energy levels, (ii) square root versus linear in magnetic field dependence, (iii) square root versus linear in Landau index $n$ dependence, (iv) $n$ versus $n+1/2$ dependence as a consequence of the existence or not of a zero energy LL, (v) degeneracy of $2N_\phi$ versus $N_\phi$ due to the presence/absence of valley degeneracy. Differences (ii) and (iii) are due to the parabolic versus linear zero-field spectrum. Arguably, (iv) is the most important difference as we will see when discussing graphene's integer quantum Hall effect. It is related to the $\pi$ Berry phase that shifts $\gamma$ from $\frac{1}{2}$ to $0$. The essential fact is that the $n=0$ LL has a magnetic-field independent energy, which is quite anomalous for a quantized cyclotron orbit. Point (v) is related to the fact that Dirac points appear in pairs (fermion doubling).

	\subsection{Trigonal warping and magneto-optical transmission spectroscopy}
	A powerful way of probing the Landau levels in graphene is magneto-optical transmission spectroscopy \cite{Sadowski}. The idea is to measure the transmission of light of tunable frequency across a graphene sheet in a fixed perpendicular magnetic field. The light is sent parallel to the magnetic field and can be absorbed when its frequency $\nu$ -- typically in the $\sim 100$ meV $\sim 10^{13}$ Hz range -- coincides with that of an inter-Landau level transition $n,\alpha \to n', \alpha'$ of energy $\Delta \vep = v_F\sqrt{2eB\hbar}(\alpha'\sqrt{n'}-\alpha\sqrt{n})$. The transition happens only if the initial LL $n$ contains electrons, the final LL $n'$ has available states and the Landau index of the initial and final LL differ by $\pm 1$ so that $n'=n\pm 1$. The latter condition results from the conservation of momentum and the fact that the photon has a very small momentum compared to that of electrons (the optical dipole selection rule permits only vertical transitions). This also implies that transitions can only occur within a valley, either $K$ or $K'$, and not between valleys which are far apart in reciprocal space as $|\vec{K}-\vec{K}'|\sim 1/a \sim 10^{10}$ m$^{-1}$. However, transitions can occur within the same band (intra-band transition $\alpha'=\alpha$) or between the valence and conduction bands (inter-band transition $\alpha'=-\alpha$). There are also special transitions involving the $n=0$ Landau level. Examples of intra-LL transition is ($n,+\to n+1,+$) with energy $\Delta \vep = v_F\sqrt{2eB\hbar}(\sqrt{n+1}-\sqrt{n})$, of inter-LL transition is ($n,-\to n+1,+$) with energy $\Delta \vep = v_F\sqrt{2eB\hbar}(\sqrt{n+1}+\sqrt{n})$ and of transition involving the zero-energy LL is ($0 \to 1,+$) with energy $\Delta \vep =v_F \sqrt{2eB\hbar}$. Magneto-spectroscopy allowed Sadowski and coworkers to observe inter-LL transitions in epitaxial graphene as a function of the magnetic field and therefore to reveal the $\sqrt{B}$ dependence of Landau levels \cite{Sadowski}. Fitting the slope allowed them to extract the Fermi velocity $v_F\approx 1.03\times 10^6$ m/s in fair agreement with $v_F=\frac{3ta}{2\hbar}=0.97\times 10^6$ m/s with $t=3$ eV and $a=0.14$ nm. Similar results have also been obtained on exfoliated graphene \cite{Jiang2007}.
	
	In a further set of experiments, measurements were extended to much larger energies (between 0.5 and 1.25 eV) and magnetic fields (up to 32 T) in order to probe the limits of graphene's description in terms of ideal linear Dirac cones \cite{Plochocka}. Let us start from a tight-binding description of graphene in zero magnetic field including nearest $t$ and next-nearest $t'$ neighbor hopping amplitudes. Typically $t'\sim 0.1 t$. In the $(A,B)$ subspace, the Hamiltonian reads
	\be
	H(\vec{k})=\left(\begin{array}{cc}f'(\vk)&f(\vk)^*\\ f(\vk)&f'(\vk)\end{array}\right)
	\ee
	where $f(\vk)=-t\sum_{j=1}^3 e^{i\vk \cdot \vec{\delta}_j}$ describes nearest neighbor hopping and $f'(\vk)=t'\sum_{j=1}^3 2\cos(\vk \cdot \vec{\tau}_j)$ describes next-nearest neighbor hopping. The vectors $\vec{\delta}_j$ connect nearest neighbor atoms (separated by a distance $a\approx 0.14$ nm) and $\pm \vec{\tau}_j$ connect next-nearest neighbor atoms (separated by a distance $a\sqrt{3}$) on the honeycomb lattice, see fig.~\ref{fig:graphenebandstructure}(a) ($\vec{\tau}_1=\vec{a}_1$, $\vec{\tau}_2=\vec{a}_2$ and $\vec{\tau}_3=\vec{a}_1-\vec{a}_2$). The Dirac points are located at $\pm \vec{K}=\pm \frac{4\pi}{3\sqrt{3}a}\vec{e}_x$, see fig.~\ref{fig:graphenebandstructure}(b). Close to each valley, we expand the Hamiltonian as a function of $\vec{q}=\vk-(\pm\vec{K})$ at third order to obtain
	\be
	f(\vk)\approx \hbar v_F\left(q-\frac{a}{4}q^{*2}-\frac{a^2}{8}\vec{q}^2 q\right) \textrm{ and } f'(\vk) \approx -3t'+\frac{9t'a^2}{4}\vec{q}^2
	\ee
	in the $K$ valley where $v_F\equiv \frac{3ta}{2\hbar}$ and $q\equiv q_x+iq_y=|\vq|e^{i\phi_q}$ here. The other valley $K'$ gives
	\be
	f(\vk)\approx -\hbar v_F\left(q^*+\frac{a}{4}q^{2}-\frac{a^2}{8}\vec{q}^2 q^*\right) \textrm{ and } f'(\vk) \approx -3t'+\frac{9t'a^2}{4}\vec{q}^2
	\ee
	The dispersion relation $\vep=\alpha |f(\vk)|+f'(\vk)$ becomes:
	\be
	\vep\approx \alpha \hbar v_F |\vec{q}|\left(1-\xi \frac{a}{4}|\vec{q}|\cos(3\phi_q)-\frac{a^2}{32}|\vq|^2 (3+\cos^2 3 \phi_q)\right)+\frac{9t'a^2}{4}\vec{q}^2 
	\ee
	where $\xi=\pm$ is the valley and $\alpha=\pm$ the band index. One notices that: (i) the spectrum is no more isotropic but trigonally warped because of $\cos(3\phi_q)$, see the iso-energy curves in fig.~\ref{fig:graphenebandstructure}(b), (ii) particle-hole symmetry is lost when $t'\neq 0$ (i.e. $f'(\vk)\neq 0$) and (iii) the two valleys are no longer degenerate due to opposite trigonal warping $\xi \cos(3 \phi_q)$. As a remark, we note that non-zero overlap between $2p_z$ atomic orbitals on neighboring carbon atoms has an effect on the low energy spectrum that is similar to that of next-nearest neighbor hopping, see for example \cite{Dresselhaus}. Consequences on Landau levels are easily obtained by making the Peierls substitution and then solving the eigen-value problem approximatively in the large $n$ (semi-classical) limit. One obtains
	\be
	\vep\approx \alpha v_F\sqrt{2e\hbar B n}\left(1-\frac{3a^2}{8 l_B^2}n \right) + \frac{9}{2}t'\frac{a^2}{l_B^2}n
	\ee
	where $l_B\equiv \sqrt{\frac{\hbar}{eB}}$ is the magnetic length. Magneto-optical transmission spectroscopy measurements qualitatively confirmed the importance of trigonal warping and other higher-order band corrections for energies above $0.5$ eV. The sensitivity of the experiment did not allow one to detect any electron-hole asymmetry up to the highest explored energies $\sim 1.25$ eV \cite{Plochocka}. The experiment therefore established the high-energy limit of validity of the massless Dirac fermion effective description of graphene, around $500$ meV $\sim t/6$. Recent experiments on high mobility ($> 10^6$ cm$^2$/V.s) suspended graphene samples have shown a limit in the low energy direction: close to the Dirac point, logarithmic deviations from linearity were found around $20$ meV and attributed to electron-electron interactions \cite{DiracConeReshaping}. In addition, these measurements set an upper bound for a zero-field gap in the band structure of graphene at $0.5$ meV \cite{Mayorov2012}.

\subsection{Tilted and anisotropic Dirac cones}
\begin{figure}[htb]
\begin{center}
\subfigure[]{\includegraphics[height=4cm]{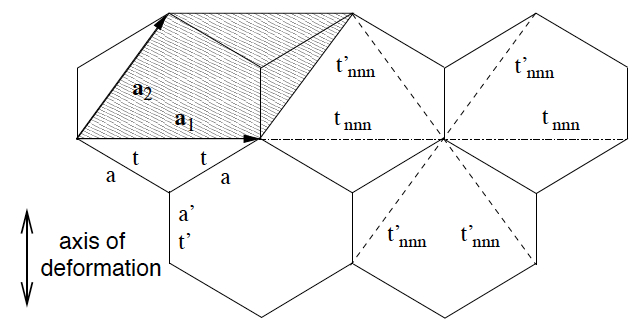}}
\subfigure[]{\includegraphics[height=5cm]{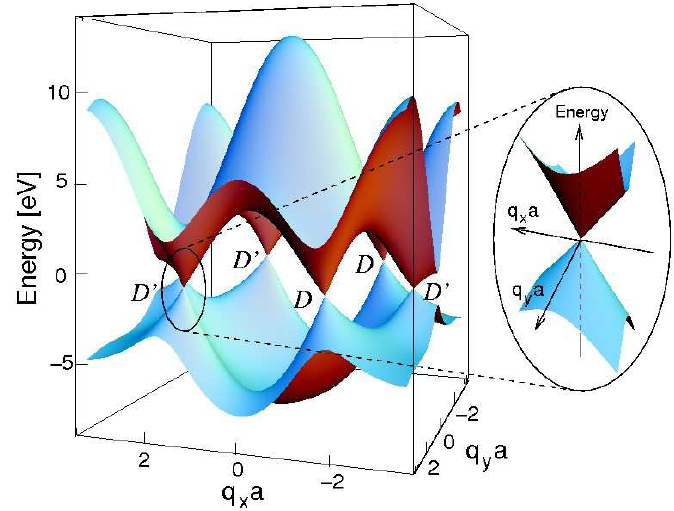}}
\caption{\label{fig:prb2008}(a) Uniaxial compression of the honeycomb lattice of graphene: the distance $a'$ is shorter than $a$ whereas the hopping amplitude $t'>t$. Next nearest neighbor hopping amplitudes $t_{nnn}$ and $t_{nnn}'$ are also indicated. (b) Dispersion relation of uniaxially deformed graphene featuring anisotropic and tilted Dirac cones at the $D$ and $D'$ points, which do not coincide with the $K$ and $K'$ corners of the Brillouin zone. From Goerbig et al. \cite{Goerbig2008}.}
\end{center}
\end{figure}

We now discuss a generalization of the previous results and study Landau levels of anisotropic and tilted Dirac cones. A tilted cone means that the dispersion relation has a conical shape in the momentum-energy ($p_x,p_y,\vep$) representation, as in graphene, but with an axis that is not parallel to the energy axis, see fig.~\ref{fig:prb2008}(b). Anisotropy refers to the fact that, even in the absence of a tilt, a constant energy contour may not be a circle (isotropic case) but an ellipse. Tilted and anisotropic Dirac cones occur for example in mechanically deformed graphene under uniaxial strain \cite{Goerbig2008}, see fig.~\ref{fig:prb2008}(a), or in quasi-2D organic salts $\alpha$-(BEDT-TTF)$_2$I$_3$, which, under pressure, can enter a so-called zero-gap state hosting massless Dirac fermions \cite{Katayama2006, Goerbig2008}. Generically, when deforming a graphene sheet, the band structure is changed in the following way. At moderate deformation, the Dirac points still exist. However they move in reciprocal space and do not coincide with the corners of the Brillouin zone anymore. The corresponding Dirac cones become anisotropic and tilted. And the cones in the two valleys are tilted in opposite directions, see fig.~\ref{fig:prb2008}(b). Upon further increasing the deformation, the Dirac cones start to couple and, eventually, if the strain is strong enough, they may encounter and annihilate in a topological merging transition. As this is the subject of the following chapter, we now concentrate on small deformations and on the effects of anisotropy and tilt of the Dirac cones on Landau levels. 

	\subsubsection{Landau levels of anisotropic and tilted Dirac cones}\label{sectiontilt}
	We consider a generalized Weyl Hamiltonian \cite{Fukuyama} in order to describe the low energy effective properties of such a tilted Dirac cone in a single valley\footnote{The other valley is described by the Hamiltonian $-H$. Generally $H_\xi =\xi H$ where $\xi=\pm $ is the valley index. In order to obtain such a concise form, we have chosen the bispinor representation $(A,B)$ in valley $K$ and $(B,A)$ in valley $K'$. Note, in particular, that the two Dirac cones have opposite tilts.}:
	\be
	H=\sum_{\mu=0}^3 \vec{v}_\mu \cdot \vec{p}\,  \sigma_\mu = \vec{\mathbf{v}}\cdot \vec{p}\, \boldsymbol{\sigma}
	\ee
where $\sigma_0=\mathbb{I}$ is the $2\times 2$ unit matrix, $\boldsymbol{\sigma}=(\sigma_1,\sigma_2,\sigma_3)=(\sigma_x,\sigma_y,\sigma_z)$ is the three-dimensional vector of Pauli matrices, $\vec{p}=(p_x,p_y)$ is the two-dimensional momentum and we take $\hbar\equiv 1$ in this section. This is the most general two bands Hamiltonian producing a linear dispersion relation. The velocities $\vec{v}_\mu=(v_\mu^x,v_\mu^y)$ with $\mu=0,...,3$ correspond to eight real parameters, which is over-specified as we show in the following. First, performing a rotation in spin space such that $\sigma_3$ is perpendicular to $\mathbf{v}^x=(v_1^x,v_2^x,v_3^x)$ and $\mathbf{v}^y=(v_1^y,v_2^y,v_3^y)$, we obtain $H=\vec{v}_0\cdot \vec{p}\sigma_0 + \vec{v}_1\cdot \vp \sigma_1 + \vec{v}_2 \cdot \vp \sigma_2$. Second, we rotate the vector $\vp \to \vq$ together with a rotation in spin space around the direction $\sigma_3$ so that, in the end \cite{Goerbig2008}
	\be
	H=\vec{w}_0\cdot \vq \, \mathbb{I} +w_x q_x \sigma_x + w_y q_y \sigma_y
	\label{gw}
	\ee
	which depends on only four real parameters or effective velocities $\vec{w}_0=(w_{0x},w_{0y})$, $w_x$ and $w_y$. The corresponding energy spectrum is $\vep_\alpha=\vec{w}_0\cdot \vec{q}+\alpha \sqrt{w_x^2q_x^2+w_y^2q_y^2}$, where $\alpha=\pm$ is the band index. It has a linear dispersion relation around the Dirac point at $\vec{q}=0$. The cone axis is tilted if $\vec{w}_0\neq 0$. The tilt means that there is a preferred direction of motion given by $\vec{w}_0$. It is similar to performing a boost to a frame of reference moving at the constant velocity $\vec{w}_0$. Even if the tilt is absent, the constant energy contours are not circles but ellipses if $w_x\neq w_y$. This anisotropy is quantified by the dimensionless number $\sqrt{\frac{w_x}{w_y}}$. If $\vec{w}_0\neq 0$, the energy spectrum is no more particle-hole symmetric $\vep_{-}(\vec{q})\neq -\vep_{+}(\vec{q})$ but still has the property $\vep_{-}(\vec{q})=-\vep_{+}(-\vec{q})$. The case of undeformed graphene corresponds to $\vec{w}_0=0$ and $w_x=w_y=v_F$.

A first approach to obtain the Landau levels is to use the semi-classical quantization condition of Onsager and Lifshitz, which we discussed above. The surface enclosed by the cyclotron orbit in reciprocal space is $S(\vep)=\pi \left(\frac{\vep}{v_F^*}\right)^2$, where the effective Fermi velocity $v_F^*$ is defined by
\be
\frac{1}{v_F^{*2}}=
\frac{1}{w_x w_y} \int_0^{2\pi } \frac{d \phi}{2\pi } \frac{1}{(1+\tilde{w}_0 \cos \phi)^2}= \frac{1}{w_x w_y}\frac{1}{(1-\tilde{w}_0^2)^{3/2}}
\ee
and where $\tilde{w}_0\equiv \sqrt{\left(\frac{w_{0x}}{w_x}\right)^2+\left(\frac{w_{0y}}{w_y}\right)^2}$ is the effective tilt parameter. This parameter is assumed to be $\tilde{w}_0<1$ in order for the orbits to be closed. The corresponding density of states (per unit area) is $\rho(\vep)=\frac{|\vep|}{2\pi v_F^{*2}}$, which is identical to that of a single cone in undeformed graphene with the replacement $v_F \to v_F^*=\sqrt{w_x w_y}(1-\tilde{w}_0^2)^{3/4}$. This effective velocity can simply be understood as the geometric mean of the two velocities $\sqrt{w_x w_y}$ corrected by a factor $(1-\tilde{w_0}^2)^{3/4}$ that takes the tilt into account. Later, we will see that this factor can also be understood as resulting from a boost. Now, the semi-classical quantization condition $S(\vep)l_B^2=2\pi(n+\gamma)$ gives the Landau levels $\vep_{\alpha,n}\approx \alpha v_F^*\sqrt{2eB(n+\gamma)}$. As for un-tilted and isotropic Dirac cones, the phase shift $\gamma=\frac{1}{2}-\frac{\Gamma}{2\pi}=0$  as a result of the cancellation between the usual $\frac{1}{2}$ factor and the $\Gamma=\pi$ Berry phase. In the end, the semiclassical LLs are $\vep_{\alpha,n} \approx \alpha v_F^*\sqrt{2eBn}$ when $n\gg 1$, which is the validity condition of the semiclassical approximation \cite{Goerbig2008}. The full quantum solution for $n=0$ confirms that $\vep=0$ is a LL; however, the corresponding eigenstate now has a finite weight on both sublattices \cite{Goerbig2008}. The existence of this zero energy Landau level can be related to a generalized chiral symmetry \cite{HatsugaiPRB2011}.

It is actually possible to compute the Landau levels exactly for all $n$ \cite{Morinari}. We start from the Hamiltonian (\ref{gw}) and perform the Peierls substitution $\vec{q} \to \vec{\Pi}=\vec{q}+e\vec{A}$ with $[\Pi_x,\Pi_y]=-ieB$. Then we introduce the following creation and annihilation operators
\be
a^\dagger=\frac{w_x \Pi_x+i w_y \Pi_y}{\sqrt{2eB w_xw_y}} \textrm{ and } a=\frac{w_x \Pi_x-i w_y \Pi_y}{\sqrt{2eB w_xw_y}} 
\ee
such that $[a,a^\dagger]=1$. This allows us to rewrite the Hamiltonian as
\be
H=\sqrt{2eB w_xw_y}\left(\begin{array}{cc}\frac{\tilde{w}_0}{2}(ae^{i\varphi}+a^\dagger e^{-i\varphi})&a\\a^\dagger &\frac{\tilde{w}_0}{2}(ae^{i\varphi}+a^\dagger e^{-i\varphi}) \end{array} \right)
\label{hbtilt}
\ee
where $\tilde{w}_0e^{i\varphi} \equiv \frac{w_{0x}}{w_x}+i\frac{w_{0y}}{w_y}$. The quantity $\tilde{w}_0$ introduced above is a dimensionless measure of the tilt and $\varphi$ is the angle between the $q_x$ axis and the tilt axis. Such a Hamiltonian can be diagonalized using algebraic methods \cite{PeresCastro}, which are too long to be exposed here. We note however that we will encounter this structure again when discussing the effect of an in-plane electric field in addition to the perpendicular magnetic field. The exact Landau levels are \cite{Morinari}
\be
\vep_{\alpha,n}=\alpha v_F^* \sqrt{2\hbar e Bn} \textrm{ where } v_F^{*}=\sqrt{w_x w_y}(1-\tilde{w_0}^2)^{3/4}
\ee
in agreement with the semiclassical calculation. This is the same LL spectrum as that of undeformed graphene with the replacement $v_F\to v_F^*$. In particular, as already noted, there is a zero-energy LL. The effect of the tilt is to reduce the Fermi velocity and therefore the spacing between Landau levels. In particular if the cones are too tilted there is a collapse of Landau levels when $\tilde{w}_0 \to 1$, which corresponds to the situation where the cyclotron orbits become open and therefore no more quantized. It is easy to get the LL spectrum for the other valley as the Hamiltonian is simply obtained by making $\vec{w}_0,w_x,w_y$ $\to$ $-\vec{w}_0,-w_x,-w_y$ (the two cones are tilted in opposite directions). This shows that the LL spectrum does not depend on the valley index. Therefore, the degeneracy of each LL is $4N_\phi$ when taking valley and spin into account.

\subsubsection{Valley splitting of tilted cones in crossed electric and magnetic fields}
We now make a small parenthesis to discuss the problem of massless Dirac fermions in crossed electric $\mathcal{E}$ and magnetic $B$ fields, which is quite interesting and was investigated by Lukose et al. \cite{Lukose}. It will present an interesting connection to that of LLs of tilted Dirac cones. These authors found that LLs are affected in an unusual way by an in-plane electric field and could eventually lead to their collapse. They obtained the following LL spectrum
\be
\vep_{\alpha,n}=\alpha v_F(1-(v_D/v_F)^2)^{3/4} \sqrt{2 e Bn} - k v_D
\label{lllukose}
\ee
where $v_D\equiv \frac{\mathcal{E}}{B}$ is the drift velocity\footnote{Indeed, the classical motion of a charged particle in crossed electric and magnetic field is helicoidal with a drift velocity given by $\vec{v}_D=\vec{\mathcal{E}}\times \vec{B}/B^2$.}. The effect of an in-plane electric field on LLs is therefore reminiscent of that of a tilt of the Dirac cones in that it induces a downward renormalization of the effective Fermi velocity $v_F\to v_F(1-(v_D/v_F)^2)^{3/4}$. Let us see how this comes about. Starting from graphene's massless Dirac Hamiltonian (in a single valley) in a perpendicular magnetic field $H=v_F[(p_x+eA_x) \sigma_x +(p_y+eA_y) \sigma_y]$, we add an in-plane electric field $\vec{\mathcal{E}}=\mathcal{E}\vec{e}_x$ and take a vector potential in the Landau gauge $\vec{A}=Bx\vec{e}_y$ so that the full Hamiltonian becomes $H=v_F[p_x \sigma_x + (p_y+eBx)\sigma_y]+e\mathcal{E}x\sigma_0$. The virtue of this gauge is that the Hamiltonian still commutes with $p_y$, which is therefore a conserved quantity, which we call $k$ in the following. We now shift the position operator $x+\frac{k}{eB}\to x$ and obtain the following one-dimensional Hamiltonian $H=v_F(p_x\sigma_x+eBx\sigma_y)+e\mathcal{E}x\sigma_0-v_Dk$. Defining the ladder operators
\be
a^\dagger=\frac{p_x+ieBx}{v_F\sqrt{2eB}} \textrm{ and } a=\frac{p_x-ieBx}{v_F\sqrt{2eB}} 
\ee
such that $[a,a^\dagger]=1$, the Hamiltonian can be re-written as
\be
H+v_D k=v_F\sqrt{2eB}\left(\begin{array}{cc}\frac{v_D}{2v_F}(ia-ia^\dagger)&a\\a^\dagger &\frac{v_D}{2v_F}(ia-i a^\dagger) \end{array} \right)
\ee
This is exactly the same form as the Hamiltonian (\ref{hbtilt}) with the replacements $H\to H+v_D k$, $\tilde{w}_0e^{i\varphi}\to \frac{v_D}{v_F}e^{i\pi/2}$ and $\sqrt{w_x w_y} \to v_F$. From the exact solution of such a Hamiltonian \cite{PeresCastro}, we therefore recover the LLs of eq. (\ref{lllukose}). This approach is quite different from that of Lukose et al. They obtained the effect of downward renormalization of the Fermi velocity $v_F \to v_F(1-(v_D/v_F)^2)^{3/4}$ by considerations of ``relativistic'' boost transformations. Indeed, if $v_D<v_F$, it is possible to boost to a reference frame in which the electric field vanishes and the magnetic field is changed from $B$ to $B\sqrt{1-(v_D/v_F)^2}$. In the boosted frame, the LL spectrum is therefore $\vep'=\pm v_F\sqrt{2eBn}(1-(v_D/v_F)^2)^{1/4}$. Now boosting back in the original (laboratory) frame, the energy is changed from $\vep'$ to $\vep'\sqrt{1-(v_D/v_F)^2}-v_D k$ so that in the end $\vep+v_Dk=\pm v_F\sqrt{2eBn}(1-(v_D/v_F)^2)^{3/4}$. The boosted frame moves with a velocity $\vec{v}_D=\frac{\vec{\mathcal{E}}\times \vec{B}}{B^2}=-v_D \vec{u}_y$ compared to the lab frame. This is just the usual drift velocity of a classical electron in crossed electric and magnetic fields. Therefore, in some sense, a tilted cone corresponds to a preferred reference frame moving at velocity $\vec{w}_0$. Note, however, that the drift velocity is the same for both valleys.

\begin{figure}[htb]
\begin{center}
\includegraphics[height=5cm]{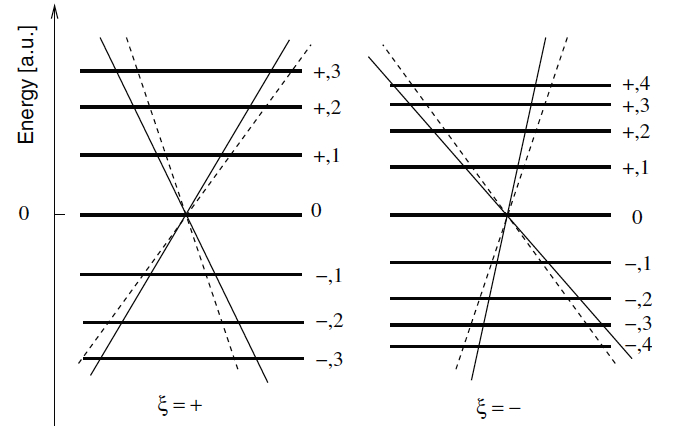}
\caption{\label{fig:epl2009}Sketch of the valley-dependent LL spectrum for tilted Dirac cones in the presence of an electric field (thick lines). We omitted the inclination of the LLs due to the term $v_D k$, which lifts the LL degeneracy. The dashed lines schematically represent the tilted cones in the two valleys ($\xi=\pm$) in the absence of an electric field. The cones in the two valleys are tilted in opposite directions in the momentum-energy space, whereas the electric field acts in the same direction. The LL spectrum in the presence of an electric field in $\xi=+$ (resp. $-$) is that of a cone with a decreased (resp. increased) tilt (full lines). From Goerbig et al. \cite{Goerbig2009}.}
\end{center}
\end{figure}
Consider now a situation with tilted Dirac cones in both a perpendicular magnetic field and an in-plane electric field. It is likely that the drift velocity and the tilt velocity will combine in an interesting fashion as the two effects do not behave the same under valley exchange \cite{Goerbig2009}. The electric field couples to electrons in both valley in the same way such that $\vec{v}_D$ does not depend on $\xi$. However the tilt velocity is not the same in both valleys. It is actually $\xi \vec{w}_0$, where $\xi=\pm$ is the valley index. For simplicity we consider the isotropic case $w_x=w_y=v_F$ although the anisotropic case can be treated \cite{Goerbig2009}. The Hamiltonian reads $H_\xi =\xi[\vec{w}_0\cdot (\vp+e\vec{A}) \sigma_0 +v_F(p_x\sigma_x+(p_y+eBx) \sigma_y)]+e\mathcal{E}x\sigma_0$. The $p_y$ momentum is conserved, $p_y=k$, and after shifting the position operator $x+kl_B^2\to x$ we obtain $H_\xi+v_Dk=\xi[(w_{0x}p_x + w_{0y}eBx)\sigma_0 +v_F(p_x \sigma_x+eBx \sigma_y)]+e\mathcal{E}x\sigma_0$. 
Introducing ladder operators $a$ and $a^\dagger$ such that $\sqrt{2eB}a^\dagger =p_x+ieBx$, it can be rewritten as
\be
H_\xi+v_Dk=v_F\sqrt{2eB}\left(\begin{array}{cc}\frac{\tilde{w}_\xi}{2} (ae^{i\varphi_\xi}+a^\dagger e^{-i\varphi_\xi})&a\\a^\dagger &\frac{\tilde{w}_\xi}{2} (ae^{i\varphi_\xi}+a^\dagger e^{-i\varphi_\xi}) \end{array} \right) \textrm{ where } \vec{w}_\xi \equiv \xi \vec{w}_0 - \vec{v}_D
\ee
with $\tilde{w}_\xi e^{i\varphi_\xi}\equiv \frac{w_{\xi x}}{v_F}+ i\frac{w_{\xi y}}{v_F}=\frac{\xi w_{0x}}{v_F}+i\frac{\xi w_{0y}+ v_D}{v_F}$ the effective tilt parameter, which depends on the valley index. This has the same structure as the Hamiltonian for tilted cones in a perpendicular magnetic field in the absence of an electric field provided $\tilde{w}_0e^{i \varphi}$ is replaced by $\tilde{w}_\xi e^{i\varphi_\xi}$ \cite{Goerbig2009}. The LL spectrum is therefore
\be
\vep_{\alpha,n,\xi} =\alpha v_F \sqrt{2eBn}(1-\tilde{w}_\xi^2)^{3/4}- v_D k
\ee
The most important feature is that the LL spectrum now depends on the valley index. Therefore the valley degeneracy can be lifted and controlled by the application of an in-plane electric field in addition to a perpendicular magnetic field, if the cones are tilted. This is easy to understand as the effective tilt velocity $\vec{w}_\xi=\xi \vec{w}_0- \vec{v}_D$ combines the (valley dependent) tilt velocity $\xi \vec{w}_0$ and the (valley independent) drift velocity $\vec{v}_D$. In particular, it is possible to imagine a situation in which the drift and the tilt velocities are aligned and cooperate in one valley $\vec{w}_-=-\vec{w}_0-\vec{v}_D=-2\vec{w}_0$, while they cancel in the other $\vec{w}_+=\vec{w}_0-\vec{v}_D=0$, see fig.~\ref{fig:epl2009}.
	
\subsubsection{Experiments}
We briefly compare deformed graphene to the organic salt $\alpha$-(BEDT-TTF)$_2$I$_3$. Undeformed graphene has $t\approx 3$~eV and $a\approx 0.14$ nm such that $v_F=\frac{3}{2}ta \approx 10^6$ m/s. In practice, local doping is rarely smaller than $\vep_F\sim 10$~meV due to the presence of inhomogeneities (electron-hole puddles). We imagine deforming graphene with an uniaxial compression such that two of the hopping amplitudes remain equal to $t$ and the third $t'$ is increased. If we call $\varepsilon=\delta a /a$ the strain, we have $t'\approx t(1-2\varepsilon)$ with $0<-\varepsilon \ll 1$. We find that both the anisotropy $\sqrt{\frac{w_x}{w_y}}\approx 1+\varepsilon$ and the tilt $\tilde{w}_0\approx 0.6 \varepsilon$ are small.  In the organic salt, there are four large molecules per unit cell and the four resulting bands are 3/4 filled so that only the two upper bands are considered here. These two bands touch at two Dirac points where the Fermi level is to a very good precision $\vep_F \sim 0.1$ meV. The molecules are much further apart such that the lattice spacing $a\approx 1$ nm and the order of magnitude of the hopping amplitude is $0.05$ eV so that $v_F \sim 10^5$ m/s. However, the anisotropy $\sqrt{\frac{w_x}{w_y}}\sim 3$ and the tilt $\tilde{w}_0\sim 0.3$ are expected to be large. Note that this numbers are rough estimates as, experimentally, it is not so easy distinguishing between the two effects of tilting and velocity anisotropy. Other authors give quite different estimates.

It should therefore be much easier to probe these effects -- such as the tilting-induced LL collapse or the valley degeneracy lifting -- in the organic salts. One could think of magneto-optical transmission spectroscopy, as was done in graphene (see a previous section), to detect the inter-LL transitions. This has not been done yet. Magneto-transport measurements -- Shubnikov-de Haas oscillations and quantum Hall effects -- were very recently performed \cite{Tajima2012}. The main difficulty was in doping the samples, which are naturally undoped, featuring a very small Fermi energy $\lesssim 0.1$~meV. These measurements revealed the expected ``relativistic'' quantum Hall effect due to the presence of a zero-mode, which was also seen as a $\pi$-shift in the phase of the Shubnikov-de Haas oscillations. However this does not directly probe the tilt of the cones. Earlier negative inter-layer magneto-resistance due to the existence of a zero-energy LL was detected \cite{Tajima2009} and compared to a theory neglecting the tilt \cite{Osada}. A theory of this effect including the tilt was developed \cite{Morinari} but the corresponding measurements have not yet been performed. At this point, we can say that massless Dirac fermions have been detected in the organic salts, but the anisotropy and the tilt have not been directly seen experimentally. Ongoing experiments in Orsay show that there is actually another family of charge carriers in this system, which are massive and not of the Dirac type \cite{Monteverde2013}, as stipulated by recent band structure calculations \cite{Alemany}.

\section{``Relativistic'' quantum Hall effects}
\subsection{Integer quantum Hall effect}
An important consequence of the peculiar Landau levels of graphene is the quantum Hall (QH) effect. Graphene has an integer quantum Hall effect that is different from that of the standard (parabolic) two-dimensional electron gases (2DEGs). The relevant difference is that QH states occur at filling factors that are shifted compared to standard 2DEGs. To distinguish it we call it the ``relativistic'' quantum Hall effect, which emphasizes that it is related to the underlying massless Dirac rather than to the Schr\"odinger Hamiltonian. Let us define the LL filling factor $\nu$ as
\be
\nu=\frac{n_c}{n_\phi}=\frac{h n_c}{eB}
\ee
where $n_c$ is the density of charge carriers, which vanishes in undoped graphene, and $n_\phi=\frac{eB}{h}=\frac{N_\phi}{\mathcal{A}}$ is the density of flux tubes, where $\mathcal{A}$ is the total area. Due to the presence of a zero-energy Landau level and of particle-hole symmetry, undoped graphene has $\nu=0$, which corresponds to a $n=0$ Landau level that is half filled, whereas all negative energy LLs are filled and those at positive energy are empty. This is quite peculiar, as in usual 2DEGs, $\nu=0$ corresponds to completely empty LLs. Here the $n=0$ LL is empty when $\nu=-2$, which means that the filling factor appears to be shifted by 2. This shift is related to the presence of a zero-energy LL and can be related to a Berry phase of $\pi$ due to the sublattice pseudo-spin 1/2. Incompressible states -- corresponding to situations where the Fermi energy is in-between LLs -- are expected for filling fractions $\nu=4(n+\frac{1}{2})=4n+2$, where the factor 4 accounts for valley and real spin degeneracies, which are assumed not to be lifted at this point. Consequences in magneto-transport are that when the filling factor is close to $4(n+\frac{1}{2})$, and if translational invariance is broken, there should be a quantized plateau in the Hall resistance $R_H=\frac{h}{e^2}\frac{1}{4n+2}$ and a simultaneous zero in the longitudinal resistivity $R_L\to 0$, see e.g. \cite{Schakel, ZA, GS, Peres}. This behavior is due to the bulk of the system being incompressible (as an insulator) -- there is a bulk gap related to the gap between LLs and to disorder -- while the edges are ideal one-dimensional conductors due to the presence of chiral edge states. See e.g. \cite{DP} for a clear explanation, using different point of views, of the origin of QH plateaus in general.

\begin{figure}[htb]
	\begin{center}
	\subfigure[]{\includegraphics[height=4.2cm]{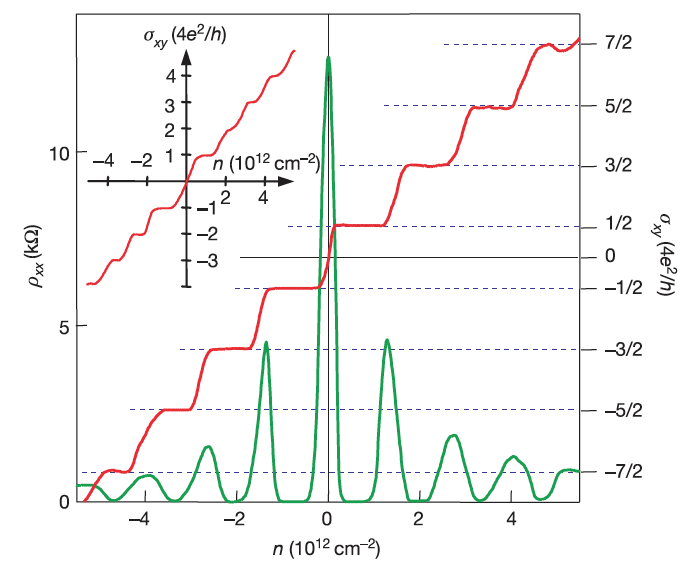}}
	\subfigure[]{\includegraphics[height=4.8cm]{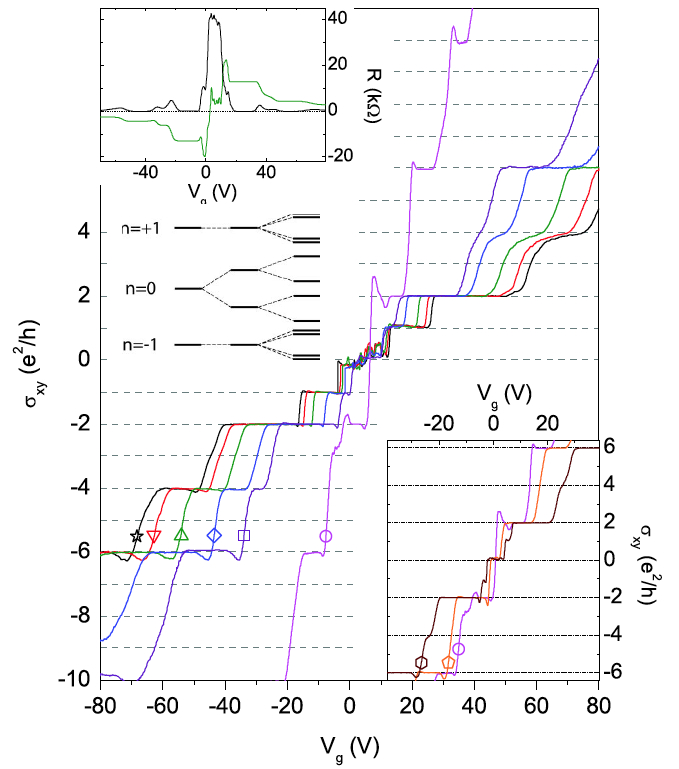}}
	\subfigure[]{\includegraphics[height=4.2cm]{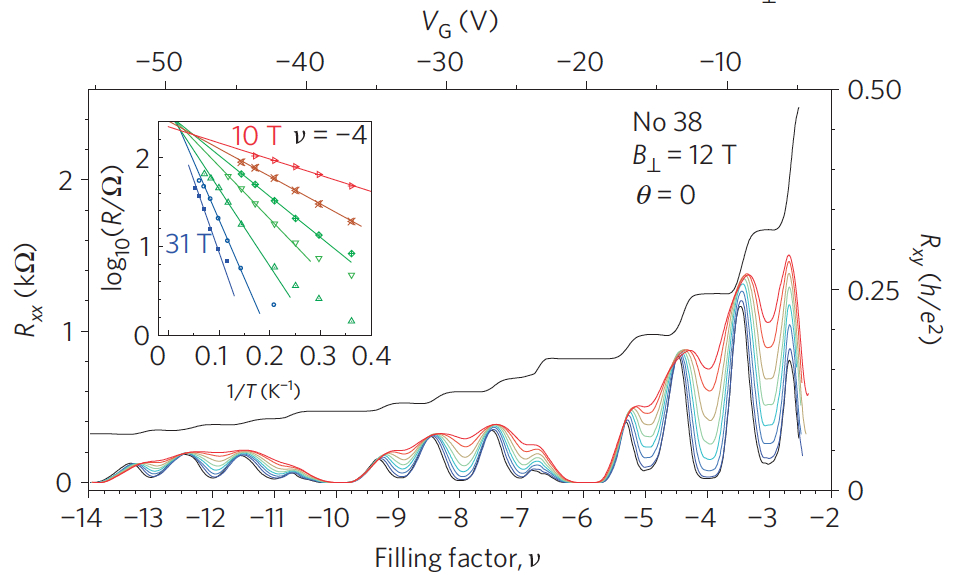}}
	\caption{\label{fig:qhe}Integer quantum Hall effects in graphene. (a) Hall conductivity $\sigma_{xy}$ and longitudinal resistivity $\rho_{xx}$ as a function of the carrier density $n_c$ in a sample with mobility $\mu \sim 10^4$ cm$^2$/V.s at $B=14$~T and $T=4$~K. Plateaux correspond to $\sigma_{xy}=4(n+\frac{1}{2}) \frac{e^2}{h}$ as expected for non-interacting massless Dirac fermions. From Novoselov et al. \cite{Novoselov2005}. (b) Hall conductivity $\sigma_{xy}$ as a function of the backgate voltage $V_g$ in a sample with $\mu < 5.10^4$ cm$^2$/V.s at $T=1.4$~K and in several magnetic fields from $9$ to $45$~T. At larger magnetic fields extra integer plateaus are found at $\sigma_{xy}= (0,\pm 1,\pm 4,...)\frac{e^2}{h}$. From Y. Zhang et al. \cite{Zhang2006}. (c) Hall $R_{xy}$ and longitudinal $R_{xx}$ resistances as a function of the filling factor $\nu$ in a high-mobility sample (graphene on boron nitride) $\mu\sim 10^5$~cm$^2$/V.s at $B=12$~T and for several temperatures between $2$ and $10$~K. At low temperature, every integer quantum Hall plateaus is seen $\nu=n$. From Young et al. \cite{Young2012}.}
	\end{center}
	\end{figure}

These theoretical expectations were fulfilled experimentally in 2005 by two groups working on exfoliated graphene in a Hall bar geometry \cite{Novoselov2005,Zhang2005}, see fig.~\ref{fig:qhe}(a). In these experiments, two knobs were available: the magnetic field $B$ and the carrier density $n_c$, which could be controlled by an electric field effect (as in a capacitor) via a backgate potential $V_g$ so that the filling factor $\nu \propto \frac{V_g}{B}$. Experiments in the quantum Hall regime confirmed the presence of a zero-energy level and the sequence of plateaus separated by $\Delta \nu=4$. In addition, measurement of the amplitude of magnetic oscillations in the longitudinal resistance (Shubnikov-de Haas oscillations) in smaller magnetic fields, enabled experimentalists to extract the cyclotron mass $m_c\equiv \hbar^2 k \frac{\partial k}{\partial \vep}|_F$ as a function of the carrier density $n_c$, which agreed with the expectation $m_c=\frac{\hbar k_F}{v_F}\propto \sqrt{n_c}$ and indirectly confirmed the zero-field linear energy dispersion \cite{Novoselov2005,Zhang2005}. These experiments launched the field of graphene as a new two-dimensional electron gas with massless Dirac fermions as carriers.

\subsection{Interaction-induced integer quantum Hall effect}
Following experiments, in larger magnetic field, revealed extra integer quantum Hall plateaus at $\nu=0;\pm 1;\pm 4;\ldots$ but not at $\nu=\pm 3; \pm 5; \ldots$ \cite{Zhang2006}, see fig.~\ref{fig:qhe}(b).\footnote{The latter have only recently been observed in graphene on boron nitride and via scanning tunneling microscopy in epitaxial graphene. We discuss them below. See also fig.~\ref{fig:qhe}(c).} These correspond to partial lifting of spin or/and valley degeneracy. The latter degeneracies can be thought of as an internal $SU(4)$ symmetry combining the real spin and the valley isospin. Such a symmetry can be broken either explicitly by single-particle effects or spontaneously by interactions. The simplest possibility is the Zeeman effect, which fully lifts the spin degeneracy of Landau levels by $\Delta_Z=g^*\mu_B B$, where $g^*\approx 2$ is the experimentally determined g-factor in graphene and $\mu_B\equiv \frac{e\hbar}{2m_0}$ is the Bohr magneton of the electron, leaving only the valley degeneracy. It can not explain alone the observed sequence of QH plateaus. Indeed, it would predict plateaus at every even integer but not at $\nu=\pm 1$, in contradiction with the experiments. Therefore, one needs to provide a mechanism to lift, at least partially, the valley degeneracy as well. Many such mechanisms have been proposed (for a recent review see \cite{MarkRMP,Yang,Barlas2012}). Two broad classes of mechanisms are: (1) quantum Hall ferromagnetism in which the spontaneous symmetry breaking is due to exchange interaction, see e.g. \cite{NM2006,GDM2006,AF2006,YDM2006} and (2) spontaneous mass generation leading to charge density (CDW), spin density (SDW) or bond density (BDW) waves groundstates. The latter scenario comes in different flavors depending on what is the relevant microscopic interaction: long-range Coulomb interaction (magnetic catalysis of the excitonic instability \cite{Khveshchenko,GMSS}), lattice scale Coulomb interaction (Hubbard on-site $U$ and nearest-neighbor $V$ terms \cite{AF2006, Herbut}) or electron-phonon interaction (either out-of-plane distortion resulting in a CDW order \cite{VAA, FL} or in-plane Kekul\'e distortion leading to a BDW order \cite{VAA, Nomura2009,Mudry}). Some rare scenarios stand apart from this classification as they do not involve interactions: for example, Ref.~\cite{LB} relies on subtle lattices effects in order to break the valley degeneracy, while Ref.~\cite{KA} invoke bond disorder.

\begin{figure}[htb]
	\begin{center}
	\subfigure[]{\includegraphics[height=1.5cm]{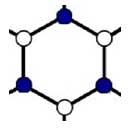}}
	\subfigure[]{\includegraphics[height=1.5cm]{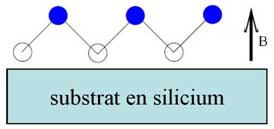}}
	\subfigure[]{\includegraphics[height=4cm]{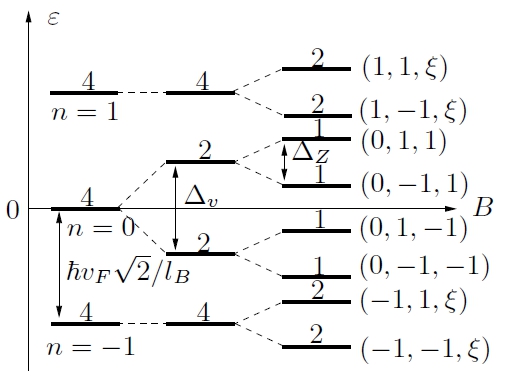}}
	\caption{\label{fig:peierls}Magnetic field induced Peierls instability. The honeycomb lattice spontaneously deforms out of plane with $A$, in blue, (resp. $B$, in white) atoms moving closer to (resp. away from) the silicon substrate. (a) Top view of the honeycomb lattice. (b) Lateral view of the buckled structure with indication of the silicon dioxide substrate and the perpendicular magnetic field. (c) Energy $\vep$ and splittings of the first Landau levels in increasing magnetic field $B$. The degeneracy in units of the flux number $N_\phi$ appears on the levels. The ``cyclotron'' $v_F\sqrt{2eB\hbar}$, valley $\Delta_v$ and Zeeman $\Delta_Z$ gaps are also specified. At large $B$, the levels are tagged by the LL $n$, the spin $s$ and the valley $\xi$ indices: $(n,s,\xi)$. Here, negative Landau indices indicate LLs in the valence band. From Fuchs et al. \cite{FL}.}
	\end{center}
	\end{figure}

We provided a simple explanation in terms of a magnetic field induced Peierls instability, which belongs to the second class of mechanisms and can be seen as resulting from electron-phonon interactions \cite{FL}. It relies on the following observation. In the $n=0$ LL, breaking valley degeneracy is equivalent to breaking sublattice degeneracy, or in other words, breaking inversion symmetry $A\leftrightarrow B$. This is related to the fact that the $n=0$ LL eigenstate for a single valley has the peculiarity of residing only on one of the sublattices (see the above discussion). All other $n \neq 0$ LL have equal weight on both sublattices. Therefore breaking inversion symmetry lifts the valley degeneracy in the $n=0$ LL and not in the others. Together with spin splitting provided by the Zeeman effect, this gives the observed sequence $\nu=2n$ and $\nu=\pm 1$ of QH states \cite{Zhang2006}. In order to break sublattice symmetry, we propose that the graphene honeycomb lattice spontaneously crumbles out-of-plane such that every $A$ atom comes closer to the substrate and every $B$ atom moves away (see fig.~\ref{fig:peierls}). Such a deformation corresponds to a frozen out-of-plane optical phonon, known as a ZO phonon. The presence of the substrate is crucial: it breaks the mirror symmetry with respect to the graphene plane, which in the end results in breaking the $A\leftrightarrow B$ symmetry. Indeed the on-site energies for $A$ and $B$ atoms are now different due to their different environment, i.e. the interaction of a type $A$ atom with the substrate is not the same as that of a type $B$ because the distance of $A$ to the substrate is shorter than that of $B$. This leads to the appearance of a staggered on-site potential equivalent to a Semenoff type mass $m$ for the Dirac fermions \cite{Semenoff}, as in boron nitride. The zero-field Hamiltonian reads
\be
H_\xi=v_F (\xi\sigma_x p_x +  \sigma_y p_y)+mv_F^2\sigma_z
\ee
at low energy, where $2mv_F^2$ is the on-site energy difference between $A$ and $B$ atoms\footnote{Using the basis with $A$ and $B$ exchanged in valley $\xi=-$, the Hamiltonian would become $H_\xi=\xi v_F (\sigma_x p_x +  \sigma_y p_y+mv_F\sigma_z)$.}. The corresponding dispersion relation is $\vep=\pm \sqrt{m^2v_F^4+v_F^2p^2}$. A lattice deformation can spontaneously occur via a Peierls distortion if the cost in elastic energy due to the distortion is balanced by a gain in kinetic energy. For two-dimensional massless Dirac fermions close to zero-doping, such a kinetic energy gain is only substantial in the presence of a strong magnetic field leading to Landau levels. In other words, the Peierls instability does not occur in zero-magnetic field in this system and is catalyzed by the magnetic field. The role of the magnetic field is to increase the density of states at zero energy: in undoped graphene, the density of states at zero magnetic field $\rho(\vep_F=0)= \frac{2k_F}{\pi \hbar v_F}=0$ becomes $\frac{\sqrt{2}}{\pi \hbar v_F l_B}\propto \sqrt{B}$. Landau levels for massive Dirac fermions are given by \cite{Haldane}:
\be
\vep_{\alpha,n}=\alpha \sqrt{m^2v_F^4+2\hbar v_F^2 eBn} \textrm{ when } n\neq 0 \textrm{ and } \vep_{\xi,n}=\xi mv_F^2 \textrm{ when } n=0
\ee
where $\alpha=\pm$ is the band index, and $\xi=\pm $ is the valley index. The important point is that the valley degeneracy is only lifted in the $n=0$ LL, and that for small $m$ the $n\neq 0$ are almost unaffected. Imagine that $A$ atoms move in the vertical direction (perpendicular to the sheet plane) by a distance $-\eta$ and that $B$ atoms move by a distance $+\eta$. Then the total energy change is $\Delta E=-N_\phi (2-|\nu|)m(\eta)v_F^2+N_{uc} G \eta^2$, where the first term is the kinetic energy gain and the second the elastic energy cost of the distortion. In the preceding expression, $N_{uc}$ is the number of unit cells in the sample, $G$ is an elastic constant, $N_\phi=B \mathcal{A}/\phi_0$ is the total number of flux tubes piercing the sample and the mass $m v_F^2=D\eta$ is assumed to be a linear function of the distance, where $D$ is a deformation potential. As the kinetic energy gain is linear in $\eta$ and the elastic energy cost is quadratic, it is always favorable to have a small distortion at non-zero magnetic field. Minimizing the total energy with respect to the distortion distance $\eta$ we find a valley splitting of the $n=0$ LL of: 
\be
\Delta_v=2 mv_F^2 = (2-|\nu|)\frac{N_\phi}{N_{uc}}\frac{D^2}{G}\propto (2-|\nu|)B
\ee
The constants $D$ and $G$ can were estimated in Ref.~\cite{FL}, and we find that $\Delta_v\approx 4$ K $\times B$[T], which is larger than the bare Zeeman splitting $\Delta_Z \approx 1.5$ K $\times B$[T], and an out-of-plane deformation of $\eta \approx 2\times 10^{-5} a \times B$[T], where the magnetic fields are in teslas and the energies in kelvins. 

The consequences of this scenario are as follows. This instability should only be present if the mirror symmetry is explicitly broken by having a different substrate and ``superstrate'', otherwise the electron-ZO phonon coupling is identically zero. Hence it should not occur in suspended samples (gravity and electrostatic coupling to the backgate are mirror symmetry breaking effects, which are too small). The valley gap scales linearly with the magnetic field. There are no gapless edge states (see the discussion below about the $\nu=0$ QH edge states). The resistivity should be very large as in a true insulator. Some of these predictions agree with recent experiments but not all (see below).

\subsection{The $\nu=0$ quantum Hall effect and edge states}
It is worth spending some time discussing the QH state occurring around $\nu=0$ (see e.g. \cite{SarmaYang}). It stands apart among QH states as its Hall conductivity has a peculiar plateau at $\sigma_{xy}=0\times \frac{e^2}{h}$ that is not as well quantized as other QH plateaus. In addition, its longitudinal resistivity does not vanish but is typically $\sim \frac{h}{e^2}$ or larger \cite{Zhang2006}. 

\begin{figure}[htb]
	\begin{center}
	\includegraphics[height=3cm]{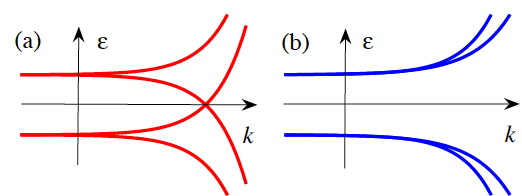}\caption{\label{fig:qhes} Dispersion of the $n=0$ Landau level of graphene in the vicinity of an edge. (a) QH metal: When spin splitting is larger than valley splitting, there are counter-propagating spin filtered edge states near $\nu \sim 0$. (b) QH insulator: When spin splitting is smaller than valley splitting, there are no edge states close to half-filling. From Abanin et al. \cite{Abanin2007}.}
	\end{center}
	\end{figure}

One way of understanding this behavior is to consider QH edge states originating from the $n=0$ LL. This LL is fourfold degenerate and near the edges of the sample, the degeneracy is lifted. Generically, one can mimic the edge potential effect in the Dirac equation by including a position dependent mass term $m(x,y) v_F^2 \sigma_z$ that is zero in the bulk of the sample ($0<x<W$) and grows to infinity at its edges ($x\sim 0$ and $W$). This is a pragmatic way of confining massless Dirac electrons in a finite width $W$ geometry. Such a term implies that edge states in one valley (or one sublattice, as $n=0$) move up in energy, while edge states corresponding to the other valley (or other sublattice) move down in energy. Now two situations arise depending on bulk valley splitting being larger or smaller than bulk spin splitting in the $n=0$ LL \cite{Abanin2007}: (i) First, imagine that valley splitting is larger. Then the edges states do not cross and the Fermi energy corresponding to $\nu \approx 0$ is in a gap both in the bulk and at the edges of the sample (see fig.~\ref{fig:qhes}(b)). In other words, there are no gapless edge states and the sample is a true insulator, known as a QH insulator. In this case, the longitudinal resistivity diverges and the Hall conductivity vanishes as there is no edge conduction. (ii) Second, assume that spin splitting is larger than valley splitting. In this case, among the four edge states ($K,\uparrow$; $K,\downarrow$; $K',\uparrow$; $K',\downarrow$), two crosses with opposite slopes. This means that, when $\nu \approx 0$, there are two counter-propagating edge states residing on the same edge (see fig.~\ref{fig:qhes}(a)). These two edge states also have opposite spin directions. This is known as a helical liquid or spin-filtered chiral edge states \cite{Abanin2006} and is similar to the edge channels of a quantum spin Hall insulator \cite{KM}. In this case, the Hall conductivity is zero because of a compensation between spin channels, while the spin Hall conductivity is quantized in units of $\frac{e}{4\pi}$, and the longitudinal resistance is $\sim \frac{h}{e^2}$ (in the absence of backscattering between adjacent counter-propagating edge states) \cite{Abanin2006}. This state is known as a QH metal \cite{Abanin2007}. 

Recent measurements revealed a magnetic field driven quantum phase transition between a QH metal (at low field) and a QH insulator (at high field) \cite{Ong}. The critical field $B_c$ was found to depend on disorder. In the low field ($B<B_c$, disorder dominated) regime the system is a QH metal with a resistance $\sim \frac{h}{2e^2} \sim 10$ k$\Omega$. Increasing the field, there is a transition toward a QH insulator in the high field (clean) regime, with a divergence of the longitudinal resistivity $\rho$ up to $\sim 40$ M$\Omega$, which is well described by a Kosterlitz-Thouless type of scaling $\rho(B)\propto e^{\#/\sqrt{B_c-B}}$. Such a high-field insulating state has also been observed in suspended graphene samples \cite{Du2009,Bolotin2009}.

\subsection{Quantum Hall $SU(4)$ ferromagnet and anisotropies}
In a previous section, we classified the different internal symmetry breaking mechanisms in two broad classes: quantum Hall ferromagnetism or spontaneous generation of a mass for Dirac fermions. There is another point of view, advocated e.g. in ref.~\cite{MarkRMP, Kharitonov}, which we now briefly present. 

Among the different energy scales involved, two are by far the largest: these are the Coulomb interaction energy $\frac{e^2}{\epsilon_b l_B}$ and the ``cyclotron'' energy $\frac{\hbar v_F}{l_B}$, which are actually of the same order as their ratio is $\frac{e^2}{\epsilon_b \hbar v_F}\sim 1$. If we forget about other smaller energy scales (such as the Zeeman effect) and consider massless Dirac fermion in an perpendicular magnetic field interacting via the long-range Coulomb interaction, the problem has an exact $SU(4)$ internal symmetry (due to the four internal states corresponding to the real spin $1/2$ and the valley isospin $1/2$). In this case, the symmetry is spontaneously broken at the mean field level by the exchange interaction and the ground state is a $SU(4)$ ferromagnet \cite{NM2006,GDM2006,AF2006,YDM2006}. Energetically, all directions in this internal space are equivalent. However, physically some directions correspond to spin ferromagnetism, some other to valley ferromagnetism, or even to spin anti-ferromagnetism. Now, what is the effect on this state of small perturbations such as the Zeeman effect or the electron coupling to optical phonons (either in-plane or out-of-plane), the lattice scale Coulomb interactions (for example described by $U$ and $V$ Hubbard-like terms), etc. Collectively, putting the Zeeman effect aside, these perturbations may be seen as anisotropies for the $SU(4)$ ferromagnet (somewhat similar to easy plane or easy axis anisotropies in usual ferromagnets). In typical magnetic fields these anisotropies (or the Zeeman effect) have much smaller characteristic energies than the Coulomb or cyclotron energies. As an order of magnitude, the ratio between the cyclotron and Zeeman energies is $\sim \frac{5.10^2}{\sqrt{B[\rm T]}}\gg 1$, where the magnetic field is in teslas. In other words, the Coulomb and cyclotron energies win the competition by far to produce a QH ferromagnet. However, much smaller energy scales compete to decide in which direction of the internal $SU(4)$ space does the ferromagnet point. In this picture, most of the spontaneous mass generation mechanisms in the $n=0$ LL lead to ground states -- such as CDW, SDW, Kekul\'e distorsion, spin ferromagnet or canted (spin) anti-ferromagnet \cite{Kharitonov} groundstates, e.g. -- that can be seen as some particular direction for the QH ferromagnet. For example, the CDW corresponds to a valley ferromagnet in the $n=0$ LL. It is not yet clear, which of these groundstates (if any) is realized in experiments and how these change when varying the magnetic field strength and direction, the carrier density or the temperature \cite{Kharitonov}. The complete phase diagram is likely to be quite complicated.

In this picture of an anisotropic $SU(4)$ QH ferromagnet, the reason that not every integer QH plateau is seen at low magnetic field is attributed to disorder \cite{NM2006}, which is an ingredient we have not yet discussed. Indeed, schematically, disorder provides a finite bandwidth for the LLs and, as in the familiar Stoner mechanism of ferromagnetism in a metal, the exchange interaction needs to overcome the cost in ``kinetic energy'' within this bandwidth for ferromagnetism to occur. Therefore, the complete phase diagram also involves the disorder strength, as measured by the mobility of the sample, for example.

\subsection{Experimental status}
We conclude this section on the QH effects in graphene with a short review of the current experimental status. With better samples -- suspended graphene or graphene on hexagonal boron nitride (hBN) -- extra plateaus have recently been measured. Every integer QH state is now being observed \cite{Young2012} and also fractional QH states \cite{Du2009,Bolotin2009,Dean2011}. There is a hierarchy of QH states with increasing magnetic field: first, the integer QH effect with spin and valley degeneracy ($\nu=4n+2$) is observed in ``low'' field (see fig.~\ref{fig:qhe}(a)). It corresponds to massless Dirac fermions with full spin and valley degeneracy. Increasing the magnetic field, interaction-induced QH plateaus appear, first with full spin degeneracy lifting but only partial valley splitting ($\nu=2n$ and $\pm 1$, see fig.~\ref{fig:qhe}(b)). Then upon increasing further the magnetic field (or equivalently improving the mobility of the samples), every integer ($\nu=n$) QH plateau is detected, meaning full spin and valley splitting (see fig.~\ref{fig:qhe}(c)). Finally, fractional QH states are observed with certain fractions such as $\nu=\pm \frac{1}{3}$, $\pm \frac{2}{3}$, $\pm \frac{4}{3}$ or $\pm \frac{7}{3}$ in addition to $\nu=n$ (not shown).

What is the correct explanation for the extra integer QH plateaus? This depends on the magnetic field regime. At moderate magnetic field, not all integer are being observed in agreement with the dynamical generation of a mass or with QH ferromagnetism in disordered samples. At higher field every integer is obtained giving support to QH ferromagnetism. The activation gap above these QH states increases linearly with the magnetic field rather than as a square root. This QH states are observed whether on a substrate or not (suspended samples), which appears to rule out any mechanism related to substrate coupling. It is not yet clear what is the microscopic mechanism behind the field driven QH metal to QH insulator transition seen around $\nu=0$: is it a bulk Kosterlitz-Thouless transition or an edge effect?

\section{Particle-hole excitations of doped graphene}
In this section, we study the particle-hole excitations in doped graphene\footnote{Undoped graphene has its own peculiarities and we do not discuss them here.}, restricting to its low-energy description in terms of massless Dirac fermions\footnote{In addition, we do not consider spin effects for the moment and assume that the two valleys are decoupled. We therefore study a single Dirac cone and merely take a fourfold degeneracy into account when needed.}. In the following, we take $\hbar \equiv 1$ and use $\la=\pm$ rather than $\alpha$ for the band index in this section. 

\subsection{Zero-field particle-hole excitation spectrum}
\begin{figure}[ht]
\begin{center}
\subfigure[]{\includegraphics[width=6cm]{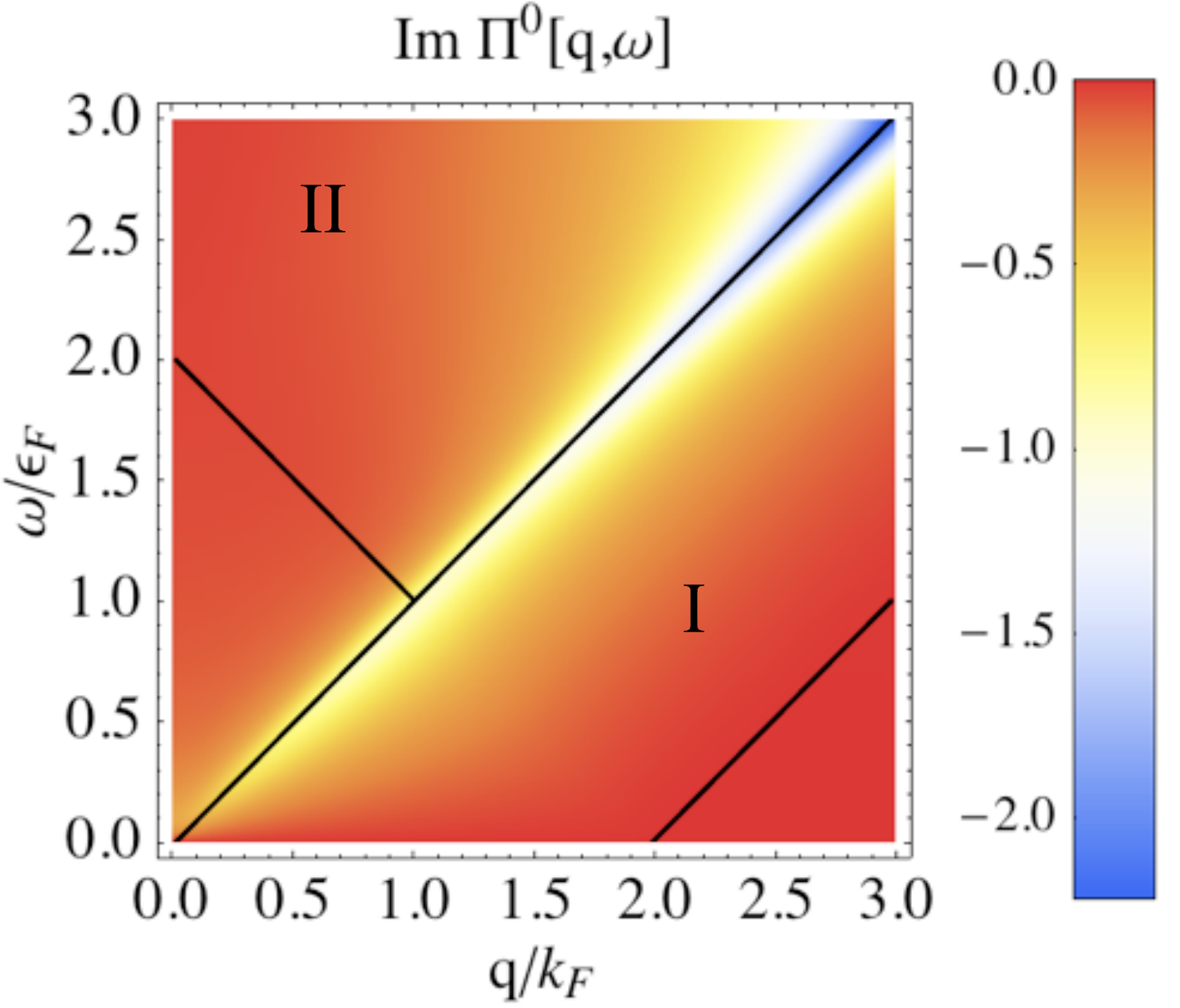}}
\subfigure[]{\includegraphics[width=6cm]{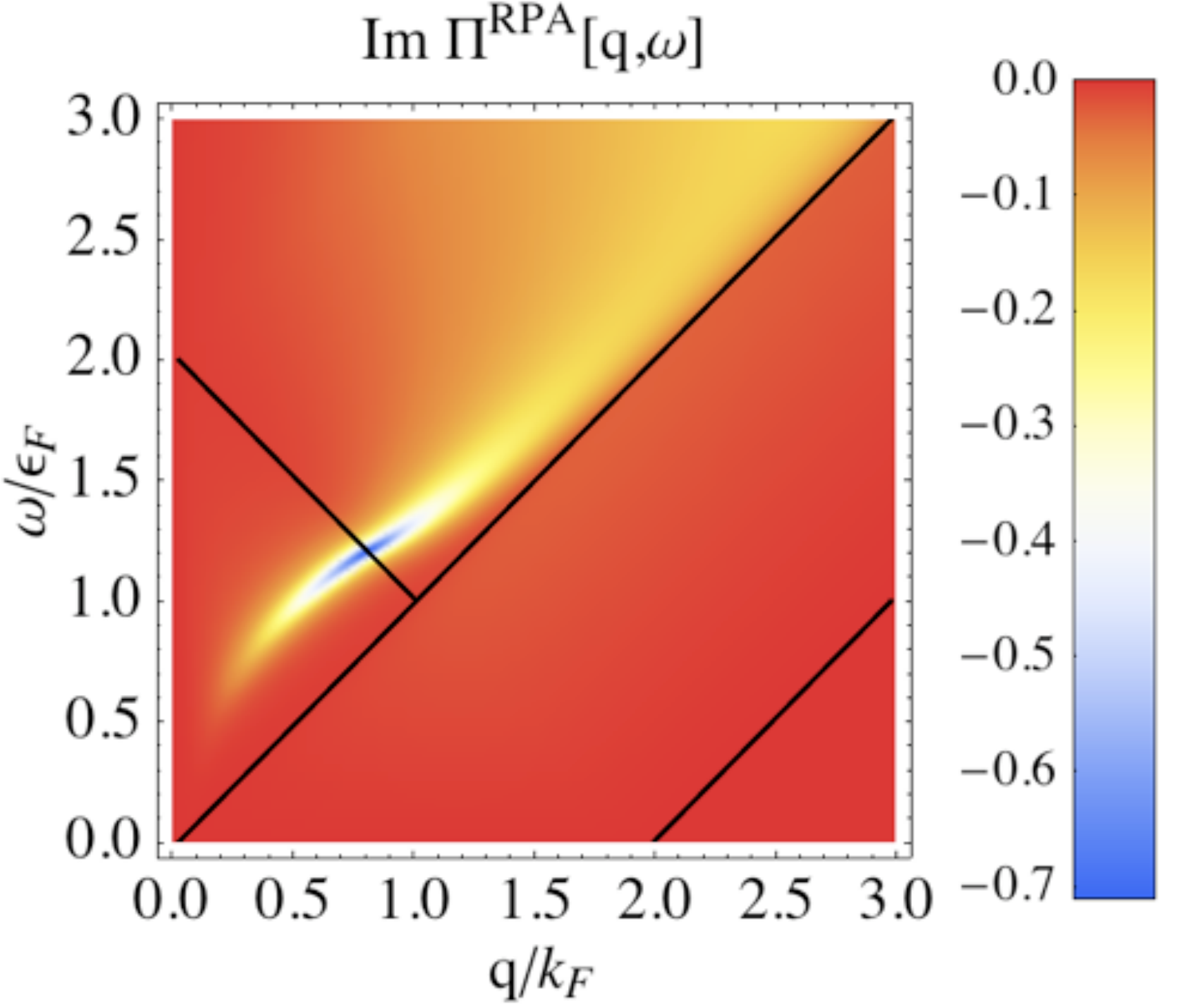}}
\caption{\label{fig:ImPiB0}Particle-hole excitation spectrum of doped graphene in zero field. Color plot of the imaginary part of the polarizability as a function of the energy $\omega$ and the momentum $q$ of a particle-hole pair. The broadening level was taken as $\delta=0.1\varepsilon_F$. (a): non-interacting electrons. Region I (resp. II) corresponds to intra-band (resp. inter-band) processes. Limits of regions are indicated by black lines. Other regions are forbidden. (b): random phase approximation for interacting electrons with $r_s=1$. The plasmon, with its square root dispersion relation in the previously forbidden region, is clearly visible.}
\end{center}
\end{figure}

We first briefly review the zero magnetic field case, which was studied in \cite{Shung1986,Ando2006phes,Wunsch2006,Hwang2007}, and then turn to our work on the finite field case \cite{Roldan2009,Roldan2010a}. A particle-hole excitation is a charge neutral excitation, which consists of removing an electron below the Fermi level at $(\lambda,\vk)$, leaving a hole behind, and promoting it to an empty state above the Fermi level at $(\lambda',\vk')$. In doped graphene, there are two families of such processes $\lambda,\vk \to \lambda',\vk'$. Let us assume that the Fermi level is in the conduction band, such that $\lambda'=+$. Then, the electron $(+,\vk')$ and the hole\footnote{Everywhere we say hole, but actually we mean ``missing electron''. Indeed, a missing electron at $\lambda,\vk$ has an energy, measured with respect to the Fermi energy, $\lambda v_F k-\varepsilon_F$, while the corresponding hole has momentum $\vk_h=-\vk$ and energy $\varepsilon_F-\lambda v_F k$.} $(\lambda,\vk)$ may be either both in the conduction band $\lambda=\lambda'$-- this is known as an intra-band electron-hole pair -- or the hole may be in the valence band $\lambda'=-\lambda$ while the electron is in the conduction band -- this is known as an inter-band electron hole pair. When plotting, the pair excitation energy $\omega=v_F(k'-\lambda k)$ as a function of its momentum $\vq=\vk'-\vk$, one realizes that there is some freedom in the relative momentum between the electron and the hole. This makes the particle-hole excitation spectrum (PHES) a continuum rather than a well defined excitation with a dispersion relation $\omega(q)$. However, the spectral weight in the continuum is not uniform but shows some structure. This weight is measured by the imaginary part of the polarizability $\Pi(\vq,\omega)$. The latter is a density-density response function and may be viewed as the electron-hole pair propagator. Its poles yield the dispersion and damping of collective excitations. The polarizability for non-interacting electrons is \cite{Shung1986,Ando2006phes,Wunsch2006,Hwang2007}
\be
\Pi^0(\vq,\omega)=\frac{4}{\mathcal{A}}\sum_{\vk,\lambda,\lambda'}\frac{\Theta(\xi_{\lambda',\vk+\vq})-\Theta(\xi_{\lambda,\vk})}{\omega+i\delta-(\xi_{\lambda',\vk+\vq}-\xi_{\lambda,\vq})}\times \frac{1+\lambda \lambda' \cos(\phi_{\vk,\vk+\vq})}{2}
\label{eq:pi0}
\ee
where $\xi_{\lambda,\vk}\equiv \lambda v_F k-\varepsilon_F$ is the single-particle energy as measured from the Fermi energy $\varepsilon_F$, $\Theta$ is the Heaviside (zero temperature Fermi-Dirac) step function, $\delta\to 0^+$ is an infinitesimal level broadening, $\phi_{\vk,\vk+\vq}$ is the angle between $\vk$ and $\vk+\vq$ and the factor of $4$ accounts for spin and valley degeneracy. The corresponding PHES, i.e. the regions of non-zero spectral weight $\textrm{Im } \Pi^0(\vq,\omega)$, is plotted in fig. \ref{fig:ImPiB0}(a). It has several peculiarities when compared to that of a standard 2DEG. First, as already mentioned, it is made of two continua: one for intra-band (region I and similar to a single band 2DEG) and one for inter-band processes (region II) separated by $\omega=v_Fq$. Between them is a forbidden region, that hosts no excitations, in the low momentum $q<k_F$ and low energy $\omega < 2\varepsilon_F$ sector. Second, because of the linear dispersion relation of massless Dirac fermions, the edges of the continuum are made of straight lines such as $\omega=v_Fq$, $\omega=v_F(q-2k_F)$ or $\omega=-v_F(q-2k_F)$, in contrast to the curves bounding the PHES of a standard 2DEG. Third, because of the chirality of massless Dirac fermions, the spectral weight in the PHES is strongly concentrated around the diagonal $\omega=v_F q$ \cite{Polini2008}. This is related to the presence of the chirality factor $\frac{1+\lambda \lambda' \cos(\phi_{\vk,\vk+\vq})}{2}$ in the polarizability, see eq.~(\ref{eq:pi0}). The chirality factor is the square of the overlap between the electron and hole bispinors\footnote{Indeed, $|\langle \lambda, \vk|\lambda', \vk' \rangle|^2=\frac{1+\lambda \lambda' \cos(\phi_{\vk,\vk+\vq})}{2}$ as $|\lambda,\vk\rangle=\frac{1}{\sqrt{2}}\left(\begin{array}{c}1\\ \lambda e^{i\phi_{\vk}}\end{array}\right)$ where $k_x+ik_y=ke^{i\phi_{\vk}}$ and $\phi_{\vk,\vk+\vq}\equiv \phi_{\vk+\vq}-\phi_{\vk}$.}. It vanishes for intra-band processes $\lambda=\lambda'$ when $\phi_{\vk,\vk+\vq}=\pi$ (this is known as the absence of backscattering) and for inter-band processes $\lambda'=-\lambda$ when $\phi_{\vk,\vk+\vq} =0$ (which is also the absence of backscattering but in its inter-band version\footnote{Indeed, backscattering is defined as a process in which the removed electron ($\lambda,\vk$) is converted in an electron ($\lambda',\vk'$) moving in the opposite direction $\vec{v}'=-\vec{v}$. However, one should remember that the relation between velocity $\vec{v}$ and momentum $\vk$ depends both on $\vk$ and on the band index $\lambda$: $\vec{v}=\langle v_F \vec{\sigma}\rangle = v_F\lambda \vk/k$. Therefore, for an intra-band process, backscattering means $\phi_{\vk',\vk}=\pi$ but for an inter-band process, it means $\phi_{\vk',\vk}=0$.}). This means that the spectral weight vanishes in region I close to $q=2k_F$ and in region II close to $q=0$.

Let us now include Coulomb interaction between electrons. The 3D Coulomb interaction potential is $V(\vec{r})=\frac{e^2}{\epsilon_b r}$, where $\epsilon_b$ is the background dielectric constant due to the medium surrounding the graphene sheet. Its 2D Fourrier transform is $V(\vq)=\frac{2\pi e^2}{\epsilon_b q}$. A convenient measure of the strength of interactions is provided by the dimensionless parameter $r_s$ which is the ratio of the typical interaction energy between two electrons $\sim \frac{2\pi e^2}{\epsilon_b \lambda_F}$ -- where $\la_F=\frac{2\pi}{k_F}=\frac{2\pi}{\sqrt{\pi n_c}}\sim n_c^{-1/2}$ is the average distance between electrons -- and of their typical kinetic energy $\varepsilon_F=v_Fk_F$. One finds (temporarily reintroducing $\hbar$)
\be
r_s=\frac{e^2}{\epsilon_b \hbar v_F}=\frac{c}{v_F}\frac{\alpha_0}{\epsilon_b}\approx \frac{2.2}{\epsilon_b}
\ee 
where $\alpha_0\equiv \frac{e^2}{\hbar c}\approx \frac{1}{137}$ is the fine structure constant. Because of this connection, $r_s$ is sometimes called graphene's fine structure constant $\alpha_g=\frac{e^2}{\epsilon_b \hbar v_F}$.\footnote{Both notations $\alpha_g$ and $r_s$ exist in the literature and point to different historical contexts. Whereas $\alpha_g$ refers to the fine structure constant of quantum electrodynamics, the notation $r_s$ comes from the dimensionless Wigner-Seitz radius in solid state physics. It describes the average distance between carriers $\sim 1/k_F \sim n_c^{-1/d}$ in units of the effective Bohr radius $a_0^*\equiv m e^2/(\ep_b \hbar^2)$, where $m$ is the band mass. For a linear dispersion relation $\vep=\hbar v_F k$, the concept of mass is ambiguous: the relativistic rest mass $\sqrt{\vep^2-\hbar^2 v_F^2 k^2}/v_F^2$ vanishes; the band mass, i.e. the inverse of the curvature of the dispersion relation $\hbar^2/(\partial^2 \vep / \partial k^2)$ is infinite; and the cyclotron mass $m_c\equiv \hbar^2 k \partial k/\partial \vep=\hbar k/v_F$ is $k$ depend. In order, to recover the correct dimensionless measure of the interaction strength $\alpha_g = r_s = e^2/(\ep_b \hbar v_F)$, one should consider that the effective Bohr radius is given by the cyclotron mass at the Fermi surface $m_c=\hbar k_F/v_F$ so that $a_0^*=m_c e^2/(\ep_b \hbar^2)$ and $r_s=1/(k_F a_0^*)$. Later, we will that this effective Bohr radius is actually the Thomas-Fermi screening radius.} It is typically of order 1 or smaller and can only be tuned by varying the dielectric constant $\epsilon_b$. Indeed, contrary to $r_s=\frac{m e^2}{\epsilon_b \hbar^2k_F}\propto n_c^{-1/2}$ in a standard 2DEG, where $m$ is the band mass, it does not depend on the electronic density $n_c$ (the density of charge carriers, which is zero in undoped graphene). In other words, it is scale invariant: naive dimensional analysis predicts that the Coulomb interaction is marginal for massless Dirac fermions\footnote{Actually, a perturbative RG analysis shows that it is marginally irrelevant and that it flows to zero as $r_s(k)=\frac{r_s}{1-\frac{r_s}{4}\ln(ka)}$ when $ka$ flows to zero in undoped graphene. In doped graphene, the flow is stopped at $k_F$.}.

Adding Coulomb interactions between electrons is easily done in the random phase approximation (RPA). This is an approximation, which amounts to keeping only bubble diagrams in the perturbative expansion of the polarizability and in resuming the geometrical series. It is known to work pretty well for doped graphene, but not so for undoped graphene \cite{Kotov2012}. The RPA polarizability is:
\be 
\Pi^{RPA}(\vq,\omega)= \Pi^0(\vq,\omega)[1-V(\vq)\Pi^0(\vq,\omega)]^{-1}
\label{RPA}
\ee 
Interactions reorganize the PHES by modifying the spectral weight, see fig.~\ref{fig:ImPiB0}(b). Most saliently, a coherent excitation -- with a well defined dispersion relation and almost no damping -- is pushed out of the continuum into the previously forbidden region, by concentrating most of the weight that was in the intra-band region I. This mode is known as the plasmon. In doped graphene, its long-wavelength dispersion relation is $\omega_{pl} \approx \sqrt{2\varepsilon_F e^2 q/\epsilon_b}$. It remains long-lived until it enters the continuum of incoherent particle-hole excitations in region II.

\subsection{Strong field particle-hole excitation spectrum}
We now turn to the case of a strong magnetic field and restrict to the integer quantum Hall regime of completely filled or empty Landau levels. In other words, we do not consider intra-LL excitations that would occur in partially filled LLs -- and which are typical of the fractional quantum Hall regime -- and concentrate on inter-LL excitations such as $\lambda,n \to \lambda',n'$ with $(\lambda,n)\neq(\lambda',n')$. 

The polarizability for noninteracting electrons is given by \cite{Roldan2009,Shizuya2007}
\be
\Pi^0(\vq,\omega)=\sum_{\lambda,\lambda',n,n'}\frac{\Theta(\xi_{\lambda',n})-\Theta(\xi_{\lambda',n'})}{\omega+i\delta-(\xi_{\lambda',n'}-\xi_{\lambda,n})}\mathcal{\overline{F}}^{\lambda \lambda'}_{n n'}(\vq)
\ee
where $\xi_{\lambda,n}\equiv \lambda v_F \sqrt{2eBn}-\varepsilon_F$ is the LL energy measured with respect to the Fermi energy, $\delta$ is the level broadening, and 
\begin{eqnarray}\label{FF}
{\cal
\overline{F}}_{nn^{\prime}}^{\lambda\lambda^{\prime}}(\vq)&=&\frac{e^{-l_B^2q^2/2}}{2\pi l_B^2}\left
( \frac{l_B^2q^2}{2}\right)^{\np-\nm}\left \{
\lambda 1_n^{*}1_{n'}^{*}\sqrt{\frac{(\nm-1)!}{(\np-1)!}}\left
[L_{\nm-1}^{\np-\nm}\left (\frac{l_B^2q^2}{2}\right )\right ]\right.\nonumber \\
&+&\left.\lambda'2_n^{*}2_{n'}^{*}\sqrt{\frac{\nm!}{\np!}}\left [L_{\nm}^{\np-\nm}\left (\frac{l_B^2q^2}{2}\right )\right ]
\right \}^2.
\end{eqnarray}
is the (square of the) form factor, i.e. the equivalent of the chirality factor in the presence of a magnetic field. In the preceding equation we used the short hand notations $1^*_n\equiv \sqrt{(1-\delta_{n,0})/2}$, $2^*_n\equiv \sqrt{(1+\delta_{n,0})/2}$, $n_>\equiv \max(n,n')$, $n_<\equiv \min(n,n')$ and $L_n^m$ are associated Laguerre polynomials. Without loss of generality, let us suppose that the Fermi level is in the conduction band, such that $\lambda'=+$. We call $N_F$ the index of the last completely filled LL, so that the Fermi energy $\vep_F$ satisfies $v_F\sqrt{2eBN_F} <\varepsilon_F< v_F\sqrt{2eB(N_F+1)}$. The filling factor $\nu$ is $\nu=4N_F+2$. Then, the polarizability contains two separate contributions depending on $\lambda$
\be
\Pi^0(\vq,\omega)=\sum_{n=1}^{N_F}\Pi_{+n}(\vq,\omega)+\Pi^{vac}(\vq,\omega)
\ee
and related to intra-band ($\lambda=+$, partially filled conduction band, similar to a metal) and inter-band ($\lambda=-$, vacuum contribution, due to the filled valence band, similar to a dielectric) processes. We defined
\begin{equation}
\Pi_{\la n,\lap n'}(\vq,\omega)\equiv \frac{{\cal
\overline{F}}_{nn^{\prime}}^{\lambda\lambda^{\prime}}(\vq)}{\xi_{\la,n}-\xi_{\lap,n'}
+\omega+i\delta }+(\omega^+\rightarrow-\omega^-)
\end{equation}
where $\omega^+\rightarrow\omega^-$ indicates the replacement $\omega+i\delta\rightarrow-\omega-i\delta$ and
\begin{eqnarray}
\Pi_{\la n}(\vq,\omega)\equiv \sum_{\lap}\sum_{n'=0}^{n-1}\Pi_{\la n,\lap n'}(\vq,\omega)+\sum_{\lap}\sum_{n'=n+1}^{N_c}\Pi_{\la n,\lap n'}(\vq,\omega)+\Pi_{\la n, -\la n}(\vq,\omega)
\end{eqnarray}
which verifies
$\Pi_{\la n}(\vq,\omega)=-\Pi_{-\la n}(\vq,\omega)$. The
vacuum polarization is defined as
\begin{equation}
\Pi^{vac}(\vq,\omega)=-\sum_{n=1}^{N_c}\Pi_{+n}(\vq,\omega)
\end{equation}
where $N_c$ is a cutoff (the index of the last LL). Taking into account that, already in the absence of magnetic field, the validity of the continuum approximation is only up to energies $\sim t$, then $v_F\sqrt{2eBN_c}\sim t$, which leads to $N_c\sim \frac{h/e}{B a^2}\sim 10^4/B[T]$, which is very large even for strong magnetic fields. However, due to the fact that the separation between LL decreases with the index $n$, it is always possible to have good semi-quantitative results from smaller values of $N_c$. We typically use $N_c=70$. The RPA polarizability is obtained as in the zero-field case, using equation (\ref{RPA}). 
\begin{figure}[ht]
\begin{center}
\subfigure[]{\includegraphics[width=6cm]{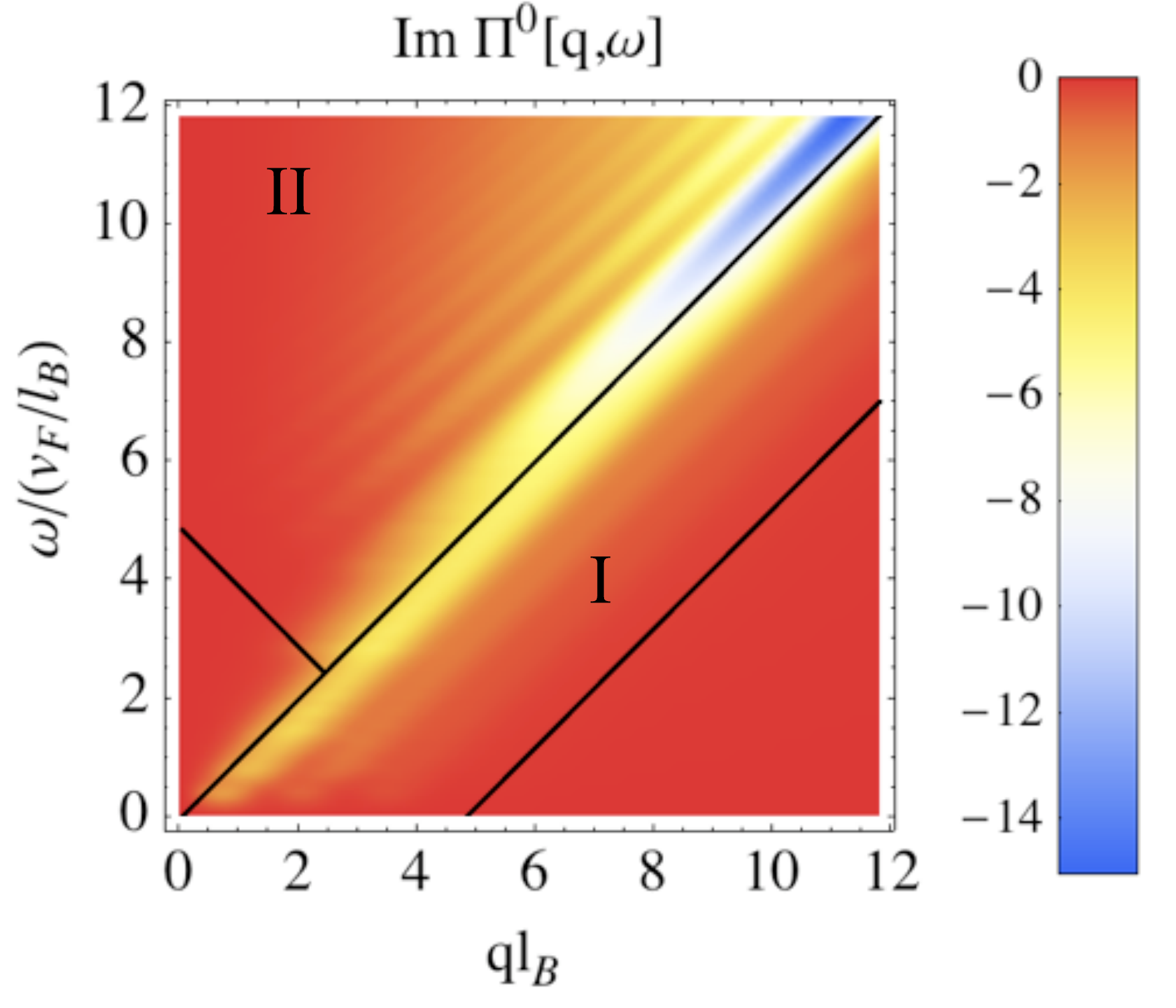}}
\subfigure[]{\includegraphics[width=6cm]{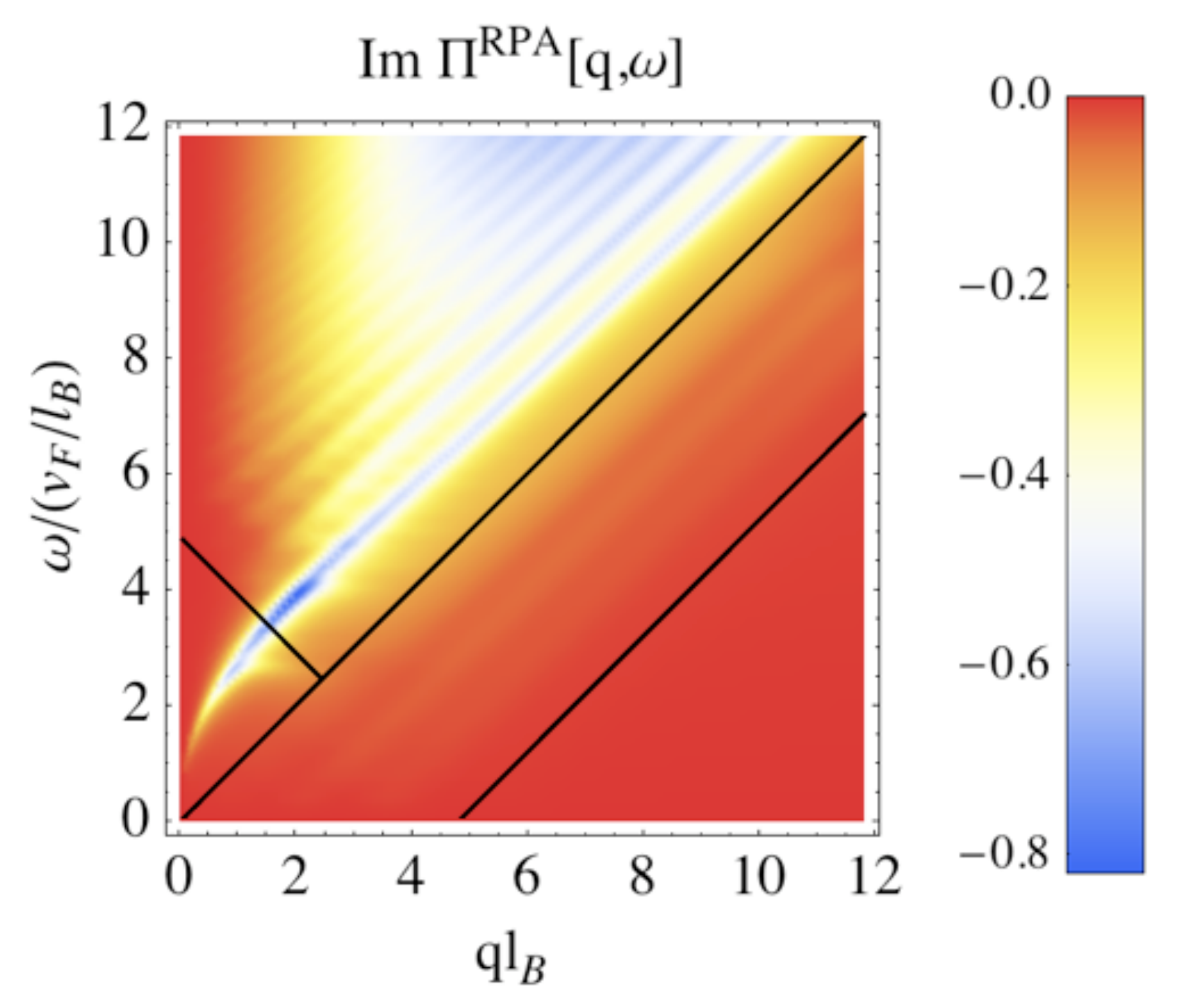}}
\caption{\label{fig:ImPiB}Particle-hole excitation spectrum of doped graphene in strong magnetic field. Color plot of the imaginary part of the polarizability as a function of the energy $\omega$ and the momentum $q$ of a particle-hole pair. The Fermi level corresponds to $N_F=3$, the cutoff is $N_c=70$ and the broadening $\delta=0.2v_F/l_B$. (a): non-interacting electrons. (b): random phase approximation for interacting electrons with $r_s=1$. The upper-hybrid mode is clearly visible, as well as the linear magneto-plasmons.}
\end{center}
\end{figure}

The non-interacting PHES is plotted in fig.~\ref{fig:ImPiB}(a) and the one obtained in the RPA in fig.~\ref{fig:ImPiB}(b). The most salient feature of the non-interacting PHES is that it is composed of diagonal lines parallel to $\omega=v_F q$ and not of horizontal non-dispersing lines as in a standard 2DEG. Once electron-electron interactions are added, these modes acquire coherence. We refer to these clearly defined diagonal lines as {\it linear magneto-plasmons} to distinguish them from almost non-dispersing magneto-excitons \cite{KallinHalperin}, which are found in a standard 2DEG. The magneto-excitons are actually also present but they are much weaker for graphene than for a standard 2DEG \cite{Iyengar2007,Bychkov2008}. In addition to these modes present in the regions I and II, there is an upper-hybrid mode, which is the descendant of the plasmon in the presence of a magnetic field. It is a mixed plasmon-cyclotron mode. It disperses in the forbidden region and is visible in fig.~\ref{fig:ImPiB}(b). Its approximate dispersion relation can be easily obtained in a collisionless hydrodynamic approach and is $\omega_{uh}=\sqrt{\omega_c^2+\omega_{pl}^2}$ where $\omega_c=eBv_F/k_F$ is the cyclotron frequency and $\omega_{pl} \approx \sqrt{2\varepsilon_F e^2 q/\epsilon_b+v_F^2q^2/2}$ is the zero-field plasmon frequency in the long wavelength limit \cite{Roldan2013}. As it will be useful later, we briefly comment on the zero-field dynamic polarizability and the plasmon dispersion in the long wavelength limit (see e.g. \cite{Wunsch2006}):
\be
\Pi^0(\vq\to 0,\omega<2\vep_F)=\frac{q^2}{\pi \omega^2}\left[\vep_F-\frac{\omega}{4}\ln \left|\frac{\omega+2\vep_F}{\omega-2\vep_F}\right|\right]\approx \frac{q^2\vep_F}{\pi \omega^2}\left(1-\frac{\omega^2}{4\vep_F^2}\right)
\textrm{ when }\omega \ll \vep_F
\label{eq:dynpol}
\ee
From the pole of the particle-hole propagator at $1=V(q)\Pi^0(\vq\to 0,\omega\ll \vep_F)$, the zero-field plasmon in the RPA is obtained as $\omega_{pl}^2\approx 2\varepsilon_F e^2 q/\epsilon_b+(3/4-r_s^2)v_F^2q^2$ \cite{Principi2009}, where $r_s\equiv e^2/(\epsilon_b v_F)$.\footnote{The discrepancy in the numerical factor in front of $v_F^2q^2$ in $\omega_{pl}^2$ with the previously cited result ($1/2$ versus $3/4-r_s^2$) comes from the fact that hydrodynamics is only heuristically valid in the collisionless regime, as it neglects the deformation of the Fermi surface and incorrectly finds the first sound velocity ($v_F/\sqrt{2}$) \cite{Roldan2013}.} When discussing the magneto-phonon resonance in the next section, we will encounter the logarithmic structure in equation~(\ref{eq:dynpol}) again.

To finish, we mention a few topics that we worked on and which are not covered here. The particle-hole excitations can also be studied in the presence of the spin degree of freedom and of the Zeeman effect, see Ref.~\cite{Roldan2010b}. This gives rise to spin-flip and spin wave modes in addition to the modes that were discussed up to now. One of the most striking features of these modes and of magneto-excitons in graphene is found in their long wavelength behavior. Indeed, their energy is renormalized by electron-electron interaction here, in contrast to the usual 2DEG with a parabolic dispersion, where such a renormalization is prohibited by Kohn's theorem. We discussed the non-applicability of Kohn's theorem to graphene electrons in Ref.~\cite{Roldan2010b}. Another extension is in the study of the avoided crossings between the upper hybrid mode and the magneto-excitons. These are known as Bernstein modes and we studied them for graphene in \cite{Roldan2011}. 

\subsection{Static screening in doped graphene}
We briefly discuss how interactions between electrons in doped graphene are screened. Photons propagate in 3D and move at velocity $c$ much greater than that $v_F\approx c/300$ of electrons which are restricted to a 2D plane. The electron-electron interaction can then be assumed to be the non-retarded 3D Coulomb interaction in vacuum $V(r)=\frac{e^2}{r}$. Electrostatic field lines pervade the 3D space: above, below and within the graphene sheet. Dielectric screening by the surrounding medium is taken care of by a dielectric constant $\epsilon_b$ which is the average of the substrate and ``superstrate'' dielectric constants, e.g. $\epsilon_b=\frac{1+\epsilon_{SiO2}}{2}\approx 2.5$ if graphene is lying on a silicon dioxide substrate. There may also be some metallic screening by nearby gates -- such as the backgate commonly used for the electric field effect tuning the carrier density --, but we do not consider this possibility here. At this point, without yet considering the screening effects of the graphene sheet, the interaction strength is measured by the dimensionless parameter $r_s=\frac{e^2}{\epsilon_b v_F}\approx \frac{2.2}{\epsilon_b}$.

The graphene sheet itself, although only two dimensional, also screens the Coulomb interaction. In electron doped graphene, there is dielectric screening due to the filled valence band and two-dimensional metallic screening due to the conduction band electrons. To see this, let us introduce the screened Coulomb potential $V_{scr}(\vq)=\frac{V(\vq)}{\epsilon(\vq)}$ and the RPA static dielectric function $\epsilon(\vq)=1-V(\vq)\Pi^0(\vq,\omega=0)$ in terms of the static polarization function. The latter can be shown to be (see e.g. \cite{Wunsch2006}):
\be
\Pi^0(\vq,0)=-\frac{q}{4v_F}-\frac{2k_F}{\pi v_F}(1-\frac{\pi q}{8 k_F})+\Theta(q-2k_F)\left[\frac{k_F}{\pi v_F}\sqrt{1-(\frac{2k_F}{q})^2} - \frac{q}{2\pi v_F}\textrm{Arccos}(\frac{2k_F}{q})\right]
\ee
\begin{figure}[ht]
\begin{center}
\subfigure[]{\includegraphics[width=5.2cm]{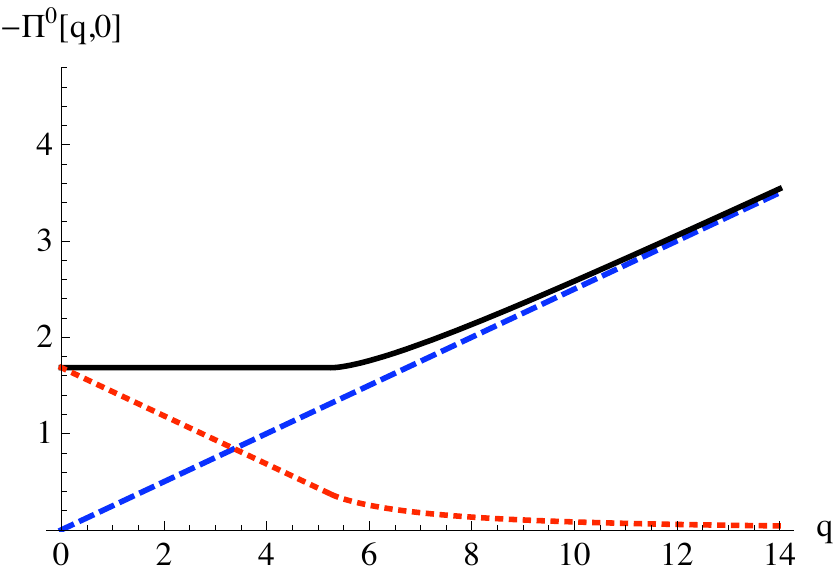}}
\subfigure[]{\includegraphics[width=5.2cm]{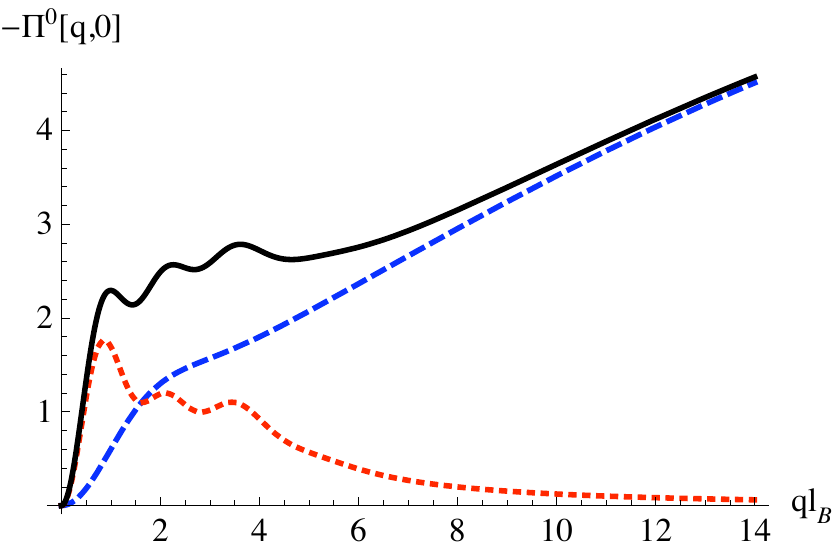}}
\subfigure[]{\includegraphics[width=5.2cm]{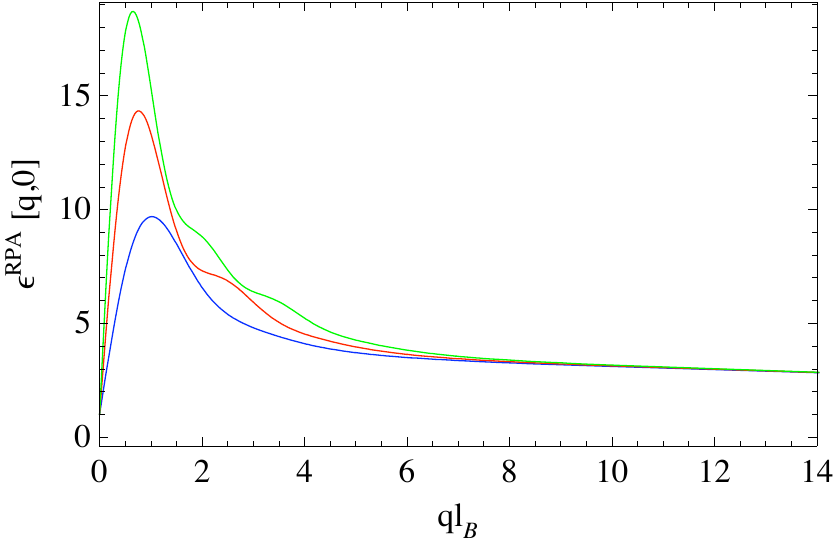}}
\caption{\label{fig:screening}Static polarization function $\Pi^0(\vq,0)=\Pi^{val}+\Pi^{cond}$ for non-interacting electrons in doped graphene (a) without and (b) with a magnetic field. The full black curve is the total polarization function, while the dotted red curve is the conduction band contribution (intra-band processes) and the dashed blue curve is the valence band contribution (inter-band processes). In (b) $N_F=3$ and $N_c=350$ and in (a) $q$ is in units of $k_F/\sqrt{7}\leftrightarrow 1/l_B$, which was chosen for comparison purposes as $k_F\leftrightarrow \sqrt{2N_F+1}/l_B$. (c): Static dielectric function $\epsilon(\vq)=1-V(\vq)\Pi^0(\vq,0)$ computed in the RPA in the presence of a magnetic field and for $r_s=1$. The three curves correspond to $N_F=1$ (blue), $2$ (red) and $3$ (green).}
\end{center}
\end{figure}
It is plotted in fig.~\ref{fig:screening}(a) and is the sum of a contribution from the filled valence band $\Pi^{val}(\vq,0)=-\frac{q}{4v_F}$ (inter-band processes) and from the partially filled conduction band $\Pi^{cond}(\vq,0)=-\frac{2k_F}{\pi v_F}(1-\frac{\pi q}{8 v_F})+\Theta(q-2k_F)\left[...\right]$ (intra-band processes). The corresponding dielectric function shows two different behaviors depending on $q$. At small wavelength $q\leq 2k_F$, $\Pi^0(\vq,0)=-\frac{2k_F}{\pi v_F}=-\rho(\varepsilon_F)$ is a constant (equal to minus the density of states per unit area) and
\be
\epsilon(q\leq 2k_F)=1+\frac{q_{TF}}{q}\geq 1+2r_s
\ee
has the typical Thomas-Fermi form, with inverse screening radius $q_{TF}=4 r_s k_F$.\footnote{To make the connection to a previous footnote, we note that the dimensionless measure of the interaction strength $r_s=q_{TF}/(4k_F)=1/(k_F a_0^*)$ so that the effective Bohr radius $a_0^*=4/q_{TF}$ is essentially the Thomas-Fermi screening radius $1/q_{TF}$.} This is metallic screening, changing the shape of the Coulomb potential into $V_{scr}(\vq)=\frac{2\pi e^2}{\epsilon_b [q+q_{TF}]}$ and avoiding the $\vq=0$ divergence of the bare potential as $V_{scr}(0)=\frac{2\pi e^2}{\epsilon_b q_{TF}}=\frac{1}{\rho(\varepsilon_F)}$ \cite{NM2006,Ando2006phes}. At short distance $q\gg 2k_F$, however, the contribution from the conduction band vanishes and only that of the valence band remains $\Pi^0(\vq,0)\approx -\frac{q}{4v_F}$, giving
\be
\epsilon(q\gg 2k_F)\approx 1+\frac{\pi}{2}r_s\equiv \epsilon_\infty
\ee
which is the behavior of an insulator with dielectric constant $\epsilon_\infty$ \cite{Ando2006phes}. 
An approximate form that captures both limits, but is only qualitatively correct in between, is $\epsilon(q)\approx \epsilon_\infty +\frac{q_{TF}}{q}$. It gives an approximate screened potential
\be
V_{scr}(\vq)\approx \frac{2\pi e^2}{\epsilon^*q+\epsilon_b q_{TF}} 
\ee
where all the dielectric screening (coming from the substrate, the superstrate and the filled valence band) is summarized in a single dielectric constant $\epsilon^*\equiv \epsilon_b \epsilon_\infty =\epsilon_b+\frac{\pi}{2}\frac{e^2}{v_F}\approx \epsilon_b+3.4$ and the metallic screening is taken care of by the screening radius $1/q_{TF}$. Physically, metallic screening can not occur on distances shorter than the average distance $1/k_F$ between mobile carriers. This means that $q_{TF}<k_F$ and therefore that $4r_s<1$ for the validity of this approach (RPA). Good metallic screening $V_{scr}(q)\approx \frac{1}{\rho(\varepsilon_F)}$ is present at long distances $r\sim \frac{1}{q}\gg \frac{1}{q_{TF}}>\frac{1}{k_F}$, whereas dielectric screening $V_{scr}(q)\approx \frac{V(q)}{\epsilon_\infty}=\frac{2\pi e^2}{\epsilon^* q}$ occurs at short distances $r\sim \frac{1}{q} \ll \frac{1}{q_{TF}}$. 

Using our results on the polarizability, we computed the way the magnetic field modifies the static screening, see fig.~\ref{fig:screening}(b) \cite{Roldan2010a}. The main differences with the zero field case are that the static polarizability $\Pi^0$ features oscillations related to the LLs in the conduction band and that it vanishes at small momentum $q\ll 1/l_B$. Indeed, the magnetic field introduces a new length scale $l_B$ in addition to the screening radius $1/q_{TF}$. This vanishing is related to the finite compressibility of the system in the integer quantum Hall regime, due to the gap between the last filled LL $N_F$ and the first empty one $N_F+1$. The dielectric function $\epsilon(q)$ computed in the RPA is shown in fig.~\ref{fig:screening}(c). It can be understood as follows. In the short wavelength limit $q\gg q_{TF}$ it behaves as an insulator with $\epsilon \to \epsilon_\infty=1+\frac{\pi}{2}r_s$ due to the filled valence band. In the long wavelength limit $q\ll 1/l_B$ it almost does not screen as $\epsilon(q)\approx 1+2\sqrt{2}r_sN_F^{3/2}ql_B\to 1$ due to the gap between LLs. However, at intermediate distances $q_{TF} \gg q\gg 1/l_B$ and apart from oscillations, it screens roughly as a metal with a dielectric function $\epsilon(q)\sim \frac{q_{TF}}{q}$, the long wavelength divergence of which is cut at $q\sim 1/(l_B\sqrt{2N_F+1})$.

We conclude this section by discussing the validity of our treatment of interactions. The random phase approximation is usually believed to be valid when (1) bubbles dominate the perturbation expansion, which occurs when the number of fermion flavors $N$ is large -- in graphene, $N=4$ due to spin and valley degeneracy --; (2) the interaction strength is not too large $Nr_s<1$ (because the screening radius should be smaller than the distance between mobile carriers) -- in graphene $Nr_s=8.8/\epsilon_b$ -- and (3) in the long wavelength limit $q<k_F$. However, it is expected to be qualitatively valid even when these conditions are not strictly satisfied \cite{Kotov2012}. In the present context of a strong magnetic field and despite all these restrictions, the RPA has the merit of allowing us to include LL mixing, which seems to be important here as it leads to the appearance of the linear magneto-plasmons instead of the more familiar magneto-excitons \cite{KallinHalperin}. The drawback is that it does not capture short-range $q\gg k_F$ physics such as the exchange hole and the excitonic effects.

\section{Magneto-phonon resonance in graphene}
We now turn to electron-phonon effects in graphene. In-plane optical phonons at the center of the Brillouin zone ($E_{2g}$ phonons at the $\Gamma$ point) are detected by Raman spectroscopy as the $G$-peak at about $0.2$ eV $\approx 1580$ cm$^{-1}$ \cite{Ferrari2006}. These phonons interact with electrons and, in zero magnetic field, the phonon frequency is renormalized by its coupling to electron-hole pairs. This effect is best revealed by the optical phonon frequency dependance as a function of the electron doping, which in graphene can be continuously tuned via an electric field effect. This was predicted in \cite{Ando2006,Lazzeri2006,CastroNetoGuinea2007} and measured in \cite{Pisana2007,Yan2007}. In particular, there is a logarithmic divergence of the phonon frequency -- akin to a Kohn anomaly-- when the phonon frequency (at zero-doping) coincides with $2\varepsilon_F$, which is the threshold for $\vq=0$ inter-band electron-hole pairs in doped graphene \cite{Ando2006,Lazzeri2006,CastroNetoGuinea2007}. 

A perpendicular magnetic field quantizes the orbital motion of electrons into Landau levels, while phonons, which are charge neutral, are not directly affected. The spectrum of electron-hole pairs becomes discrete in a magnetic field -- the so-called magneto-excitons or inter-LL transitions. Now, due to electron-phonon coupling, the dressed optical phonon frequency is expected to oscillate \cite{Ando2007}, with strong renormalization each time the undressed frequency matches that of an inter-LL transition \cite{Goerbig2007}. This effect is known as a magneto-phonon resonance. In the following, we present a peculiar fine structure of this resonance in graphene \cite{Goerbig2007}.

We start from a Hamiltonian made of three parts $H=H_{el}+H_{ph}+H_{c}$. In the following $\hbar\equiv 1$, $\lambda=\pm$ is the band index, and we do not take the electron spin into account (except for the twofold spin degeneracy, but no Zeeman effect). The electron part in the presence of a magnetic field is
\be
H_{el}=\int d^2 r \sum_{\xi} \psi_\xi^\dagger (\vec{r}) [\xi v_F \vec{\sigma}\cdot \vec{\Pi}] \psi_\xi (\vec{r})=\sum_{\lambda,n,m,\xi}\varepsilon_{\lambda,n}c_{\lambda,n,m,\xi}^\dagger c_{\lambda,n,m,\xi}^\dagger
\label{eq:hel}
\ee
where $\psi_\xi(\vec{r})=\sum_{\lambda,n,m}\varphi_{\lambda,n,m,\xi}(\vec{r})c_{\lambda,n,m,\xi}$ is the bispinor annihilation operator of an electron at position $\vec{r}$ in valley $\xi$, $c_{\lambda,n,m,\xi}$ is that for an electron in a Landau level with energy $\varepsilon_{\lambda,n}=\lambda v_F\sqrt{2eBn}$, where $n$ is the Landau index and $m=0,1,...,N_\phi-1$ an additional index related to the guiding center degree of freedom and which accounts for the macroscopic degeneracy of LLs. The corresponding bispinor wavefunction is $\varphi_{\lambda,n,m,\xi}(\vec{r})=\frac{1}{\sqrt{2}}\left(\begin{array}{c}(1-\delta_{n,0})\langle \vec{r}|n-1,m\rangle\\ \sqrt{1+\delta_{n,0}}\lambda \xi \langle \vec{r}|n,m\rangle\end{array} \right)$ where $\langle \vec{r}|n,m\rangle$ is the usual LL wavefunction in the symmetric gauge $\vec{A}=\frac{B}{2}(-y,x,0)$. 

The phonon part (restricted to $\vq=0$, i.e. the $\Gamma$ point) is 
\be
H_{ph}=\omega_{ph} \sum_{\mu}b_{\mu}^\dagger b_{\mu}  
\label{eq:hph}
\ee
where $\vec{u}(\vec{r})=\sum_{\mu}\frac{1}{\sqrt{2N_{uc}M\omega_{ph}}}(b_{\mu}^\dagger + b_{\mu})\vec{e}_{\mu}$ is the relative displacement between the two sublattices and $\vec{e}_{\mu}$ denotes the two possible linear polarizations ($\mu=$ TO or LO) of in-plane optical phonons.  As the two optical phonons are degenerate, instead of working with linear polarization, one can define circular polarizations: $u_\circlearrowleft\equiv (u_x+i u_y)/\sqrt{2}$ and $u_\circlearrowright\equiv u_\circlearrowleft^*$, where the corresponding index is $\mathpzc{A}=\circlearrowleft,\circlearrowright$. 

The coupling between electron and phonons is described by  \cite{IshikawaAndo2006}
\be
H_{c}=g\sqrt{2M\omega_{ph}}\sum_\xi \int d^2 r  \psi_\xi^\dagger (\vec{r}) [\sigma_x u_y - \sigma_y u_x] \psi_\xi (\vec{r})
\label{eq:hc}
\ee
where $g$ is the electron-phonon coupling constant, which can be estimated as $g= -\frac{3}{2}\frac{\partial t}{\partial a}\frac{1}{\sqrt{M\omega_{ph}}}\approx 0.26$ eV with the help of Harrison's law $t\propto \frac{1}{a^2}$.

To proceed, we note that the only non-zero matrix elements of the coupling Hamiltonian involve inter-LL transitions (or magneto-excitons) of the type  $\lambda, n,m,\xi$ $\to$ $\lambda',n\pm 1,m,\xi$. These selection rules are identical to those for magneto-optical absorption \cite{Sadowski}. Now, if we consider electron doping such that the Fermi level is in the conduction band and the last filled LL has index $N_F$, the optical phonons can essentially couple to {\it inter-band} magneto-excitons of energy $\Delta_n\equiv v_F\sqrt{2eB} (\sqrt{n+1}+\sqrt{n})$, where $n>N_F$ \footnote{There is also one possible {\it intra-band} magneto-exciton of energy $\Delta_{+,N_F\to +,N_F+1}=v_F\sqrt{2eB} (\sqrt{N_F+1}-\sqrt{N_F})$ but it plays a minor role as typically its energy is much smaller than that of the phonons $0.2$ eV, apart from the limiting case $N_F=0$, which is taken into account in the above description.}. At fixed energy $\Delta_n$, there are two such inter-band magneto-excitons: $-,n+1 \to +,n$ and $-,n\to +,n+1$. They can be distinguished by the circular polarization of the phonon they couple to. The coupling Hamiltonian shows that $-,n+1 \to +,n$ is coupled to $u_\circlearrowleft$ and $-,n\to +,n+1$ to $u_\circlearrowright$. All this suggests to define the following creation operators for inter-band magneto-excitons:
\begin{eqnarray}
\phi_\circlearrowleft^\dagger(n,\xi)&=&\frac{i}{\sqrt{2N_\phi (\bar{\nu}_{-,n+1}-\bar{\nu}_{+,n})}}\sum_m c^\dagger_{+,n,m,\xi} c_{-,n+1,m,\xi}\\
\phi_\circlearrowright^\dagger(n,\xi)&=&\frac{i}{\sqrt{2N_\phi (\bar{\nu}_{-,n}-\bar{\nu}_{+,n+1})}}\sum_m c^\dagger_{+,n+1,m,\xi} c_{-,n,m,\xi}
\end{eqnarray}
The normalization factor in the preceding equations comes from the requirement of bosonic commutation relations $[\phi_\mathpzc{A} (n,\xi),\phi^\dagger_{\mathpzc{A}'}(n',\xi')]=\delta_{\mathpzc{A},\mathpzc{A}'} \delta_{\xi,\xi'}\delta_{n,n'}$. These commutation relations are obtained in a mean-field approximation with $\langle c^\dagger_{\lambda,n,m,\xi}c_{\lambda',n',m',\xi'}\rangle=\delta_{\lambda,\lambda'}\delta_{n,n'}\delta_{m,m'}\delta_{\xi,\xi'}(\delta_{\lambda,-}+\delta_{\lambda,+}\bar{\nu}_{\lambda,n})$ where $0\leq \bar{\nu}_{\lambda,n}\leq 1$ is the partial filling factor of the $n$th LL, normalized to one. Another interesting feature is that because of a relative sign $\xi$ between the electron Hamiltonian (\ref{eq:hel}) and the coupling Hamiltonian (\ref{eq:hc}), the phonons only couple to the {\it valley-antisymmetric} inter-band magnetoexciton $\phi_{\mathpzc{A},as}(n)\equiv [\phi_{\mathpzc{A}}(n,\xi=+)-\phi_{\mathpzc{A}}(n,\xi=-)]/\sqrt{2}$ and not to the {\it valley-symmetric} inter-band magnetoexciton $\phi_{\mathpzc{A},s}(n)\equiv [\phi_{\mathpzc{A}}(n,\xi=+)+\phi_{\mathpzc{A}}(n,\xi=-)]/\sqrt{2}$. \footnote{In the case of magneto-optical transmission spectroscopy, the selection rules are the same but the coupling is only to the valley-symmetric magneto-excitons. Therefore the avoided crossings between optical phonons and inter-band magneto-excitons can not be detected via optical absorption.} Therefore, the electron-phonon coupling becomes:
\be
H_c=\sum_{\mathpzc{A},n}g_\mathpzc{A}(n)[b_\mathpzc{A}^\dagger\phi_{\mathpzc{A},as}+b_\mathpzc{A}\phi_{\mathpzc{A},as}^\dagger]
\ee
where $g_\circlearrowleft(n)\equiv g\sqrt{(1+\delta_{n,0})\gamma}\sqrt{\bar{\nu}_{-,n+1}-\bar{\nu}_{+,n}}$ and $g_\circlearrowright(n)\equiv g\sqrt{(1+\delta_{n,0})\gamma}\sqrt{\bar{\nu}_{-,n}-\bar{\nu}_{+,n+1}}$ are effective coupling constants with $\gamma\equiv 3\sqrt{3}a^2/(2\pi l_B^2)$. The phonon Hamiltonian is just $H_{ph}=\omega_{ph}\sum_{\mathpzc{A}}b_\mathpzc{A}^\dagger b_\mathpzc{A}$ and the electron one is replaced by its bosonized form in terms of valley-antisymmetric inter-band magneto-excitons
\be
H_{el}\to H_{bo}=\sum_{\mathpzc{A},n}\Delta_n \phi_{\mathpzc{A},as}^\dagger(n)\phi_{\mathpzc{A},as}(n)
\ee
Other magneto-excitons (such as valley-symmetric inter-band magneto-excitons) are simply ignored as they are not coupled to the phonons. The total Hamiltonian we arrive at, $H_{bo}+H_{ph}+H_{c}$, describes the coupling of two types of bosonic modes -- phonons and magneto-excitons -- and is quadratic in these operators. Therefore it can be solved exactly. Before doing so, we summarize a few approximations that were made: we restricted ourselves to zero-momentum, considered only inter-LL excitations and discarded intra-LL transitions, and used a mean-field approximation to bosonize the electronic Hamiltonian. In a diagrammatic language, in the perturbative expansion of the phonon propagator, we only kept the electron-hole bubbles (as in the random phase approximation) but not the vertex corrections or the electron self-energy terms. 

We now compute the dressed phonon propagator $D_\mathpzc{A}(\omega)$ using Dyson's equation $D_\mathpzc{A}(\omega)=D_0(\omega)+D_0(\omega)\Pi_\mathpzc{A}(\omega)D_\mathpzc{A}(\omega)$, where $D_0(\omega)=\frac{2\omega_{ph}}{\omega^2-\omega_{ph}^2}$ is the bare propagator and $\Pi_\mathpzc{A}=\sum_{n=N_F+1}^{N_c}g_\mathpzc{A}^2(n)\frac{2\Delta_n}{\omega^2-\Delta_n^2}$ is the phonon self energy, which is roughly given by the magneto-exciton propagator multiplied twice by the magneto-exciton to phonon coupling $g_\mathpzc{A}$. The integer $N_c$ is a high-energy cutoff defined by $v_F\sqrt{2eB N_c}\sim t$ that takes care of the finite bandwidth -- $N_c$ can be thought-of as the index of the top most LL. The pole of the dressed propagator gives the renormalized phonon frequency $\tilde{\omega}_\mathpzc{A}$ as
\be
\tilde{\omega}_\mathpzc{A}^2=\omega_{ph}^2+2\omega_{ph}\sum_{n=N_F+1}^{N_c}g_\mathpzc{A}^2(n)\frac{2\Delta_n}{\tilde{\omega}_{\mathpzc{A}}^2-\Delta_n^2}
\label{dpd}
\ee
Note that the degeneracy between the two circular polarizations of phonons is generally lifted by the coupling to the two circular polarizations of magneto-excitons. In the following, we study the preceding equation in different limits.

\subsection{Zero magnetic field}
It is important to realize that the bare phonon frequency $\omega_{ph}$ is not the measured one. Indeed, in undoped graphene in zero magnetic field, there is already renormalization of the phonon frequency due to coupling to inter-band magneto-excitons, as the valence band is full of electrons (the vacuum is actually a Dirac sea, a filled valence band). We now define $\tilde{\omega}_0\approx 200$ meV as the phonon frequency measured in undoped graphene $\varepsilon_F=0$ in zero field. The bare phonon frequency $\omega_{ph}$ would correspond to having completely empty $\pi$ bands. To relate these two frequencies, we take the zero field limit of eq. (\ref{dpd}) such that $n\gg 1$, $\sum_n \to \int dn$, $\tilde{\omega}_\mathpzc{A}\to \tilde{\omega}$, $g_\mathpzc{A}\to g\sqrt{\gamma}$, $\Delta_n\approx 2 v_F \sqrt{2eBn}\to 2\varepsilon$ to obtain:
\be
\tilde{\omega}^2-\omega_{ph}^2=8\omega_{ph}\lambda_{ep} \int_{\varepsilon_F}^t d\varepsilon \frac{\varepsilon^2}{\tilde{\omega}^2-4\varepsilon^2}
= 2\omega_{ph}\lambda_{ep}\left[\varepsilon_F-t -\frac{\tilde{\omega}}{4}\ln \left|\frac{\varepsilon_F+\frac{\tilde{\omega}}{2}}{\varepsilon_F-\frac{\tilde{\omega}}{2}}\frac{t-\frac{\tilde{\omega}}{2}}{t+\frac{\tilde{\omega}}{2}}\right|\right]
\label{zf}
\ee
where $\lambda_{ep}\equiv \frac{2}{\pi\sqrt{3}}\frac{g^2}{t^2}\sim 4.10^{-3}$ is the dimensionless electron-phonon coupling constant. In the undoped case we find
\be
\tilde{\omega}_0^2-\omega_{ph}^2= 2\omega_{ph}\lambda_{ep}\left[-t -\frac{\tilde{\omega}_0}{4}\ln \left|\frac{t-\frac{\tilde{\omega}_0}{2}}{t+\frac{\tilde{\omega}_0}{2}}\right|\right]\approx -2\omega_{ph}\lambda_{ep}t
\label{or}
\ee
where in the last step we used that $t\gg \tilde{\omega}_{0}$. This gives the relation between the bare frequency $\omega_{ph}$ and the measured one $\tilde{\omega}_0$ in undoped graphene in zero field. As $\lambda_{ep}\ll 1$, we linearize eq. (\ref{zf}) to obtain
\begin{eqnarray}
\tilde{\omega}&\approx&\tilde{\omega}_0 + \lambda_{ep}\left[\varepsilon_F-\frac{\tilde{\omega}_0}{4}\ln \left|\frac{\tilde{\omega}_0+2\varepsilon_F}{\tilde{\omega}_0-2\varepsilon_F}\right|\right]
 \end{eqnarray}
where $\tilde{\omega}_0\approx \omega_{ph}-\lambda_{ep}t$. Here we recover the result of previous calculations at zero-field  \cite{Ando2006,Lazzeri2006,CastroNetoGuinea2007}, including the logarithmic singularity at $\tilde{\omega}_{0}=2\varepsilon_F$, which was already encountered in the dynamical polarizability in equation~(\ref{eq:dynpol}).

\subsection{Strong magnetic field}
\begin{figure}[ht]
\begin{center}
\subfigure[]{\includegraphics[width=8cm]{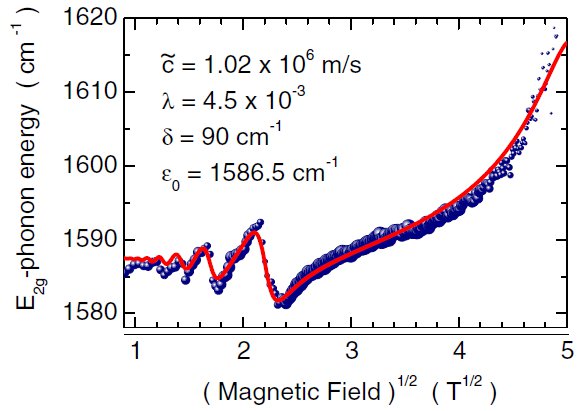}}
\subfigure[]{\includegraphics[width=5cm]{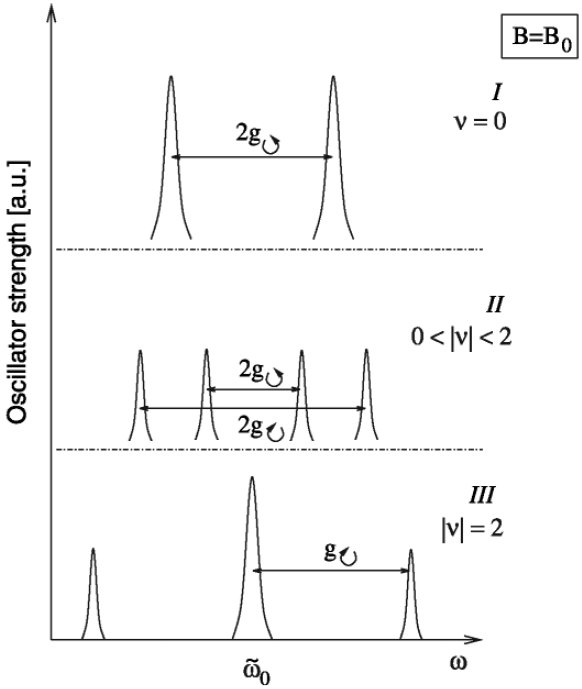}}
\caption{\label{fig:magnetophonon}Magnetophonon resonance in graphene. (a) Measured frequency of the optical phonon at the $\Gamma$ point as a function of the magnetic field. The fit is with eq.~(\ref{eq:fit}) with the following parameters: Fermi velocity $v_F=1.02\times 10^6$~m/s, dimensionless electron-phonon coupling parameter $\lambda_{ep}=4.5\times 10^{-3}$, level broadening $\delta=90$~cm$^{-1}$ and zero-field frequency in undoped graphene $\tilde{\omega}_0=1586.5$~cm$^{-1}$. From Faugeras et al. \cite{Faugeras2009}. (b) On resonance fine structure of the magneto-Raman spectra as a function of the filling factor $\nu$ for the strongest magneto-phonon resonance $B=B_0$. Oscillator strength as a function of the mixed phonon - magneto-exciton frequency. From Goerbig et al. \cite{Goerbig2007}.}
\end{center}
\end{figure}

In a finite magnetic field, the self-consistent equation (\ref{dpd}) can be solved numerically to obtain $\omega_\mathpzc{A}$. This gives rise to magnetic oscillations in the phonon frequency \cite{Ando2007}. For example, in undoped graphene, the dressed frequency does not depend on the circular polarization $\tilde{\omega}_\mathpzc{A}=\tilde{\omega}$ and eliminating $\omega_{ph}$ for $\tilde{\omega}_0$, \footnote{Using $\tilde{\omega}_0^2-\omega_{ph}^2\approx -2\tilde{\omega}_0\lambda_{ep}t\approx -2\tilde{\omega}_0\lambda_{ep}\sum_{n=0}^{N_c}\frac{\Delta_0^2}{\Delta_n}$.} we find
\be
\tilde{\omega}^2-\tilde{\omega}_0^2\approx 2\tilde{\omega}_0 \lambda_{ep}\Delta_0^2\sum_{n=0}^{N_c}\left[\frac{\Delta_n}{\tilde{\omega}^2-\Delta_n^2}+\frac{1}{\Delta_n}\right]
\label{eq:fit}
\ee
to lowest order in $\lambda_{ep}$. This equation was used to fit the oscillations in the phonon frequency measured in magneto-Raman experiments in epitaxial graphene \cite{Faugeras2009}, see fig.~\ref{fig:magnetophonon}(a). 

We now concentrate on the on-resonance situation, when the phonon frequency coincides with that of a particular magneto-exciton $\tilde{\omega}_0\sim \Delta_n$. In that case, we may restrict the sum in (\ref{dpd}) to a single term and eliminate $\omega_{ph}$ for $\tilde{\omega}_0$ using (\ref{or}) to obtain
\be
\tilde{\omega}_\mathpzc{A}^2-\tilde{\omega}_{0}^2\approx 2\tilde{\omega}_{0}\left[\sum_{n=N_F+1}^{N_c}g_\mathpzc{A}^2(n)\frac{2\Delta_n}{\tilde{\omega}_{\mathpzc{A}}^2-\Delta_n^2}+\lambda_{ep}t\right]\approx 2\tilde{\omega}_{0}g_\mathpzc{A}^2(n)\frac{2\Delta_n}{\tilde{\omega}_{\mathpzc{A}}^2-\Delta_n^2}
\ee
at lowest order in $\lambda_{ep}$ or equivalently $g_\mathpzc{A}^2$. This equation describes the coupling between four modes: a particular magneto-exciton of energy $\Delta_n$ and the phonon of energy $\tilde{\omega}_0$, each coming in two circular polarizations $\mathpzc{A}=\circlearrowleft,\circlearrowright$. It can be easily solved to give the frequencies of the mixed magneto-exciton-phonon modes \cite{Goerbig2007}:
\be
\tilde{\omega}_\mathpzc{A}^\pm (n)\approx \frac{\tilde{\omega}_0+\Delta_n}{2}\pm \sqrt{\frac{(\Delta_n-\tilde{\omega}_0)^2}{4}+\tilde{g}_\mathpzc{A}^2(n)}
\ee
As Raman spectroscopy is only affected by the phonon component of the mixed mode, this avoided crossing appears as a magnetic oscillation in the phonon frequency (see fig.\ref{fig:magnetophonon}(a)), rather than a splitting in two or more lines. Actually, sitting exactly on resonance, where $\tilde{\omega}_\mathpzc{A}^\pm (n)\approx \tilde{\omega}_0\pm \tilde{g}_\mathpzc{A}(n)$, there are generically four frequencies when varying the filling factor. For example, in the case of the strongest magneto-phonon resonance $\tilde{\omega}_0\approx \Delta_0$ (corresponding to a magnetic field $B=B_0$ where $B_0\equiv \frac{\tilde{\omega}_0^2}{2ev_F^2}\sim 30$~T), we may play with the electron doping as follows, see fig.~\ref{fig:magnetophonon}(b). At  zero doping, the $n=0$ LL is half-filled, corresponding to a filling factor $\nu=0$, and $g_\mathpzc{A}$ does not depend on the circular polarization so that there are only two different frequencies ($\tilde{\omega}_0\pm g_\circlearrowright=\tilde{\omega}_0\pm g_\circlearrowleft$). Now, doping the system such that $0<\nu<2$, in this case $g_\circlearrowleft>g_\circlearrowright$ and four different frequencies occur ($\tilde{\omega}_0\pm g_\circlearrowright \neq \tilde{\omega}_0\pm g_\circlearrowleft$). When the $n=0$ LL is completely filled ($\nu=2$), one of the circular polarization of magneto-exciton is Pauli blocked so that $g_\circlearrowright=0$ and only three different frequencies occur ($\tilde{\omega}_0$, $\tilde{\omega}_0\pm g_\circlearrowleft$).
	
Following the predictions \cite{Ando2007,Goerbig2007}, several experiments have measured the magneto-phonon resonance in graphene \cite{Faugeras2009} and in graphite \cite{Yan2010,Kossacki2011,Smirnov2012}. The magneto-phonon resonance has also been used as a probe of the band structure of multi-layer graphene samples \cite{Faugeras2012}. Recent experiments detected the circular polarization of optical phonons coupled to magneto-excitons and the resulting fine structure of the Raman $G$ line in graphene \cite{Smirnov2012b}.	
	
\section{Magneto-transport as a probe of impurities limiting the conductivity of doped graphene}
In this section, we summarize work done in collaboration with the experimental group of H. Bouchiat in Orsay and with D. Maslov \cite{Monteverde2010}. The drive is to understand what are the impurities limiting electrical transport in typical graphene samples. The idea is to use the wealth of information that can be extracted from magneto-transport measurements to compare to different impurity models in order to better characterize the relevant impurities.

\subsection{Diffusive and incoherent transport in graphene}
In order to set the stage, we give orders of magnitude for important transport-related quantities. A recent review on transport in graphene can be found in Ref.~\cite{Peres2010}. We consider exfoliated graphene samples on a silicon dioxide substrate. The size of a rectangular sample (of width $W$ and length $L$) is typically $W\sim L\sim 1-5$~$\mu$m and the mobility is $\mu\sim 10^4$~cm$^2$/V.s. The carrier density $n_c$ can be controlled via a backgate voltage $V_g$ according to the capacitor law $e n_c=C_g V_g$, where $C_g$ is the capacitance per unit area. As an order of magnitude $n_c=k_F^2/\pi \sim 10^{12}$~cm$^{-2}$. Close to the neutrality point, the carrier density saturates at a non-zero value $\sim 10^{10}-10^{11}$~cm$^{-2}$ due to inhomogeneities in the sample (known as electron-hole puddles). The Fermi energy $\vep_F=\hbar v_F k_F \sim 100-500$~K is much larger than the temperature $T$ of usual experiments so that the gas of carriers is degenerate. The regime of undoped graphene is therefore hard to access in practice. From the above numbers, we can obtain useful length scales. The transport mean free path $l_{tr}=\hbar k_F \frac{\mu}{e}\sim 0.1$~$\mu$m and the phase coherence length $L_\varphi \sim l \sqrt{\frac{T_F}{T}}\sim 1$~$\mu$m$/\sqrt{T\textrm{[K]}}$ \cite{Morozov2006} (or the thermal diffusion length $L_T=\sqrt{\frac{\hbar D}{k_B T}}\sim 1$~$\mu$m$/\sqrt{T\textrm{[K]}}$ where $D=v_F l_{tr}/2$ \cite{FerryGoodnick}) are both smaller than the sample size so that transport occurs in a diffusive and mostly incoherent regime, described reasonably well by semi-classical kinetic theory. The Fermi wavelength (average distance between carriers) $\la_F\sim 10-50$~nm is smaller than the mean free path, which means that disorder is rather weak. Quantum corrections to transport, such as weak localization or universal conductance fluctuations, are not expected to play a major role except at lower temperature or close to the neutrality point where the conductivity becomes of the order of the conductance quantum. In the following, we stick to semi-classical diffusive transport theory to describe doped graphene.

\subsection{Gate-dependence of the conductivity: scattering time}
Transport measurements in monolayer graphene have shown that the conductivity $\sigma$ varies roughly linearly (or sub linearly) with the carrier density $n_c$ away from the neutrality point and saturates at a non-universal minimal value $\sigma\sim 4e^2/h$ at the neutrality point \cite{Peres2010}. In addition, these features are almost temperature independent from liquid helium up to room temperature. In the incoherent, diffusive and low temperature regime, the conductivity is given by the Drude-Einstein formula $\sigma=e^2 D \rho(\vep_F)$ where $D=v_F l_{tr} /2=v_F^2\tau_{tr}/2$ is the 2D diffusion constant, $\rho(\vep_F)=\frac{2k_F}{\pi\hbar v_F}$ is the density of states (per unit area) at the Fermi level and $k_F=\sqrt{\pi |n_c|}$. The conductivity is therefore
\be
\sigma=2\frac{e^2}{h}k_Fl_{tr}\propto k_F\tau_{tr}(k_F)
\ee
which is $\gg e^2/h$ in the weak disorder limit $k_Fl_{tr}\gg 1$, i.e. not too close to the neutrality point. In order to explain the experimentally found linear dependence $\sigma \propto |n_c|\propto k_F^2$, one has to find a scattering mechanism such that the transport time $\tau_{tr}\propto k_F$. To do so, we first rely on the Born approximation, valid for weak scatterers, and estimate the elastic scattering rate at the Fermi surface\footnote{We assume that the scattering on the impurity does not change spin and that the impurity range is larger than the inter-atomic distance $a$ such that it does not allow inter-valley scattering. In other words, as a result of spin and valley degeneracy, there are four parallel and independent species of massless Dirac fermions.}:
\be
\frac{1}{\tau_e}=\frac{2\pi}{\hbar}N_i \frac{\mathcal{A}\rho(\vep_F)}{4} \int \frac{d\theta}{2\pi}|\langle\vk',\alpha |\hat{V}|\vk,\alpha\rangle|^2
\ee
where $n_i=\frac{N_i}{\mathcal{A}}$ is the impurity density and $\frac{\mathcal{A}\rho(\vep_F)}{4}$ is the density of states at the Fermi level (per spin and per valley). The initial and final states are $|\vk,\alpha\rangle$ and $|\vk',\alpha'\rangle$, where $\alpha=\pm$ is the band index, and $\theta$ is the angle between $\vk'$ and $\vk$. Elastic scattering implies that $k'=k=k_F$, $\alpha'=\alpha$. The transferred momentum is $\vq=\vk'-\vk$ and its norm is $q=2k_F\sin \frac{\theta}{2}$, which is on the order of $k_F$. In the following, we assume a central potential $V(\vec{r})=V(r)$ such that $V(\vq)=V(q)$, where $V(\vq)=\int d^2 r e^{i \vq\cdot \vec{r}}V(\vec{r})$ is the Fourier transform of the potential $V(\vec{r})$ for a single impurity (it has the dimension of energy$\times$length$^2$). For the moment, we do not take the angular dependence of scattering into account and neglect the difference between different angular averages (such as that defining transport $\tau_{tr}$ and elastic scattering $\tau_e$ times to be discussed below) and simply write $\tau\sim \tau_e$. We therefore approximate the angular average involved in the calculation of the scattering rate by $\int \frac{d\theta}{2\pi}|\langle\vk',\alpha |\hat{V}|\vk,\alpha\rangle|^2\sim |V(q\sim k_F)|^2$ so that the order of magnitude of the scattering rate is:
\be
\frac{1}{\tau}\sim \frac{2\pi}{\hbar}n_i |V(q\sim k_F)|^2\frac{\rho(\vep_F)}{4}\propto |V(q\sim k_F)|^2 k_F\, .
\ee
From here, we consider different impurity models and obtain the corresponding dependence of the scattering time $\tau$ on $k_F$.

First, consider weak impurities of range $R$ short compared to $\lambda_F$. This corresponds to $V(q)=$ cst and gives a scattering rate $\tau^{-1}\propto |V(q\sim k_F)|^2 k_F \propto k_F$ that does not account for the experimentally found dependence $\tau \propto k_F$ \cite{Shon1998}. 

Second, charged impurities -- screened or not -- do produce such a dependence \cite{NM2006,Ando2006phes,NM2007}. Indeed, the screened Coulomb potential is $V_{TF}(q)=\frac{2\pi e^2}{\ep_b (q+q_{TF})}$ where $q_{TF}=4 r_s k_F$ is the Thomas-Fermi inverse screening radius $\sim 1/R$ (see the previous section for a discussion of screening in graphene). These are long ranged impurities as $R\sim \frac{1}{q_{TF}}\gg \frac{1}{k_F}$ (physically, the screening radius can not be shorter than the average distance between carriers, which are responsible for screening). Therefore $V(q\sim k_F)\sim \frac{2\pi e^2}{\ep_b k_F}=\frac{4r_s}{\rho(\vep_F)}$ if screening is neglected, or $V(q\sim k_F)\sim \frac{2\pi e^2}{\ep_b q_{TF}}=\frac{1}{\rho(\vep_F)}$ if screening is important. In both cases, as $r_s$ does not depend on carrier density, $V(q)\propto \frac{1}{k_F}$ and $\tau \propto|V(q)|^{-2} k_F^{-1} \propto k_F$ as needed. Although charged impurities do explain the linear dependence of $\sigma$ on $n_c$ and stand as a good explanation for the presence of the electron-hole puddles seen in the vicinity of the charge neutrality point, this scenario is not fully convincing. Indeed, the gate dependence of the conductivity is sometimes found to be sublinear rather than linear (see below). Also the dielectric environment does not seem to affect the conductivity as strongly as it should if charged impurities were the dominant impurities limiting transport \cite{Ponomarenko2009}. 

Third, another possible scenario is that of short range impurities that are strong and induce resonant scattering \cite{Ostrovsky2006,Novikov2007,Stauber2007,Basko2008}. This is, for example, the case of vacancies, voids or adatoms. The Born approximation is no longer valid and one should use the $\hat{T}$ matrix instead of the bare impurity potential $\hat{V}$ when computing the scattering rate: 
\be
\frac{1}{\tau_e}=\frac{2\pi}{\hbar}N_i \frac{\mathcal{A}\rho(\vep_F)}{4} \int \frac{d\theta}{2\pi} |\langle\vk',\alpha |\hat{T}|\vk,\alpha\rangle|^2
\ee
The $\hat{T}$ matrix satisfies $\hat{T}=\hat{V}+\hat{V}\hat{G}_0\hat{T}$ where $\hat{G}_0=\frac{1}{\vep-H+i0^+}$ is the bare Green function of a massless Dirac fermion with Hamiltonian $H=v_F \vp\cdot \vec{\sigma}$ (no spin or valley degeneracy for simplicity\footnote{There is a subtlety here when discussing a single-site vacancy. This is a very short range impurity and inter-valley scattering is possible, contrary to what we assumed. Therefore, one should take the two valleys into account. This means that the density of final states is twice larger than what we assumed. But as the $\hat{T}$ matrix element is found to be inversely proportional to this density of states, the scattering rate is actually half what we obtain. Here, we are only interested in the dependence on $k_F$ and such factors of $2$ are inessential for our purpose.}). To investigate the case of strong scatterers of short range, we consider a simple model for an extreme case, namely a single-site vacancy \cite{Basko2008}. A missing atom on the lattice site $\vec{r}=0,l=A$,e.g., where $l$ is the sublattice index, corresponds to a localized impurity potential $\hat{V}=V_0\delta(\vec{r}) \frac{\sigma_0+\sigma_z}{2}$ in the limit $V_0\to \infty$. Its range $R\lesssim a$. It is important to note that the $2\times 2$ matrix structure of the impurity potential is no longer $\sigma_0=\mathbb{I}_{2\times 2}$ but $\frac{\sigma_0+\sigma_z}{2}$: this will play a role when discussing the ``chirality factor''. The relevant $\hat{T}$ matrix element is easily found $\langle\vk',\alpha |\hat{T}|\vk,\alpha\rangle=\frac{1}{2\mathcal{A}}\frac{1}{\frac{1}{V_0}-\langle 0,A|G_0|0,A\rangle}$ as a function of the local Green function on the impurity site $\langle \vec{r}=0,l=A|G_0|0,A\rangle$. The latter is found to be $\langle 0,A|G_0|0,A\rangle=\int \frac{d^2 k}{(2\pi)^2}\frac{\vep}{(\vep+i0^+)^2-\hbar^2v_F^2k^2}\approx \frac{\rho(\vep_F)}{4}\left[\ln(\frac{|\vep_F|}{\hbar v_F/a})-i\frac{\pi}{2} \right]$, where $\hbar v_F/a \sim t$ is a UV cutoff corresponding to the bandwidth. In the $V_0\to \infty$ limit, $\langle\vk',\alpha |\hat{T}|\vk,\alpha\rangle\approx -\frac{1}{2\mathcal{A} \langle 0,A|G_0|0,A\rangle}$ and we find
\be
\tau\propto k_F \left(\ln^2(k_Fa)+\frac{\pi^2}{4} \right)\sim k_F\ln^2(k_Fa)\label{tau}
\ee
as $k_Fa\ll 1$. The case of strong impurities of range $R$ larger than $a$ but still smaller than $\la_F$ can also be computed \cite{Novikov2007} and gives a scattering time 
\be 
\tau\propto k_F \left(\ln^2(k_FR)+\frac{\pi^2}{4} \right)\sim k_F\ln^2(k_FR)\label{taubis}
\ee 
as $k_F R\ll 1$.
Both types of resonant scatterers ($R\lesssim a \ll \lambda_F$ or $a\ll R\ll \la_F$) give a sublinear conductivity compatible with the experiments. 
We now make a small parenthesis to comment on the unitary limit. This limit result in the strongest possible resonant scattering. It gives  the maximum scattering cross-section and corresponds to keeping only the term $\frac{\pi^2}{4}$ in eq.~(\ref{tau}) or (\ref{taubis}). It is only reached upon fine tuning the Fermi energy such that $\frac{1}{V_0}=\frac{\rho(\vep_F)}{4}\ln(\frac{|\vep_F|}{\hbar v_F/a})$ in the denominator of the $\hat{T}$ matrix element, leaving only the $i\frac{\pi}{2}$ factor. It gives $\tau \propto k_F$, but only for fine-tuned values of $k_F$. As this can not be valid for a range of $k_F$, it does not account for the experimental behavior. 

In summary, either charged long-range impurities or strong short range impurities (resonant but not unitary scatterers and of range $R\ll \la_F$ but larger or smaller than $a$) could be responsible for the experimentally found linear ($\sigma\propto |n_c|$) or sub-linear ($\sigma\propto |n_c|\ln^2(k_F R)$) behavior of the conductivity as a function of the gate voltage. The measurement of the ratio between the transport and elastic scattering times will allow us to decide.

\subsection{Transport versus elastic scattering times}
We now focus on the dependence of the transport $\tau_{tr}$ and the elastic scattering $\tau_e$ times as a function of the Fermi wavevector $k_F$ as a way of obtaining the impurity range $R$. On the one hand, the elastic scattering time (sometimes called the quantum lifetime) is the lifetime of a plane wave state for an electron. It can be interpreted as the mean time between elastic scattering events. The elastic scattering rate is
\be
\frac{1}{\tau_e}=\int_0^{2\pi} \frac{d\theta}{2\pi} W(\theta)=\langle W(\theta) \rangle
\ee
in terms of the probability $W(\theta)$ to be scattered at an angle $\theta$. The rate $\tau_e^{-1}$ is simply the flat average of the angular probability $W(\theta)$. On the other hand, the transport time is a weighted average of scattering events that favors those that most relax the electrical current. Qualitatively, it is the typical time needed for an electron to be backscattered, which may require several scattering events, so that $\tau_{tr}\geq \tau_e$. The transport scattering rate is:
\be
\frac{1}{\tau_{tr}}=\int_0^{2\pi} \frac{d\theta}{2\pi} (1-\cos \theta)W(\theta)=\langle (1-\cos\theta)W(\theta) \rangle
\ee
The $1-\cos\theta=2\sin^2\frac{\theta}{2}$ gives a high/low weight to backward/forward scattering. Its definition comes from the relaxation time approximation in the Boltzmann equation \cite{AbrikosovBook}. 

The difference between these two scattering times comes from the angular dependence of the scattering probability $W(\theta)$, which in the Born approximation is $\propto |\langle\vk',\alpha|\hat{V}|\vk,\alpha\rangle|^2$. Consider an impurity potential $\hat{V}=V(\vec{r})\hat{M}$, where $\hat{M}$ is a $2\times 2$ hermitian matrix in sublattice ($A,B$) sub-space, which is typically either $\sigma_0$ -- if the range of the impurity is much larger than $a$ so that $A$ and $B$ sublattices are equally affected -- or $\frac{\sigma_0\pm \sigma_z}{2}$ -- if the impurity is really localized on an $A$ or $B$ site and has a very short range $\lesssim a$ \cite{Shon1998}. We call $R$ the range of $V(\vec{r})$. If this range is much larger than $a$, then inter-valley scattering is suppressed as the $K$ and $K'$ valleys are too far apart $\sim 1/a$ in reciprocal space \cite{Shon1998}. 
The matrix element $|\langle\vk',\alpha|\hat{V}|\vk,\alpha\rangle|^2=\mathcal{A}^{-2}|V(q)|^2 |\langle\vk',\alpha|\hat{M}|\vk,\alpha\rangle|^2$ contains two factors with an angular dependence. The first $|V(q=2k_F\sin\frac{\theta}{2})|^2$ is the familiar Fourier transform of the potential, which contains an angular dependence via $q$ only if $q\sim k_F \gg 1/R$. Indeed, the opposite limit $k_F R\ll 1$ corresponds to isotropic low energy scattering, where $V(q)\approx $ cst and there is no angular dependence coming from $|V(q=2k_F\sin\frac{\theta}{2})|^2$. The second factor $|\langle\vk',\alpha|\hat{M}|\vk,\alpha\rangle|^2$ is related to the bi-spinor structure of the incoming and outgoing states $|\vk,\alpha\rangle=\frac{1}{\sqrt{2}}\left(\begin{array}{c}1\\ \alpha e^{i\phi_{\vk}} \end{array}\right)$, where $k_x+ik_y=ke^{i\phi_{\vk}}$, and is therefore peculiar two massless Dirac fermions. In the case of $\hat{V}\propto \sigma_0$, it gives the square of the overlap of two bi-spinors, which is known as the chirality factor $|\langle\vk',\alpha|\vk,\alpha\rangle|^2=\cos^2\frac{\theta}{2}$, where $\theta=\phi_{\vk'}-\phi_{\vk}$ is the angle between $\vk'$ and $\vk$. For $\hat{V}\propto \frac{\sigma_0\pm \sigma_z}{2}$, it gives a constant $|\langle\vk',\alpha|\hat{M}|\vk,\alpha\rangle|^2=1/4$. This second factor therefore only gives an angular dependence if $R\gg a$. For example, in the Born approximation for an impurity of range $R\gg a$ such that inter-valley scattering can be neglected, the rate of collisions at angle $\theta$ is
\be
W(\theta)=\frac{2\pi}{\hbar}N_i\frac{\mathcal{A}\rho(\vep_F)}{4}|\langle\vk',\alpha|\hat{V}|\vk,\alpha\rangle|^2=\frac{2\pi}{\hbar}n_i\frac{\rho(\vep_F)}{4}|V(q=2k_F\sin\frac{\theta}{2})|^2\cos^2\frac{\theta}{2}
\ee
where $\hat{V}=V(\vec{r})\sigma_0$, such that the angular dependence is contained in $W(\theta)\propto |V(2k_F\sin \frac{\theta}{2})|^2\cos^2\frac{\theta}{2}$.

The ratio between the transport and the elastic scattering times gives information on the range $R$ of the impurities. As the Fermi wavelength $\la_F$ is much larger than atomic spacing $a$, depending on $R$ compared to these two other length scales, three regimes need to be considered: 

(1) If $R\gg \la_F\gg a$, which is typically the case of Coulomb scatterers, the dominant effect is the divergence of $W(\theta)$ when $\theta \to 0$, i.e. in the forward direction. Indeed, $V(q=2k_F\sin\frac{\theta}{2})=\frac{2\pi e^2}{\ep_b q}\propto \frac{1}{\sin(\theta/2)}$ and $W(\theta)\propto \frac{\cos^2(\theta/2)}{\sin^2(\theta/2)}$. The transport time $\propto \int d\theta (1-\cos\theta)W(\theta)$ is finite due to the cancellation between the $(1-\cos \theta)$ weight and the $|V(q)|^2\sim1/q^2$ factor, but the inverse elastic scattering time $\propto \int d\theta W(\theta)$ diverges at $\theta \to 0$ because of $W(\theta)\sim \theta^{-2}$. This divergence in forward scattering is cut at $\theta_c\sim \frac{1}{k_FR}$ where $R$ is a long wavelength cutoff such as the screening radius, the distance to a screening gate, or the size of the sample, i.e. anything that may play the role of a potential range for a potential of infinite range. Then $\int d\theta W(\theta)\sim \theta_c^{-1} \sim k_F R$ and $\tau_{tr}/\tau_e\sim k_F R\gg1$. Physically, most collisions occur in the forward direction and are therefore inefficient to relax the flow of electrons so that $\tau_{tr}\gg \tau_e$. 

(2) In the intermediate range $a\ll R \ll \lambda_F$, the main effect comes from the chirality factor and the scattering probability $W(\theta)\propto \cos^2\frac{\theta}{2}$. Inter-valley scattering is suppressed because $R\gg a$ and intra-valley backscattering is forbidden by the chirality factor $\cos^2\frac{\pi}{2}=0$. Therefore $\tau_{tr}/\tau_e=\int d\theta \cos^2\frac{\theta}{2}/
\int d\theta \cos^2\frac{\theta}{2}(1-\cos\theta)=2$. 

(3) If the range is very short $R\leq a\ll \la_F$, inter-valley scattering is possible and there is no chirality factor, as $|\langle\vk',\alpha|\hat{M}|\vk,\alpha\rangle|^2=1/4$, as we have seen above. This means that $W(\theta)\approx $ cst is structureless and therefore $\tau_{tr}/\tau_e=\int d\theta/\int d\theta (1-\cos\theta)=1$. 

In summary, the experimental dependence of $\tau_{tr}/\tau_e$ on $k_F$ reveals the range of the relevant impurities. Three well identified limits are $\tau_{tr}/\tau_e=1$, $2$ or $k_FR\gg 1$.

\subsection{Magneto-transport experiments}
In order to obatin $\tau_{tr}$ and $\tau_e$, magneto-transport experiments were performed in Orsay as a function of magnetic field, temperature and carrier density (or more precisely, backgate voltage) for several gated graphene samples (monolayer, bilayer, two-terminal or Hall bar geometry) obtained by mechanical exfoliation on silicon dioxide \cite{Monteverde2010}. Samples have a typical rectangular shape of area $W\times L\sim 1-10$~$\mu$m$^2$ and a mobility $\mu \sim 3000-6000$~cm$^2$/V.s. The magnetic field is typically $B\sim 1$~T, the temperature is around that of liquid helium $T\sim 4$~K and the carrier density is $n_c\sim 10^{12}$~cm$^{-2}$. In the presence of a magnetic field, the density of states acquires oscillations due to the underlying presence of temperature smeared and disorder broadened Landau levels. As a consequence, many physical quantities -- such as resistance or magnetization -- acquire magnetic oscillations. In Orsay, the resistivity tensor in the presence of a magnetic field was obtained from the measurement of resistances. Keeping only the first harmonic of the quantum oscillations, the longitudinal and transverse (or Hall) resistivity are \cite{Coleridge}:
\begin{eqnarray}
\rho_{xx}&= &\rho_0+\delta \rho_{xx}=\rho_0[1+A\cos(\frac{\pi\vep_F}{\hbar\omega_c})]\\
\rho_{xy}&=&\rho_0\omega_c\tau_{tr}-\frac{\delta \rho_{xx}}{2\omega_c \tau_{tr}}
\end{eqnarray}
The cyclotron frequency is $\omega_c=\frac{eB}{m_c}$, where $m_c=\frac{\hbar k_F}{v_F}$ is the cyclotron mass. The phase of these oscillations is related to the $\pi$ Berry phase of massless Dirac fermions, i.e. to the existence of a zero energy LL. In the above equations, the longitudinal resistivity $\rho_{xx}$ is the sum of the zero field resistivity $\rho_0=1/\sigma$, where $\sigma=2\frac{e^2}{h}k_Fv_F \tau_{tr}$, and of magnetic oscillations $\delta \rho_{xx}$ due to the presence of underlying Landau levels. The amplitude $A=4A_{T}A_D$ of these ShdH oscillations is controlled by the temperature via the Lifshitz-Kosevich factor $A_{T}=\frac{2\pi^2k_BT/\hbar\omega_c}{\sinh(2\pi^2k_BT/\hbar\omega_c)}$ \cite{LK} and by the disorder via the Dingle factor $A_D=\exp(-\frac{\pi}{\omega_c \tau_e})$ \cite{Dingle}. The transverse resistivity $\rho_{xy}$ is the sum of the classical Hall resistance $\rho_0 \omega_c \tau_{tr}=\frac{B}{e n_c}$ and of magnetic oscillations $\propto \delta \rho_{xx}$.

\begin{figure}[htb]
\begin{center}
\subfigure[]{\includegraphics[width=4cm]{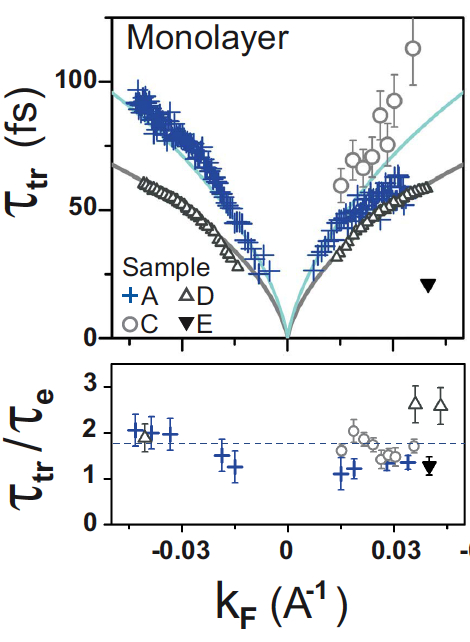}}
\subfigure[]{\includegraphics[width=6.5cm]{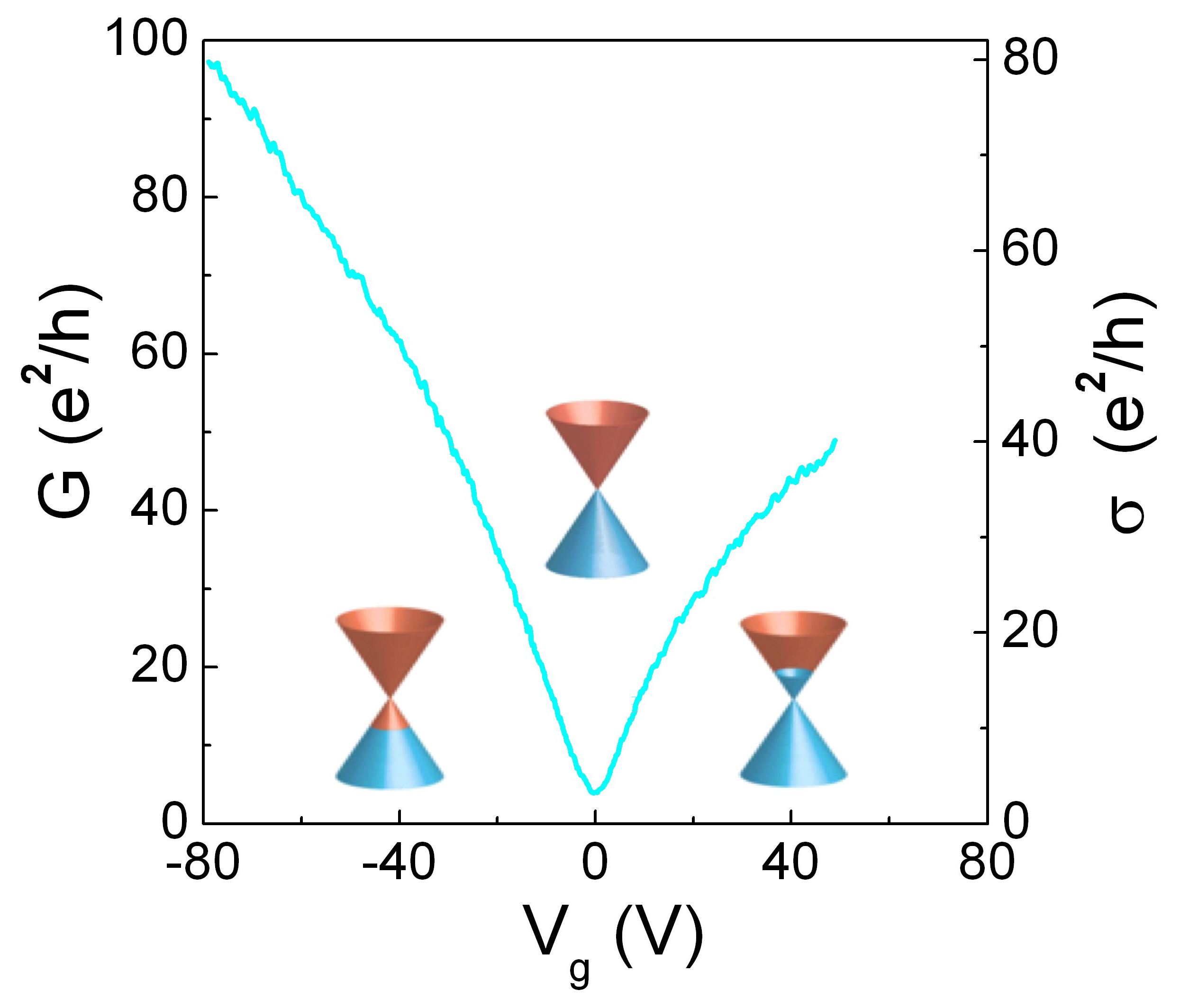}}
\caption{\label{fig:tautrtaue} (a) Upper panel: transport time $\tau_{tr}$ as a function of $k_F$ for four different samples (A, C, D, E) of monolayer graphene. Positive/negative values of $k_F$ correspond to electron/hole doping. The fit is according to the resonant scatterer model, see eq.~(\ref{taubis}). Lower panel: ratio $\tau_{tr}/\tau_e$ as a function of $k_F$ for four different samples. (b) Gate voltage $V_g$ dependence of the two-terminal conductance $G$ of sample $A$ at zero magnetic field. According to the capacitor law $k_F^2=\pi n_c\propto V_g$. The contact resistance has been subtracted. The conductivity $\sigma$ shows a sub linear behavior compatible with resonant scatterers $\sigma \propto |n_c|\ln^2 (|n_c| R^2)$. From Monteverde et al. \cite{Monteverde2010}.}
\end{center}
\end{figure}
The period of the magnetic oscillations is measured as a function of the gate voltage in order to obtain $k_F$ without relying on the capacitor model $k_F^2=\pi n_c\approx \pi C_g V_g/e$, which breaks down in the vicinity of the neutrality point because of the presence of inhomogeneities (electron-hole puddles). The amplitude $A$ of the ShdH as a function of $B$ gives access to the cyclotron mass $m_c$ at high temperature $k_B T\gg \hbar \omega_c$ (via the Lifshitz-Kosevich factor) and to the elastic scattering time $\tau_e$ at low temperature $k_B T\ll \hbar \omega_c$ (via the Dingle factor). The main result is a plot of the two scattering times $\tau_{tr}$ and $\tau_e$ as a function of $k_F$, see fig.~\ref{fig:tautrtaue}(a). Their ratio is almost independent of $k_F$ and $\tau_{tr}/\tau_e \lesssim 2$. Both features point to impurities of range shorter than the Fermi wavelength and probably of range comparable to the inter-atomic distance. This is not compatible with charged impurities. Also, the transport time is well fitted by a formula that assumes the presence of resonant scatterers (range shorter than the Fermi wavelength but not necessarily shorter than the lattice spacing), see fig.~\ref{fig:tautrtaue}(b). In addition, measurements conducted on similarly prepared bilayer samples -- which we do not discuss here -- are also compatible with strong short range impurities and not with charged impurities \cite{Monteverde2010}.

\subsection{Conclusion on impurities in graphene}
The above results indicate that the main scattering mechanism in these exfoliated graphene samples is due to strong neutral defects, with a range shorter than the Fermi wavelength and possibly on the order of the inter-atomic distance, inducing resonant (but not unitary) scattering. Likely candidates are vacancies, voids, adatoms or short-range ripples. This does not exclude the presence of long-ranged charged impurities responsible for electron-hole puddles, but their contribution to the scattering rates appears negligible in all the samples investigated \cite{Monteverde2010}. 

Raman measurements revealed the presence of atomic scale defects that contribute a small previously unnoticed D peak  \cite{Ni2010}. The density of these defects is enough to explain the limited mobilities achievable in graphene on a substrate. These defects are likely to be resonant scatterers. Also several measurements on samples on varying substrates \cite{Ponomarenko2009}, and on suspended samples \cite{Mayorov2012}, indicate that impurities limiting the transport are to be found within the graphene sheet and not in the substrate. Overall, there is growing indications that resonant scatterers are one of the limiting impurities \cite{Peres2010}.

We should mention that other, disagreeing, measurements of $\tau_{tr}/\tau_e$ were done in Ref.~\cite{Hong2009} on exfoliated graphene on silicon dioxide. These authors find that $\tau_{tr}/\tau_e$ is in between $1.5$ and $5.1$ and explain their results by the presence of two types of impurities: weak short-range impurities and charged impurities located in the substrate 2~nm below the graphene sheet.

\section{Conclusion: magnetic-field independent Landau level and parity anomaly}
\label{conclusion1}
We conclude this first part of the thesis by summarizing the most important aspects of massless Dirac fermions in a perpendicular magnetic field. Arguably, it is the existence of a magnetic field independent (zero-energy) $n=0$ Landau level \cite{McClure}. From a semi-classical perspective, this is tied to the $\pi$ Berry phase, which is itself a consequence of the sublattice pseudo-spin 1/2. When the Dirac fermions become massive, this $n=0$ Landau level still exists, has a magnetic-field independent energy but is no more at zero energy \cite{Haldane}. However, this LL has no particle-hole symmetric partner within the same valley. This is known as the parity anomaly for a single flavor (valley) of massive Dirac fermions. It means that the vacuum, corresponding to zero Fermi energy, is charged and that this vacuum charge is proportional to the number of flux tubes threading the system. This excess charge is carried by the anomalous (partner-less) $n=0$ LL. However, in lattice models such as boron nitride, Dirac fermions appear in pairs (fermion doubling) and the anomaly in one valley is canceled by that in the other \cite{Semenoff,Haldane}.  

From a solid state perspective, we may look at fermion doubling and parity anomaly cancellation in the following way. Consider an isolated (un-contacted) sample of boron nitride. By definition, it is charge neutral (undoped). Applying a perpendicular static magnetic field can not induce charges in the crystal. Therefore, even in the presence of a strong perpendicular magnetic field, the crystal remains uncharged. Hence, the parity anomaly can not occur (otherwise the ``vacuum'' would be charged). If Dirac fermions are present (whether massive or not) they must occur in pairs in order that the system remains charge neutral. 


	\chapter{Topological properties and stability of Dirac fermions}

{\it There is more to the Hamiltonian than its spectrum. (after Neil Young: ``There is more to the picture than meets the eye.'')}

\vspace{0.5cm}

In this chapter, we show that Dirac fermions are characterized by a topological quantity, which is a winding number for their sublattice pseudo-spin \footnote{In the whole chapter, we do not take the real spin into account.}. This important property of Dirac fermions is not tied to the existence or absence of a contact (Dirac) point between bands. In the case that the Dirac fermions are massless (i.e. when there is a linear contact point), this winding number is also the Berry phase (divided by $\pi$) along a trajectory encircling the Dirac point. The winding number is a topological charge and a Dirac fermion can be seen as a unit vortex -- a topological defect -- in the relative phase $\phi_{\vk}$ between components of the bispinor in reciprocal space. We discuss two aspects of this winding number (also known as a topological Berry phase). 
First, the semi-classical quantization of cyclotron orbits for Dirac fermions shows that it is the winding number that shifts the Landau index by one-half -- and not the Berry phase -- and can lead to the appearance of a peculiar Landau level, the energy of which does not dependent on the magnetic field. 
Second, it is the winding number that insures the stability/robustness of Dirac fermions with respects to perturbations such as lattice deformation. Opening of a gap in the energy spectrum is not enough to get rid of the Dirac fermions -- although they become massive. Dirac fermions typically appear in pairs -- the so-called fermion doubling. To make them disappear, one needs to reach a situation where two Dirac fermions merge in a topological transition similar to the annihilation of a vortex and an anti-vortex. This situation was recently realized in two different experiments on graphene analogues -- one made of cold atoms in an optical lattice, the other with microwaves in a lattice of dielectric resonators. It can be efficiently probed via Bloch oscillations and Landau-Zener transitions. In some cases, a St\"uckelberg interferometer is also realized with a pair of Dirac cones. Before discussing these topics, we start this chapter by giving a brief introduction to Berry phases for electrons in solids.    
	
	\section{Introduction to Berry phases in solids}
	\label{bp}
Berry phases for electrons in solids are thoroughly reviewed in \cite{Niu2010}. Here we only scratch the surface of this vast subject. In the context of Bloch electrons and band theory, Berry phases type terms occur in the semi-classical description of the electron motion when projected or restricted to a single band. Under the influence of external fields -- such as electric $\vec{\mathcal{E}}$ or magnetic $\vec{B}$ fields -- virtual transitions to other bands are possible and induce anomalous terms, loosely referred to as ``Berry phase terms'', in the semiclassical equations of motion. They are anomalous in the sense that they are not easily derived and were long neglected or forgotten\footnote{We should pay due credit to the work of pioneers of such effects in the 50-70's: Blount, Luttinger, Kohn, Roth, Zak, etc. However, we have the feeling that it is only recently, mainly through the work of Q. Niu and others, that the correct equations of motion and their generality, for example the connection to the geometrical phase of Berry, have been understood. This knowledge is only progressively diffusing in the condensed matter community and is slowly making its way in textbooks.}. The correct semiclassical equations of motion at order $\hbar$ for an electron, the wavepacket of which is restricted to the $\alpha$th band, are \cite{Niu2010}
\beqn
\hbar \dot{\vk}&=&-e[\vec{\mathcal{E}}+\dot{\vec{r}}\times \vec{B}]\\
\dot{\vec{r}}&=&\frac{1}{\hbar}\vec{\nabla}_{\vk}[\varepsilon_{\alpha,0}(\vk)-\vec{\mathcal{M}}_\alpha(\vk)\cdot \vec{B}]-\hbar \dot{\vk}\times \frac{\vec{\Omega}_\alpha(\vk)}{\hbar}
\eeqn
where $\vec{r}\equiv \vec{r}_\alpha$ is its average position, $\hbar \vk\equiv \hbar \vk_\alpha$ its average gauge invariant crystal momentum, $\varepsilon_{\alpha,0}(\vk)$ the band energy in the absence of external fields and $\alpha=1,...,N$ is the band index. In the following, we restrict ourselves to a two-dimensional system. The two Berry phase terms are (1) the Zeeman-like effect $-\vec{\mathcal{M}}_\alpha(\vk)\cdot \vec{B}$ and (2) the anomalous velocity $-\dot{\vk}\times \vec{\Omega}_\alpha(\vk)$. The first is due to the orbital magnetic moment $\vec{\mathcal{M}}_\alpha(\vk)$ -- related to the sublattice pseudo-spin and not to the real spin -- that shifts the energy $\varepsilon_{\alpha,0}\to \varepsilon_{\alpha}\equiv \varepsilon_{\alpha,0}-\vec{\mathcal{M}}_\alpha \cdot \vec{B}$ in the presence of a magnetic field $\vec{B}$. This magnetic moment has a nice semi-classical interpretation: it can be viewed as resulting from the self-rotation of the charged wavepacket of an electron restricted to a band in a revival of the Goudsmit and Uhlenbeck idea for the origin of the real spin \cite{Niu2010}. 
\begin{figure}[htb]
\begin{center}
\includegraphics[width=5cm]{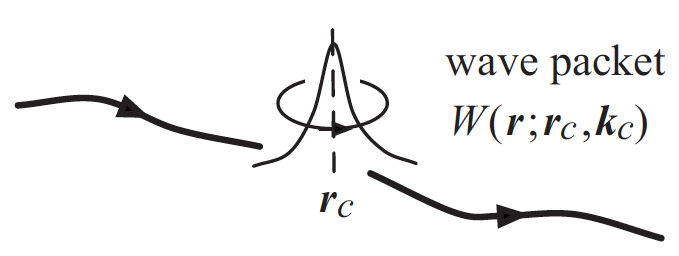}
\caption{\label{fig:wavepacket} Wavepacket for an electron restricted to a single band. Its self-rotation gives rise to an orbital magnetic moment. From Xiao et al. \cite{Niu2010}.}
\end{center}
\end{figure}
The second appears as a dual to the magnetic Lorentz force, where the role of the magnetic field in reciprocal space is played by the Berry curvature $\vec{\Omega}_\alpha(\vk)$ and that of the velocity by $\dot{\vk}$. That these two quantities are related to virtual transitions to other bands is obvious from their definition
\begin{equation}
\vec{\Omega}_\alpha (\vec{k})=(i\sum_{\alpha' \neq
\alpha}\frac{\langle
u_{\vec{k},\alpha}|\partial_{k_x}H(\vec{k})
|u_{\vec{k},\alpha'} \rangle \langle
u_{\vec{k},\alpha'}|\partial_{k_y}H(\vec{k})
|u_{\vec{k},\alpha}
\rangle}{[\varepsilon_{\alpha,0}(\vec{k})-\varepsilon_{\alpha',0}(\vec{k})]^2}
+ \textrm{c.c.}) \, \vec{e}_z
 \end{equation}
 and
   \begin{equation}
\vec{\mathcal{M}}_\alpha (\vk)=(i\frac{e}{2\hbar}\sum_{\alpha' \neq
\alpha}\frac{\langle
u_{\vec{k},\alpha}|\partial_{k_x}H(\vec{k})
|u_{\vec{k},\alpha'} \rangle \langle
u_{\vec{k},\alpha'}|\partial_{k_y}H(\vec{k})
|u_{\vec{k},\alpha}
\rangle}{\varepsilon_{\alpha,0}(\vec{k})-\varepsilon_{\alpha',0}(\vec{k})}
+ \textrm{c.c.}) \, \vec{e}_z
 \label{magnetization}
 \end{equation}
 where $H(\vk)=\sum_{\alpha}\varepsilon_{\alpha,0}(\vk)|u_{\vk,\alpha}\rangle\langle u_{\vk,\alpha} |$ is the $N\times N$ $\vk$-dependent Hamiltonian, where $N$ is the number of bands ($\alpha=1,...,N$) and $|u_{\vk,\alpha}\rangle$ is a Bloch state. Also these quantities are the first order in $\hbar$ correction to the classical equations of motion of Bloch and Peierls: $\hbar \dot{\vk}=-e[\vec{\mathcal{E}}+\dot{\vec{r}}\times \vec{B}]$ and $\dot{\vec{r}}=\frac{1}{\hbar}\vec{\nabla}_{\vk}\varepsilon_{\alpha,0}(\vk)$. This can be seen by consistently writing the equations of motion in terms of the classical gauge-invariant momentum $\hbar \vk$ and position $\vec{r}$. Then it appears that $\vec{\mathcal{M}}_\alpha (\vk)$ and $\vec{\Omega}_\alpha (\vec{k})/\hbar$ are proportional to $\hbar$ and vanish in the classical limit. In summary, Berry phase terms are a physical manifestation of the coupling between bands introduced by external fields and due to virtual (quantum) transitions to other bands.
 
As mentioned, the Berry curvature can be seen as a magnetic field in reciprocal space. The corresponding vector potential 
\begin{equation}
\vec{\mathcal{A}}_\alpha (\vec{k})=i\langle
u_{\vec{k},\alpha} | \vec{\nabla}_{\vec{k}}
 u_{\vec{k},\alpha} \rangle
 \end{equation}
 is called the Berry connection such that $\vec{\Omega}_\alpha
(\vec{k})=\vec{\nabla}_{\vec{k}}\times
\vec{\mathcal{A}}_\alpha = \Omega_\alpha \vec{e}_z$. The line integral of the Berry connection along a close path $C=\partial S$ in reciprocal space 
\be
\Gamma_\alpha(C)=\oint_C d\vk \cdot i\langle
u_{\vec{k},\alpha} | \vec{\nabla}_{\vec{k}}
 u_{\vec{k},\alpha} \rangle = \int_S d^2 k \, \Omega_\alpha
\ee
is called the Berry phase and is the equivalent of an Aharonov-Bohm phase in $\vk$-space. In the preceding equation $S$ is the area in reciprocal space enclosed by the path $C$. The Berry phase is defined modulo $2\pi$. This expression is the original one derived by Berry \cite{Berry1984}. It is the expression for the geometrical phase acquired by a quantum particle when it traces a closed trajectory in projective Hilbert space. It is non-trivial, when the path encloses a curved space. Here, curvature means Berry curvature, which is the natural curvature of the Hermitian line bundle that to each $\vk$ in the first Brillouin zone associates the ray generated by $|u_{\vk,\alpha}\rangle$ in Hilbert space, where $H(\vk)|u_{\vk,\alpha}\rangle=\varepsilon_{\alpha,0}(\vk)|u_{\vk,\alpha}\rangle$ \cite{BSimon1983}. In this language, the geometrical phase is the holonomy. The integral of the Berry curvature over the complete Brillouin zone is an integer called the Chern number.

Massless Dirac fermions are particular as they carry a Berry phase, which is not only geometrical but is topological -- it does not depend on the precise path $C$ as long as it encloses the Dirac point -- and quantized in units of $\pi$. We will show that it is actually a winding number (times $\pi$). The Bloch state of the Hamiltonian $H_\xi =v_F(\xi p_x \sigma_x+p_y \sigma_y)$ corresponding to the energy $\alpha \hbar v_F k$ is 
\be
|u_{\vk,\alpha=+}\rangle=\frac{1}{\sqrt{2}}\left(\begin{array}{c}1\\ \xi  e^{i\xi \phi_{\vk} } \end{array}\right) \textrm{ and } |u_{\vk,\alpha = -}\rangle=\frac{1}{\sqrt{2}}\left(\begin{array}{c} -\xi e^{-i\xi \phi_{\vk}} \\  1 \end{array} \right) \ee 
where $k_x+ik_y=ke^{i\phi_{\vk}}$ defines the angle $\phi_{\vk}$ and $\alpha=\pm$ is the band index. Because we will be interested in cyclotron orbits, we consider trajectories of constant energy. For example, along a constant energy contour ($C$ is a circle) that is traveled anti-clockwise, the Berry phase is
\be
\Gamma_\alpha(C)= -\frac{\xi \alpha}{2}\oint_C d\vk \cdot \vec{\nabla}_{\vec{k}} \phi_{\vk}=-\frac{\xi \alpha}{2} \oint d\phi_{\vk}=-\xi \alpha \pi\equiv \pi
\ee
as $\vec{\mathcal{A}}_\alpha=-\xi \alpha \vec{\nabla}_{\vec{k}} \phi_{\vk}/2$. Here, this is just a measure of the topological charge of the vortex in the relative phase $\pm\xi \phi_{\vk}$ of the wavefunction between the two sublattices. This topological charge, or winding number, is defined by:
\be
W_C\equiv -\alpha \xi \frac{1}{2\pi}\oint_C d\phi_{\vk}=-\alpha \xi \, .
\ee
Note that the winding number $W_C=\pm 1$ and that there is the same difference between $+1$ and $-1$ as that between a vortex and an anti-vortex. This is to be contrasted with the Berry phase, which is defined modulo $2\pi$ so that $\pi\equiv -\pi$. The winding number only depends on the relative phase between the $A$ and $B$ components of the bispinor. We will see that, in general, the Berry phase also depends on the relative weight on the two sublattices and not only on the relative phase.

The case of massive Dirac fermions $H_\xi =v_F(\xi p_x \sigma_x+p_y \sigma_y+mv_F\sigma_z)$ is quite instructive as it shows that the Berry phase is no longer topological or quantized. The Bloch spinor corresponding to the energy $\varepsilon_\alpha(\vk)=\alpha \sqrt{m^2v_F^4+ \hbar^2v_F^2k^2}$ is
\be
|u_{\vk,\alpha=+}\rangle=\left(\begin{array}{c}\cos \frac{\theta}{2}\\ \xi \sin \frac{\theta}{2}  e^{i\xi \phi} \end{array}\right) \textrm{ and } |u_{\vk,\alpha = -}\rangle=\left(\begin{array}{c} -\xi \sin \frac{\theta}{2} e^{-i\xi \phi} \\  \cos \frac{\theta}{2} \end{array} \right) \ee 
where $\cos \theta = m v_F^2/\varepsilon_+(\vk)$ and $\sin \theta = \hbar v_F k/\varepsilon_+(\vk)$, so that $\theta$ is the azimuthal angle and $\phi$ the polar angle on the Bloch sphere. The Bloch spinors are written such that they are single-valued everywhere on the Bloch sphere, except at the south pole where $\theta=\pi$ and $\phi$ is undefined (this point can not be reached in this model, anyway). For example, along a constant energy contour ($C$ is a circle) that is travelled anti-clockwise, the Berry phase is
\be
\Gamma_\alpha(C)= -\xi \alpha\oint_C d\vk \cdot \vec{\nabla}_{\vec{k}} \phi \sin^2\frac{\theta}{2}=-\xi \alpha \pi (1-\cos \theta)
\ee
and the winding number is still $W_C=-\alpha \xi$, as the relative phase $\xi\phi$ between the $A$ and $B$ components of the bispinor is unaffected by the mass term, that only changes the relative weights. Now the Berry phase depends on the precise path, and not only on whether it encircles the Dirac point or not, and is no longer quantized as it can take different values depending on the energy of the cyclotron orbit.

	
	\section{Semiclassical quantization of cyclotron orbits for two coupled bands}
	\label{sq}
\subsection{Onsager's quantization and phase mismatch}
The dispersion relation of Bloch electrons in two dimensional (2D) crystals generally exhibit regions of closed orbits of constant energy in reciprocal space. As a consequence, it is expected that applying a perpendicular magnetic field gives rise to quantized cyclotron
orbits and the corresponding Landau levels. A semiclassical approach to obtain these Landau levels consists of first computing the area
of the classical cyclotron orbits and then imposing the Bohr-Sommerfeld quantization condition in the form suggested by
Onsager and Lifshitz for Bloch electrons \cite{Onsager,LK}. The semiclassical quantization condition for a cyclotron
orbit $C$ reads:
\begin{equation}
S(C)l_B^2=2\pi[n+\gamma] \label{Onsager}
\end{equation}
where $S(C)\equiv \iint d^2 k$ is the $\vk$-space area enclosed by the cyclotron orbit, $\vk$ is the (gauge-invariant) Bloch wavevector, $l_B\equiv \sqrt{\hbar/eB}$ is the magnetic length, $-e$ is the electron charge and $n$ is an integer. The quantity $\gamma$ is called a phase mismatch ($0\leq \gamma<1$) and was not determined by Onsager or Lifshitz.  A relation between the phase mismatch $\gamma$ and the nature of the electronic Bloch functions was obtained by Roth \cite{Roth}. She
found that $\gamma$ can depend on the cyclotron orbit $C$ and that
$\gamma(C)$ can be related to a quantity $\Gamma(C)$ later
identified as a Berry phase acquired by the Bloch electron during a cyclotron orbit $C$. The relation reads
\begin{equation}
\gamma(C)=\frac{1}{2}-\frac{\Gamma(C)}{2\pi}
\label{RW}
\end{equation}
where $1/2$ is the Maslov index contribution coming from two caustics and $-\Gamma(C)/2\pi$ the Berry phase contribution. 
From the dependence of the cyclotron surface $S(C)$ on the energy $\varepsilon$ of the orbit, one can
usually rewrite the above quantization condition as
\begin{equation}
S(\varepsilon)l_B^2=2\pi[n+\gamma_L] \label{OnsagerEnergy}
\end{equation}
where we introduced $\gamma_L$ in place of $\gamma$. Then by inverting $S(\varepsilon)$, one obtains the (semiclassical)
Landau levels
\begin{equation}
\varepsilon_n=S^{-1}[\frac{2\pi}{l_B^2}(n+\gamma_L)]=\text{function}[B(n+\gamma_L)]
\label{SCLL}
\end{equation}
where $n$ is now interpreted as the Landau index and $\gamma_L$ is a shift for the Landau index or a phase offset for quantum oscillations, such as Shubnikov-de Haas oscillations in the magneto-resistance. For an isolated band, the shift $\gamma_L$ is equal to the phase mismatch $\gamma(C)$ introduced above. But this is not the case if there is coupling between several bands such that the energy of the cyclotron orbit is shifted (by Berry phase terms) compared to the constant energy curve in the absence of a magnetic field.

\subsection{Winding number versus Berry phase in a two coupled bands model}
Let us now consider a simple model for two coupled bands that will emphasize these Berry phase effects and the way they affect the semi-classical quantization of cyclotron orbits \cite{Fuchs2010}. We restrict ourselves to a particle-hole and time-reversal symmetric two-band model with a $2\times 2$ $\vk$-dependent Hamiltonian [in the following $\hbar\equiv 1$]:
 \begin{equation}
H(\vec{k})=\left(
\begin{array}{cc}\Delta & f(\vec{k})\\
f^*(\vec{k}) & -\Delta \end{array} \right)
\end{equation}
where $\vec{k}$ is the Bloch wavevector in the first
Brillouin zone (BZ). The function $f(\vec{k})$ is usually
obtained as a sum over hopping amplitudes in a tight binding
description. Time-reversal symmetry imposes
$H(-\vec{k})^*=H(\vec{k})$ and therefore
$f(-\vec{k})^*=f(\vec{k})$. Note that Bloch's theorem imposes $|f(\vec{k}+\vec{G})|=|f(\vec{k})|$ for any reciprocal lattice vector $\vec{G}$. However it does not
require that $f(\vec{k}+\vec{G})=f(\vec{k})$. An important assumption here is that the diagonal term $\Delta$ does not depend on the wavevector and can therefore be interpreted simply as an on-site energy. This term explicitly breaks the inversion
symmetry. Introducing the energy spectrum
$\varepsilon_0(\vec{k})=\alpha
\sqrt{\Delta^2+|f(\vec{k})|^2}$, where $\alpha=\pm 1$ is the
band index, and the azimuthal $\theta(\vec{k})$ and polar
$\phi(\vec{k})$ angles on the Bloch sphere, such that $\cos
\theta=\Delta/|\varepsilon_0|$, $\sin \theta=|f|/|\varepsilon_0|$ and
$\phi\equiv -\textrm{Arg} f$, the Hamiltonian can be rewritten as
\begin{equation}
H(\vec{k})=|\varepsilon_0| \left(
\begin{array}{cc}\cos \theta & \sin \theta e^{-i \phi}\\
\sin\theta e^{i\phi} & -\cos \theta \end{array} \right)
\end{equation}
The eigenfunction of energy $\varepsilon_0=\alpha |\varepsilon_0|$
is $\psi(\vec{r})=u_{\vec{k}}(\vec{r})
e^{i\vec{k}\cdot \vec{r}}$ where the Bloch spinor is
\begin{eqnarray}
|u_{\vec{k},\alpha}\rangle &=&
 \left(\begin{array}{c}\cos(\theta/2)\\
 \sin(\theta/2) e^{i\phi} \end{array}\right) \text{ if } \alpha=+1 \nonumber \\
&=&\left(\begin{array}{c} -\sin(\theta/2 )e^{-i\phi}\\
 \cos (\theta/2)\end{array}\right) \text{ if } \alpha=-1
\end{eqnarray}
The Berry phase along a cyclotron orbit is
\begin{equation}
\Gamma(C)=\oint_C d \vec{k} \cdot \vec{\mathcal{A}}=-\alpha
\sin^2\frac{\theta}{2}\oint_C d \vec{k} \cdot
\vec{\nabla}_{\vec{k}} \phi=\pi W_C[1-\cos \theta]
\end{equation}
where $W_C\equiv -\alpha\oint_C d\phi/2\pi=d(|\vep_0 \Gamma|)/d(\pi |\vep_0|)$ is the winding number. It counts the total charge of the vortices in $\phi$, which are encircled by the cyclotron orbit $C$.

Starting from the Onsager-Roth relation, see eq.
(\ref{Onsager},\ref{RW}),
\begin{equation}
S(\varepsilon_0)l_B^2=2\pi[n+\frac{1}{2}]- \Gamma(C)
\end{equation}
where $\varepsilon_0$ is the band energy in zero magnetic field, we obtain the quantization of $S(\varepsilon)$ where $\varepsilon$ is
the energy in presence of a magnetic field. Using the relation between the energy and the curvature
$\varepsilon_0=\varepsilon+\mathcal{M}B$ with
$\mathcal{M}=e\varepsilon_0 \Omega$ (valid for two bands with particle-hole symmetry \cite{Fuchs2010}), we obtain:
\begin{equation}
S(\varepsilon_0)l_B^2=S(\varepsilon)l_B^2+
\bar{\Omega}(\varepsilon_0)|\varepsilon_0| \frac{d
S}{d|\varepsilon_0|}
\end{equation}
In the previous equation, we introduced the Berry curvature
$\bar{\Omega}$ averaged over a constant energy orbit
\begin{equation}
\bar{\Omega}(\varepsilon_0)\equiv
\frac{1}{(2\pi)^2\rho(\varepsilon_0)}\frac{d \Gamma}{d|\varepsilon_0|}=\frac{d\Gamma}{dS}
\end{equation}
where $\rho(\vep_0)$ is the density of states (per unit area) in zero magnetic field. Therefore, we obtain
\begin{equation}
S(\varepsilon_0)l_B^2=S(\varepsilon)l_B^2+|\varepsilon_0|\frac{d
\Gamma}{d|\varepsilon_0|}
\end{equation}
which does not require the cyclotron orbit to be circular. The
energy quantization condition can now be rewritten as
\begin{equation}
S(\varepsilon)l_B^2=2\pi[n+\frac{1}{2}]-\frac{d (|\varepsilon_0|
\Gamma)}{d|\varepsilon_0|}=2\pi[n+\frac{1}{2}]-\pi W_C \label{main}
\end{equation}
in which we recognized the winding number. Inverting this last
relation $S(\varepsilon)l_B^2=2\pi[n+(1-W_C)/2]$ allows one to obtain the (semiclassical) Landau levels for
the whole energy band. Finally, the Landau index shift is
\begin{equation}
\gamma_L=\frac{1}{2}-\frac{W_C}{2}
\end{equation}
This last equation, which is the main result of this section, shows that the Landau index shift $\gamma_L$ is related to the topological part of the Berry phase $\pi W_C$ and not to the complete Berry phase $\Gamma(C)$. Physically, the topological Berry phase $\pi W_C$ is just the usual $\pi$ phase that a bi-spinor acquires in Hilbert space as a result of a $2\pi$ rotation in position space. Here the spin 1/2 is actually the sublattice pseudo-spin. In the expression for $\gamma_L$, the winding number only matters modulo 2. However, there is a significance of the winding number beyond its being even or odd. We discuss this point below in connection to the degeneracy of the zero energy Landau level.

In summary, the quantization of cyclotron orbits depends on the coupling to other bands via two ``Berry phase terms'': the Berry phase $\Gamma(C)$ {\it and} the energy shift $\varepsilon_0 \to \varepsilon=\varepsilon_0-\mathcal{M}B$. In the end, the Landau index shift $\gamma_L$ (or equivalently the phase offset of quantum oscillations) is related to the winding number $W_C$ and not to the complete Berry phase $\Gamma(C)$. This also shows that an important property of Dirac fermions lies in the singularities of the relative phase $\phi$, which is not tied to the existence or not of a contact point between bands. It exists both for massless and for massive Dirac fermions. We mention that similar results were obtained from a different perspective in \cite{Carmier}.

\subsection{Example: tight-binding model of boron nitride}
\begin{figure}[htb]
\begin{center}
\subfigure[]{\includegraphics[height=6cm]{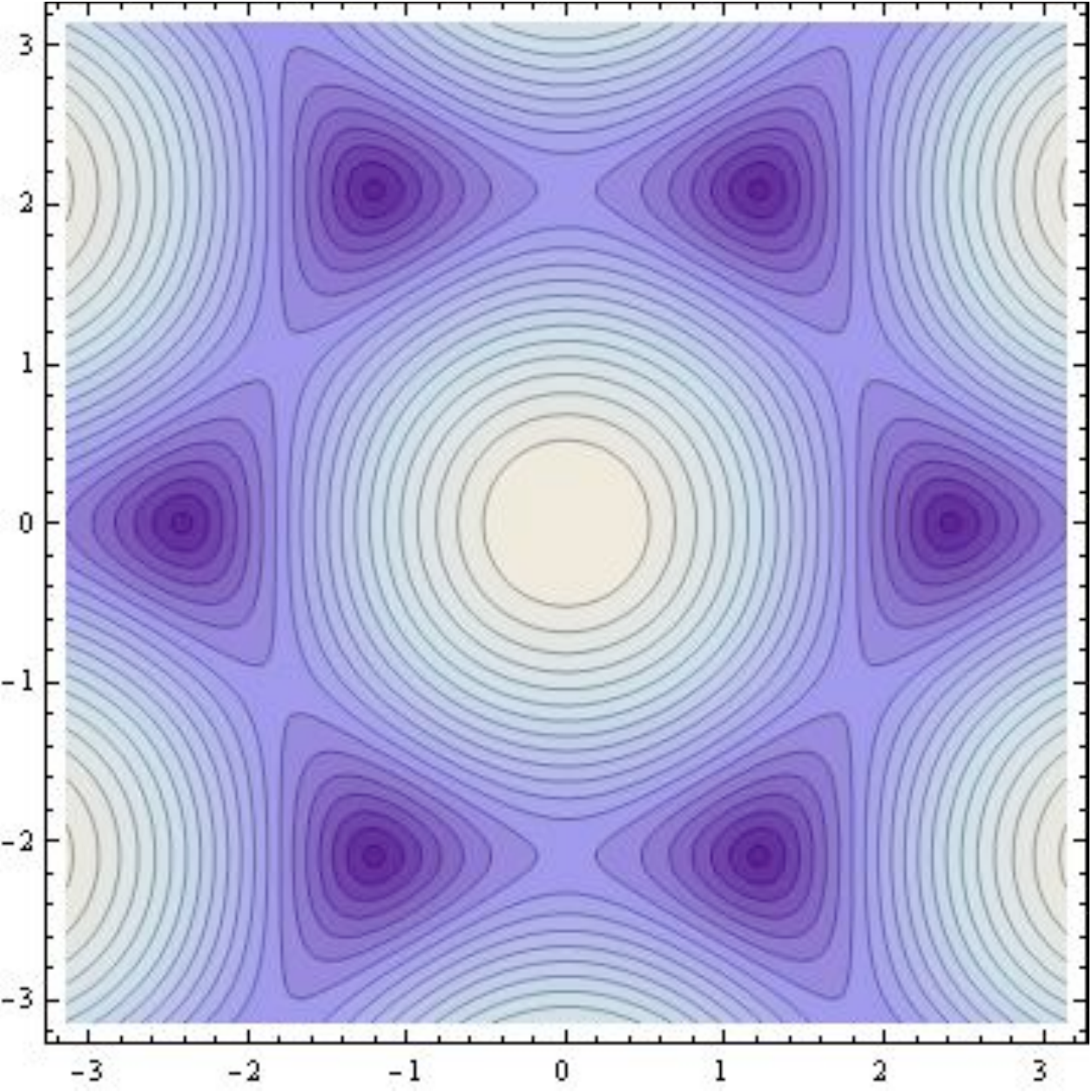}}
\subfigure[]{\includegraphics[height=6cm]{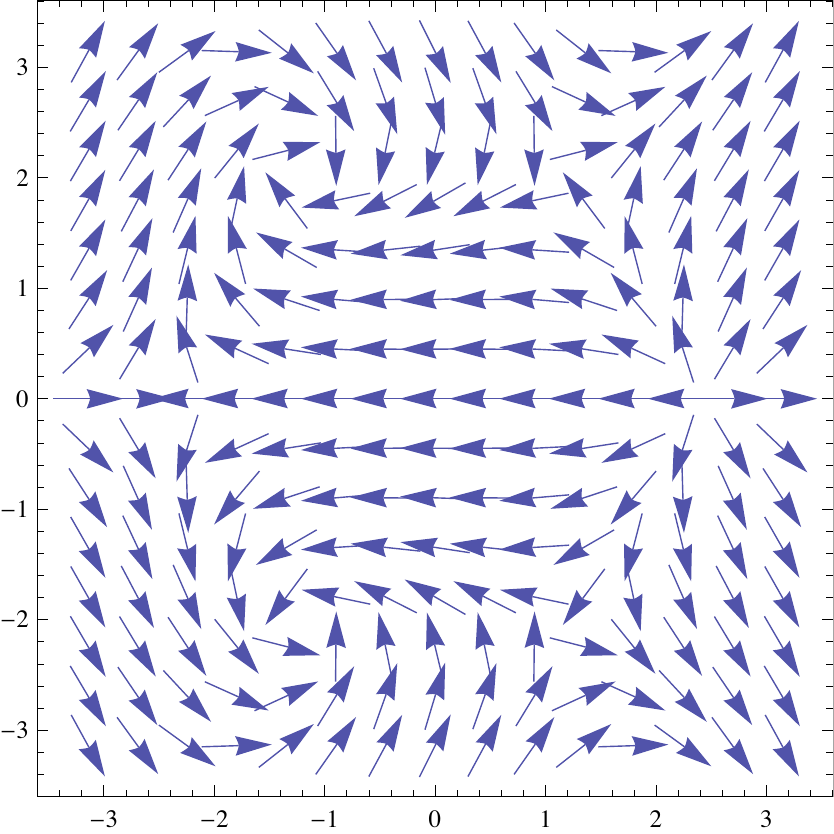}}
\caption{\label{isoenergies}(a): Isoenergy lines
($\varepsilon_0(\vk)$=constant) of boron nitride in the first
Brillouin zone for $\Delta=0.1$ [energies in units of $t$]. In the
semiclassical limit, cyclotron orbits in reciprocal space follow the
isoenergy lines. (b): Polar angle on the Bloch sphere
$\phi(\vec{k})\equiv -\textrm{Arg} f(\vec{k})$ in
the BZ. The winding number $W_C$ measures the topological charge of
vortices in the polar angle $\phi$. From Fuchs et al. \cite{Fuchs2010}.}
\end{center}
\end{figure}

As a concrete example of a band structure made of coupled bands, we consider a boron nitride monolayer, sometimes known as ``gapped graphene'' (see also the introduction). It is similar to graphene, having a honeycomb lattice, albeit with two atoms per unit cell that have different on-site energies (boron and nitrogen) so that there is a gap between the valence and the conduction bands. We use a tight binding model, with hopping amplitude $t$ and nearest-neighbour distance $a$, given by the following $2\times 2$ Hamiltonian in $(A,B)$ subspace:
\begin{equation}
H(\vec{k})=\left(
\begin{array}{cc}\Delta & f(\vec{k})\\
f^*(\vec{k}) & -\Delta \end{array} \right) \, \textrm{with}\,
f(\vec{k})=-t[e^{-i\vec{k}\cdot
\vec{\delta}_1}+e^{-i\vec{k}\cdot
\vec{\delta}_2}+e^{-i\vec{k}\cdot
\vec{\delta}_3}]
\end{equation}
where $\vec{k}$ is the wavevector in the entire Brillouin
zone and $\vec{\delta}_1, \vec{\delta}_2,
\vec{\delta}_3$ are vectors connecting an $A$ atom with its
three nearest $B$ neighbours. Note that, contrary to $|f(\vec{k})|$,
$f(\vec{k})$ does not have the periodicity of the reciprocal
lattice but satisfies
$f(\vec{k}+\vec{G})=f(\vec{k})\exp(i\vec{G}\cdot
\vec{\delta}_3)$ where $\vec{\delta}_3$ is the vector
relating the two atoms $A,B$ of the basis and $\vec{G}$ a reciprocal lattice vector. The energy spectrum is $\varepsilon_0(\vk)=\alpha \sqrt{|f(\vk)|^2+\Delta^2}$, see fig.~\ref{fig:bnbandstructure}(a), and the corresponding iso-energy lines in the BZ are plotted in fig.~\ref{isoenergies}(a). These lines correspond to cyclotron orbits. The relative phase (polar angle) $\phi$ is defined by $f(\vec{k})=|f(\vec{k})|e^{-i\phi_{\vk}}$ and is plotted in fig.~\ref{isoenergies}(b) in the BZ. The vortices and anti-vortices around the $K$ and $K'$ corners of the hexagonal BZ are clearly visible.

\begin{figure}[htb]
\begin{center}
\includegraphics[height=6cm]{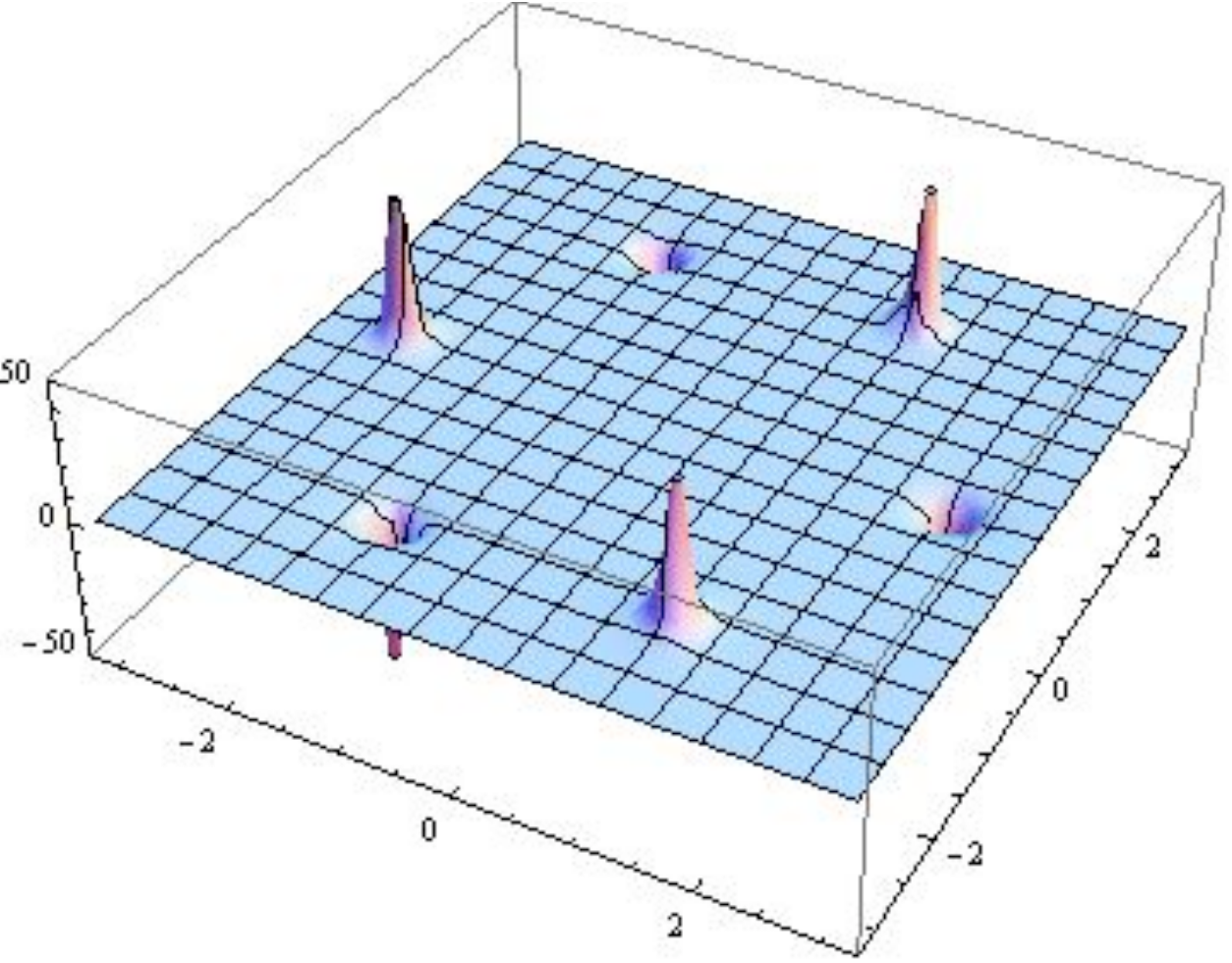}
\caption{\label{courburefig}Berry curvature $\Omega$, see eq.~(\ref{eq:omega}), [in units of
$a^2$] in the conduction band of boron nitride as a function of the
Bloch wavevector $(k_x,k_y)$ [in units of $1/a$] in the entire
Brillouin zone  for $\Delta=0.1t$. From Fuchs et al. \cite{Fuchs2010}.
}
\end{center}
\end{figure}

The Berry phase for a cyclotron orbit $C$ of constant energy
$\varepsilon_0$ is $\Gamma(C)=\pi
W_C[1-\frac{\Delta}{|\varepsilon_0|}]$ where $W_C\equiv -\alpha
\oint_C d\theta/2\pi$ is the winding number, which is $\pm 1$ when
encircling a valley (because of a vortex in $\phi$) and $0$ when
the orbit is around the $\Gamma$ point (center of the BZ), see fig.~\ref{isoenergies}. 
A saddle point in the energy dispersion at
$|\varepsilon_0|=\sqrt{\Delta^2+t^2}$ separates the cyclotron orbits
which encircle the two valleys from the cyclotron orbit which
encircle the $\Gamma$ point. As a consequence,
\begin{eqnarray}
\Gamma(C)&=&-\alpha \xi \pi [1-\Delta/|\varepsilon_0|] \,\,\,
\textrm{if} \,\,\, \Delta \leq |\varepsilon_0| <
\sqrt{\Delta^2+t^2} \,\,\, (\textrm{i.e. } W_C=-\alpha\xi=\pm 1)\nonumber \\
&=&0 \,\,\, \textrm{if} \,\,\, \sqrt{\Delta^2+t^2} < |\varepsilon_0|
\leq \sqrt{\Delta^2+(3t)^2} \,\,\, (\textrm{i.e. } W_C=0)
\end{eqnarray}
We checked this simple expression for the Berry phase along a cyclotron orbit numerically by directly computing the integral of
the curvature in $\vec{k}$ space over the area encircled by the cyclotron orbit. The analytical expression of the Berry curvature in the BZ is
\begin{equation}
\Omega(\vec{k})
=a^2
\frac{\sqrt{3}\alpha t^2
\Delta}{|\varepsilon_0(\vec{k})|^3}\sin
(\vec{k}\cdot\frac{\vec{\delta}_2-\vec{\delta}_3}{2})
\sin(\vec{k}\cdot\frac{\vec{\delta}_3-\vec{\delta}_1}{2})
\sin(\vec{k}\cdot\frac{\vec{\delta}_1-\vec{\delta}_2}{2})
\label{eq:omega}
\end{equation}
and is plotted in fig.~\ref{courburefig}. Because of time reversal symmetry, the curvature satisfies $\Omega(-\vec{k})=-\Omega(\vec{k})$ and its integral over the entire BZ vanishes. As inversion symmetry is absent ($\Delta\neq 0$) $\Omega(-\vec{k})\neq \Omega(\vec{k})$. If both inversion and time-reversal symmetries are present, this argument predicts that the Berry curvature vanishes everywhere. In fact, this is true, except at contact points between bands. This is best seen by taking the $\Delta\to 0$ limit, where boron nitride turns into graphene. In graphene, the Berry curvature is singular: it is given by a Dirac $\delta$ function carrying half a flux quantum at the position of the two contact points. 

From the energy quantization relation $S(\varepsilon)l_B^2=2\pi[n+1/2]-\pi W_C=2\pi[n+\gamma_L]$ it is now possible to obtain the (semiclassical) Landau levels for the whole energy band of boron nitride. It shows that the Landau index shift $\gamma_L=1/2
\pm 1/2=0$ (modulo 1) vanishes for cyclotron orbits encircling a single valley ($K$ or $K'$). Whereas for orbits around the $\Gamma$
point, it is $\gamma_L=1/2+0=1/2$. Close to half-filling, boron nitride is described by a massive Dirac Hamiltonian $H_\xi =v_F(\xi p_x \sigma_x+p_y \sigma_y+mv_F\sigma_z)$ with $v_F=\frac{3}{2}ta$ and $mv_F^2=\Delta$. Landau levels can be computed exactly and are $\varepsilon_{\alpha,n\neq 0}=\alpha v_F\sqrt{2eBn+m^2v_F^2}\approx \alpha[mv_F^2+\frac{eB}{m}(n+0)]$ in the low field limit
\footnote{This should be contrasted to the familiar Landau levels $\varepsilon_n=\frac{eB}{m}(n+\frac{1}{2})$ corresponding to the single band Schr\"odinger Hamiltonian $H=\frac{\vp^2}{2m}$.}. The $1/2$ shift in the Landau level index for boron nitride at low energy is due to the non-zero winding number and can not be attributed to a contact point (as there is none) or to the complete Berry phase (which is energy dependent). In contrast, close to the band bottom (at the $\Gamma$ point), $\vep \approx -\sqrt{(3t)^2+\Delta^2}$, the dispersion is parabolic and $\gamma_L=1/2$, so that usual LLs are obtained $\vep_n+\sqrt{(3t)^2+\Delta^2}\propto (n+\frac{1}{2})B$.

Our work \cite{Fuchs2010} has been extended in several directions. Park and Marzari \cite{ParkMarzari} have shown that, beyond the semi-classical limit which we assumed and which is not valid for small Landau indices, the winding number is responsible not only for the Landau index shift but also for the existence and the degeneracy of the zero-energy Landau level (in gapless systems). This is true not only for monolayer graphene but more generally for graphene multilayers with $ABC$ stacking. The degeneracy of the zero-energy LL in turn gives the magnitude of the step in the integer quantum Hall effect near the neutrality point, which permits a direct measurement of the pseudo-spin winding number. See also \cite{ECRYS}, where the number of topologically protected zero energy Landau levels is discussed. Very recently, Wright and McKenzie \cite{Wright} generalized the relation between the phase offset and the Berry phase to Dirac fermions without particle-hole symmetry, such as that found at the surface of 3D topological insulators. They have shown that in this case, the phase offset is not quantized (it can be anything between 0 and 1 and not just 0 or 1/2). In conclusion, these examples show that the Berry phase, which is geometrical by nature, can be topological or not and can be quantized or not. These two notions are to be distinguished in general.

	
	
	\section{Motion and merging of Dirac fermions}
\subsection{Introduction}
\label{mergingintro}
Dirac points are linear contact points between bands. They are also quantized vortices in reciprocal space \cite{Volovik} and are therefore characterized by a topological charge -- the pseudo-spin winding number or topological Berry phase. They usually occur in pairs with opposite winding number, which is known as fermion doubling. One way of getting rid of Dirac points is via annihilation between unlike topological charges $-1 + 1 \to 0$. Considering a situation in which the Fermi level lies exactly at the Dirac points, this merging transition can also be seen as one in which the topology of the Fermi ``surface'' changes -- here from two Fermi points to a single point at the transition and eventually to no point. In this context, it is known as a Lifshitz transition \cite{Lifshitz1960}. 

Consider the massless Dirac Hamiltonian $H=v_F(p_x\tau_z\sigma_x+p_y\sigma_y)$. It is important to realize that getting rid of Dirac fermions is not equivalent to gapping the spectrum. Indeed, there are ways of opening a gap by rendering the Dirac fermions massive without loosing the underlying topological charges. Let us consider three examples on the honeycomb lattice (two concerning spinless and one concerning spinfull fermions). First, for spinless fermions, breaking inversion symmetry with a staggered on-site potential $\propto \sigma_z$ opens a gap, as shown by Semenoff \cite{Semenoff}. This is the case of boron nitride, and it leads to a quantum valley Hall insulator or trivial band insulator. Second, another example is by breaking time reversal symmetry with a potential $\propto \sigma_z \tau_z$ but keeping the other translational symmetries as in Haldane's model \cite{Haldane}. This leads to a gapped state known as a Chern insulator or quantum anomalous Hall state\footnote{For completeness, we should note that there is also a possibility known as a Kekul\'e dimerization pattern that opens a mass gap and is related to a particular in-plane distorsion of the honeycomb lattice \cite{VAA,Ryu2009}. We mentioned that possibility previously when discussing interaction-induced quantum Hall plateaus. It gives rise to a potential $\propto \sigma_z \tau_{x,y}$ and involves a complex gap. It does not break time reversal or inversion symmetry but it couples the two valleys.}. Third, for spinfull electrons, intrinsic spin-orbit coupling creates a potential $\propto \sigma_z\tau_z s_z$, which opens a gap without breaking any symmetry\footnote{Actually it breaks time-reversal symmetry for each spin copy separately as in Haldane's model.}. This is the quantum spin Hall insulator or $Z_2$ topological insulator of Kane and Mele \cite{KM}. Therefore, gapping the Dirac fermions is easy as it does not require a threshold but can be done by an arbitrary small perturbation, such as a staggered on-site potential of infinitesimal strength. However, getting rid of the Dirac fermions is much harder as it requires moving the Dirac points in reciprocal space until they meet and annihilate. After merging, there is a gap in the spectrum and no more Dirac fermions but two uncoupled bands\footnote{Indeed, after merging, the Berry curvature is essentially zero in the whole BZ, so that the semiclassical equations of motion of an electron are identical to that of a truly isolated band.}. This explains the stability and robustness of Dirac fermions \cite{Volovik}. 

In the following, we describe in more details the merging transition of Dirac points. We start by giving an example (graphene under uniaxial strain) and then show how to construct a minimal model that captures the relevant ingredients of the transition. Then we describe a recent experiment with artificial graphene that resulted in the observation of the merging transition and the corresponding theory of Landau-Zener-St\"uckelberg transition that we developed to interpret it. 

\subsection{Graphene under uniaxial strain}
We first study a simple model to understand the different ingredients needed to describe the merging transition of two Dirac cones. Consider a tight-binding model of graphene and call $t_1$, $t_2$, $t_3$ the three nearest neighbor hopping amplitudes. The three corresponding distances between nearest neighbors are called $\delta_1$, $\delta_2$ and $\delta_3$, see fig.~\ref{fig:graphenebandstructure}(a). The undeformed honeycomb lattice corresponds to $\delta_1=\delta_2=\delta_3=a$ and $t_1=t_2=t_3=t$. Then, we deform the lattice by uniaxial strain such that one of the three nearest neighbor hopping amplitudes $t'\equiv t_3$ becomes different from the two others $t\equiv t_1=t_2$, simultaneously $\delta_3\neq \delta_1=\delta_2$. For example, making $t'>t$ and $\delta_3<\delta_1=\delta_2$ can be done either by compression in the $y$ direction or by tension in the $x$ direction, see fig.~\ref{fig:graphenebandstructure}(a). In general, under such a deformation, the physical modifications are as follows: (1) the Bravais lattice is deformed and therefore also the reciprocal lattice and the corresponding Brillouin zone (e.g. its high symmetry points such as corners $K$ and $K'$ move); (2) the hopping parameters change; (3) as a result the Dirac points move away from the high symmetry points (in other words, it becomes crucial to distinguish the position of the BZ corners and that of the Dirac points); (4) the Dirac cones become anisotropic (they have different velocities in perpendicular directions); (5) the Dirac cones become tilted; (6) eventually, if the strain is strong enough, the two Dirac points may merge, annihilate and open a gap.

\begin{figure}[htb]
\begin{center}
\includegraphics[height=5cm]{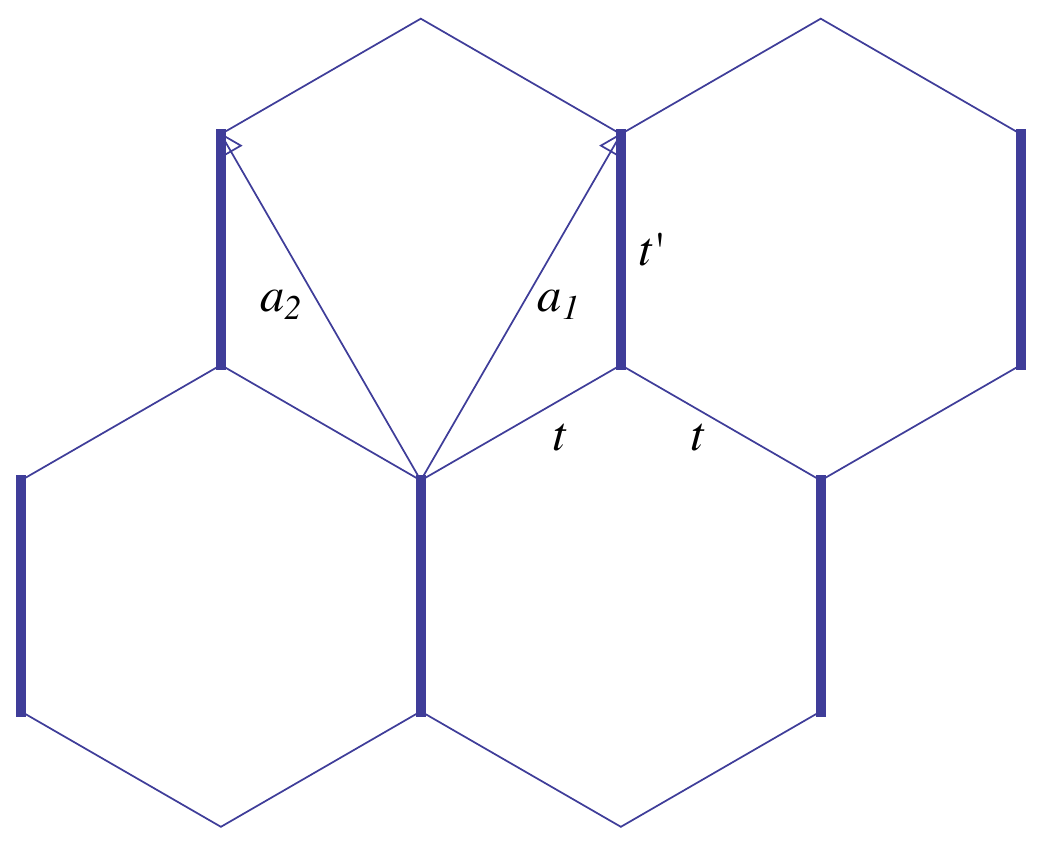}
\caption{\label{fig:ttpmodel}Anisotropic nearest neighbor tight-binding model on the honeycomb lattice (the ``$t-t'$ model''). The Bravais lattice vectors $\vec{a}_1=a(\frac{\sqrt{3}}{2},\frac{3}{2})$, $\vec{a}_2=a(-\frac{\sqrt{3}}{2},\frac{3}{2})$, with $a$ the nearest neighbor distance, and the three hopping amplitudes $t'\equiv t_3$ and $t\equiv t_1=t_2$ are indicated. From Montambaux et al. \cite{Montambaux2009b}.}
\end{center}
\end{figure}
As shown in \cite{Pereira2009}, the geometrical deformation of the lattice is unessential. What is important is that the hopping amplitudes vary. In reality both things are driven by strain but in the theoretical description they can be artificially decoupled. In the following, we assume that the honeycomb lattice is unaffected by strain ($\delta_1=\delta_2=\delta_3=a$) and that only its hopping amplitudes vary ($t'\neq t$). In such a case, the Dirac points do move away from the corners of the BZ, which is the physically important effect \cite{Pereira2009}. In section \ref{sectiontilt}, we have shown that tilting of the Dirac cones happens in the presence of diagonal terms (in the $A$, $B$ subspace) such as next-nearest neighbor hopping \cite{Goerbig2008}. The tilt can be important in some circumstances, but as it does not modify the topological properties of the merging transition\footnote{Indeed, as shown in the section \ref{sectiontilt} on tilted cones, the tilt gives a contribution to the low energy effective Hamiltonian which is proportional to the identity matrix $\sigma_0$ in $A$, $B$ subspace. Therefore the Hamiltonians with and without the tilt have identical eigenvectors and as a consequence  identical Berry curvature. In addition, the position of the Dirac points only depends on off-diagonal terms in the $H(\vk)$ and not on next-nearest neighbor hopping.}, we will also neglect it. In the end, we consider an undeformed honeycomb lattice with anisotropic nearest neighbor hoppings $t'\neq t$ (in short the ``$t-t'$ model'', see fig.~\ref{fig:ttpmodel}). This model is described by the $2\times 2$ $\vk$-dependent Hamiltonian
\be
H(\vk)=\left( \begin{array}{cc}0&f(\vk)^*\\f(\vk)&0 \end{array}\right) \textrm{ where } f(\vk)=-\left(t'+t e^{i \vk \cdot \vec{a}_1}+t e^{i \vk \cdot \vec{a}_2}\right)
\ee
where $\vec{a}_1=a(\frac{\sqrt{3}}{2},\frac{3}{2})$, $\vec{a}_2=a(-\frac{\sqrt{3}}{2},\frac{3}{2})$ are the Bravais lattice vectors of the honeycomb lattice with $a$ the nearest-neighbor distance and we take $\hbar\equiv 1$. 
The merging transition of this model was studied in several papers \cite{Hasegawa2006,Zhu2007,Dietl2008,Wunsch2008,Montambaux2009a,Montambaux2009b}. It is easy to understand its physics qualitatively. Making $t'$ larger than $t$ drives the system towards dimerization. In the limit $t'\gg t$, where links $t_1$ and $t_2$ are cut, the system consists of isolated dimers on the $t_3$ links which have a gap $\sim 2t'$. We note that in the opposite direction $t'\ll t$, when the $t_3$ links are cut, the system moves towards decoupled one-dimensional chains made of alternating $t_1$ and $t_2$ links. It becomes very anisotropic while remaining gapless. We are interested in the situation $t'>t$ where a gap eventually opens.
\begin{figure}[htb]
\begin{center}
\subfigure[]{\includegraphics[height=5cm,angle=90]{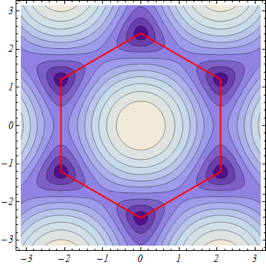}}
\subfigure[]{\includegraphics[height=5cm,angle=90]{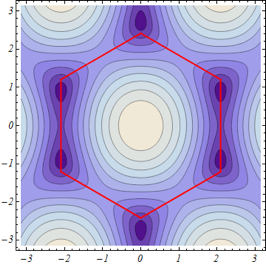}}
\subfigure[]{\includegraphics[height=5cm,angle=90]{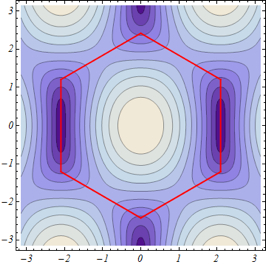}}
\caption{\label{contourstpt}Iso-energy lines of the $t-t'$ model in the first Brillouin zone (red hexagon; $k_x$,$k_y$ are in units of $1/a$) as a function of $t'/t$. The $M$ points are on the edge of the BZ at equal distance between its corners ($K$ points). (a) $t'=t$, Dirac points coincide with the $K$ points; (b) $t'=1.4t$, Dirac points move away from the $K$ corners along the edges of the BZ; (c) $t'=2t$, Dirac points merge at one of the $M$ points.}
\end{center}
\end{figure}
Here we do not enter into the details of that specific model but simply summarize the main features of the transition (i) on the spectrum $\varepsilon(\vk)=\pm|f(\vk)|=\pm |t'+te^{i\vk \cdot \vec{a}_1}+te^{i\vk \cdot \vec{a}_2}|$ (see fig.~\ref{contourstpt}) and (ii) on the quantized vortices in the relative phase $\phi$ defined by $f(\vk)=|f(\vk)|e^{-i\phi_{\vk}}$ (see fig.~\ref{vortexanni} for a qualitatively similar behavior although on a different model, to be discussed below). When $t'$ becomes slightly larger than $t$, the two Dirac points move away from the $K,K'$ corners and towards one of the three $M$ points (on the edge of the BZ, mid-way between $K$ and $K'$). The Dirac points are located at $k_x a=\pm \frac{2}{\sqrt{3}}\arccos \frac{t'}{2t}$ and $k_y a=\frac{2\pi}{3}$, which is only possible if $t'\leq 2t$. We start from $t'=t$ (see fig.~\ref{contourstpt}(a)), which corresponds to undeformed graphene. The Dirac cones are isotropic and located at the corners of the hexagonal BZ. As $2t>t'>t$, the Dirac points approach each other and the Dirac cones become anisotropic: the velocity in the approach direction decreases, while that in the perpendicular direction remains almost unchanged (see fig.~\ref{contourstpt}(b)). When $t'=2t$, the two Dirac points coincide at $k_x a=0$ and $k_y a=\frac{2\pi}{3}$, which is one of the $M$ points of the BZ (see fig.~\ref{contourstpt}(c)). The velocity in the approach direction vanishes and therefore the low energy spectrum becomes quadratic in that direction, while it remains linear in the perpendicular direction $\varepsilon \sim \pm \sqrt{q_x^4 + q_y^2}$ (where $\vq$ is defined with respect to the $M$ point), which is known as a hybrid or semi-Dirac spectrum \cite{Dietl2008}. At the transition, the vortex and the anti-vortex in $\phi$ are on top of each other and carry a total charge 0. When $t'>2t$ a gap $2(t'-2t)$ opens, the Dirac points and the quantized vortices no longer exist. 

Such a transition is probably unreachable in graphene as it requires a very large strain of $\sim 23$~\% \cite{Pereira2009}. Another crystal in which a merging transition of this kind is expected is the organic salt $\alpha$-(BEDT-TTF)$_2$I$_3$ \cite{Katayama2006,Montambaux2009a}. It is a quasi-2D crystal containing four large organic molecules per unit cell\footnote{There is an effective description in terms of a $2$ bands model with two atoms per unit cell, a tetragonal Bravais lattice and four different hopping amplitudes \cite{Katayama2006}.}. Its hopping parameters are easily changed under hydrostatic pressure. Above a critical pressure of $\sim 1.5$~GPa, it features Dirac cones which are tilted, anisotropic and within the BZ (rather than on its edges). These points are expected to move under applying extra pressure and to merge at the $\Gamma$ point at a hydrostatic pressure of $\sim 4$~GPa \cite{Fukuyama}. This has not been observed yet but there are current experimental efforts underway in Orsay \cite{Monteverde2013}. Before discussing systems where this transition was achieved, we derive a minimal Hamiltonian for the effective description of the merging transition.
	
\subsection{Minimal description of the merging transition}
We study under which general conditions a pair of Dirac points in the electronic spectrum of a 2D crystal may merge and annihilate \cite{Montambaux2009a,Montambaux2009b}. We consider a two-bands Hamiltonian for a 2D crystal with two atoms $A$ and $B$ per unit cell. As discussed before, we do not modify the geometry of the lattice and assume that only the hopping amplitudes are changed by the strain. The $2\times 2$ $\vk$-dependent Hamiltonian is
$$ H({\vec k})= \left(%
\begin{array}{cc}
h_{AA}({\vec k})  & h_{AB}({\vec k}) \\
  h_{BA}({\vec k}) &  h_{BB}({\vec k}) \\
\end{array}%
\right) \ ,  $$ with the 2D wave vector ${\vec k}$. 
Time-reversal symmetry ($H({\vec k})=H^*(-{\vec k})$) imposes
${h_{AB}}({\vec k})={h_{BA}}^*({\vec k})\equiv f({\vec k})$ and,
together with hermiticity, real symmetric diagonal terms
$h_{AA}({\vec k})=h_{AA}(-{\vec k})$ ($h_{BB}({\vec
k})=h_{BB}(-{\vec k})$). Furthermore, we consider a 2D lattice with
inversion symmetry such that $h_{AA}({\vec k})=h_{BB}({\vec k})$.
The resulting energy dispersion reads $\varepsilon_{\pm}({\vec
k})=h_{AA}({\vec k})\pm |f({\vec k})|$, and we set $h_{AA}({\vec k})=0$ because this term simply shifts the energy as a
function of the wave vector (and may tilt the Dirac cones) but does not affect the topological properties of the semi-metal-insulator phase transition as discussed previously.

From now on, we therefore discuss the Hamiltonian in its reduced form
\be  H({\vec k})= \left(%
\begin{array}{cc}
0  & f({\vec k}) \\
  f^*({\vec k}) &  0 \\
\end{array}%
\right) \ , \label{H} \ee where the off-diagonal terms have the
periodicity of the Bravais lattice and may be written quite
generally in the form
\be f({\vec k})= \sum_{m,n} t_{mn} e^{- i {\vec k} \cdot \R_{mn}} \
, \label{fofk} \ee
where the $t_{mn}$'s are real, a consequence of time-reversal
symmetry $H({\vec k})=H^*(-{\vec k})$, and   $\R_{mn}=
m \a_1 + n \a_2$ are vectors of the underlying Bravais lattice.

The energy spectrum is given by  $\vep({\vec k})=\pm |f({\vec k})|$,
and the Dirac points, that we name $\D$ and $-\D$ are solutions of
$f(\D)=0$. Since $f({\vec k})=f^*(-{\vec k})$, the Dirac points,
when they exist, necessarily come in by pairs as a result of time-reversal symmetry. The
position $\D$ of the Dirac points can be anywhere in the BZ and move
upon variation of the band parameters $t_{mn}$. Around the Dirac
points $\pm \D$, the function $f({\vec k})$ varies linearly. Writing
${\vec k} = \pm D+\vq$, we find
\be f(\pm \D+\vq)=\vq \cdot (\pm {\vec v}_1 - i {\vec v}_2)
 \label{fofq}
\ee
where the velocities ${\vec v}_1$ and ${\vec v}_2$ are given by
\begin{eqnarray}
{\vec v}_1 &=& \sum_{mn} t_{mn} \R_{mn} \sin \D \cdot \R_{mn} \nonumber \\
{\vec v}_2 &=& \sum_{mn} t_{mn} \R_{mn} \cos \D \cdot \R_{mn}
\label{v1v2}
\end{eqnarray}

Upon variation of the band parameters, the two Dirac points may
approach each other and merge into a single point $\D_0$. This
happens when $\D=-\D$ modulo a reciprocal lattice vector $\G=p
\a^*_1 + q \a_2^*$, where  $\a^*_1$ and  $\a_2^*$ span the
reciprocal lattice. Therefore, the location of this merging point is
simply the time-reversal invariant momentum $\D_0= \G/2$. 
There are then four possible inequivalent points the coordinates of which are $\D_0= ( p\a^*_1 + q \a_2^*)/2$,
with $(p,q)$ = $(0,0)$, $(1,0)$, $(0,1)$, and $(1,1)$. For example in the honeycomb lattice, these points are the $\Gamma$ point and the three $M$ points. The condition
$ f(\D_0)= \sum_{mn} (-1)^{\beta_{mn}} t_{mn} =0$, where
$\beta_{mn}= p m + q n$, defines a manifold in the space of band
parameters. As we discuss below, this manifold separates a
semi-metallic phase with two Dirac cones and a band insulator\footnote{Here, we restrict to the case of only two Dirac points. Actually, with long range hoppings, it is possible to have many pairs of Dirac points, see e.g. \cite{Sticlet2013}. Such a situation is more delicate to discuss as one should look for the sign change of $f(\D_0)$ for all the possible $\D_0$ points. This is discussed in \cite{FuKane2007}.}.

In the vicinity of the $\D_0$ point, $f$ is {\it purely imaginary}
(${\vec v}^0_1=0$), since $\sin (\G \cdot \R_{mn}/2)=0$.
Consequently, to lowest order, the linearized Hamiltonian  reduces
to ${\cal H}= \vq
 \cdot {\vec v}_2^0  \sigma^y$, where  ${\vec v}^0_2=\sum_{mn} (-1)^{\beta_{mn}} t_{mn} \R_{mn} $.
We choose the local  reference system such that ${\vec v}^0_2 \equiv
c_y \, \hat y$ defines the $y$-direction. In order
to account for the dispersion in the local $x$-direction,  we have
to expand $f(\D_0+\vq)$ to second order in $\vq$:
 \be f(\D_0+\vq)= - i \vq \cdot {\vec v}^0_2 -{1 \over 2} \sum_{mn} (-1)^{\beta_{mn}}
  t_{mn} ( \vq \cdot
 \R_{mn})^2 \ .  \label{fofq0} \ee
Keeping the quadratic term in $q_x$, the new Hamiltonian
 may be written as
\be H_0(\vq)   = \left(
  \begin{array}{cc}
    0 & {q_x^2 \over 2 m^*} -  i c_y q_y  \\
 {q_x^2 \over 2 m^*} +  i c_y q_y & 0 \\
  \end{array}
\right) \ .  \label{H0topo}\ee
where the effective mass $m^*$ is defined by
 \be {1 \over m^*}=  \sum_{mn} (-1)^{{\beta_{mn}}+1} t_{mn}
 R^2_{mn,x} \ ,  \ee
and where  $R_{mn,x}$ is the component of $\R_{mn}$ along   the {\it
local} $x$-axis (perpendicular to ${\vec v}_2^0$).  The terms of
order $q_y^2$ and $q_x q_y$ are neglected at low energy. The diagonalization of $H_0(\vq)$ is
straightforward and the energy spectrum
\be \vep = \pm \sqrt{c_y^2 q_y^2+\left({q_x^2 \over  2 m^*}\right)^2}
\ee
has a remarkable structure: it is linear in one direction and quadratic in the other. 

The merging of the Dirac points in $D_0$ marks the transition
between a semi-metallic phase and an insulating phase. The transition is driven by the parameter
\be \Delta\equiv  f(\D_0)= \sum_{mn} (-1)^{\beta_{mn}} t_{mn} \label{gap} \ee
which changes  its sign at the transition. This parameter $\Delta$
therefore drives the transition and we call it the merging parameter or merging gap. In the vicinity of the transition,
the Hamiltonian has the form
\be H_{min}(\vq)=
\left(
  \begin{array}{cc}
    0 & \Delta+ {q_x^2 \over 2 m^*} -  i c_y q_y  \\
 \Delta + {q_x^2 \over 2 m^*} +  i c_y q_y & 0 \\
  \end{array}
\right) \label{newH} \ee
with the spectrum
\be \vep= \pm \sqrt{(\Delta + {q_x^2 \over 2 m^*})^2 + q_y^2 c_y^2}
\label{Ueps1} \ee
\begin{figure}[htb]
\begin{center}
\includegraphics[height=7cm]{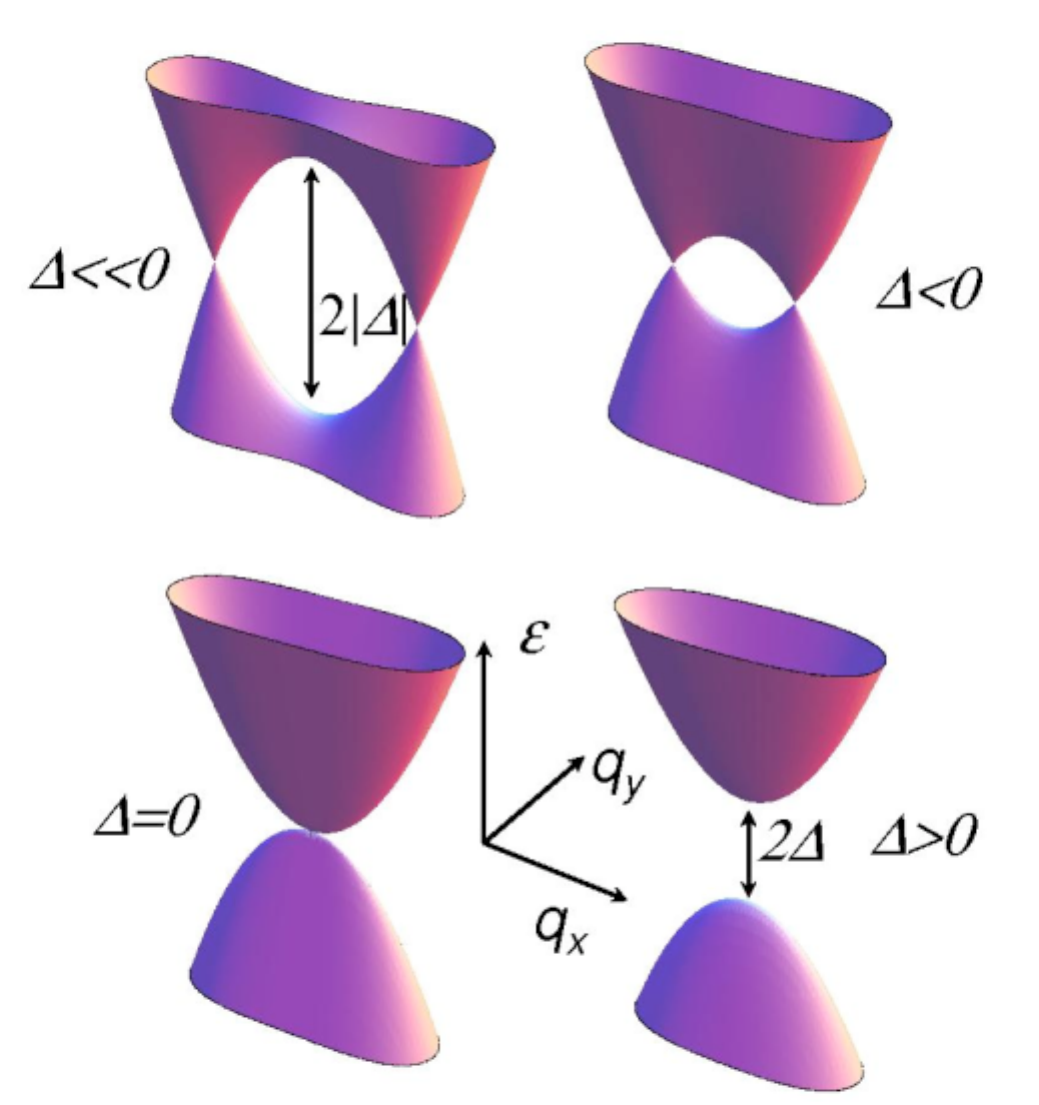}
\caption{\label{fig.bicones}Evolution of the energy spectrum of the minimal Hamiltonian as a function of the driving parameter $\Delta$ for the merging transition when it changes sign. The saddle point when $\Delta<0$ becomes a gap when $\Delta>0$. Exactly at the merging ($\Delta=0$), the spectrum is linear in $q_y$ and quadratic in $q_x$. From Montambaux et al. \cite{Montambaux2009b}.}
\end{center}
\end{figure}
\begin{figure}[htb]
\begin{center}
\subfigure[]{\includegraphics[height=4cm]{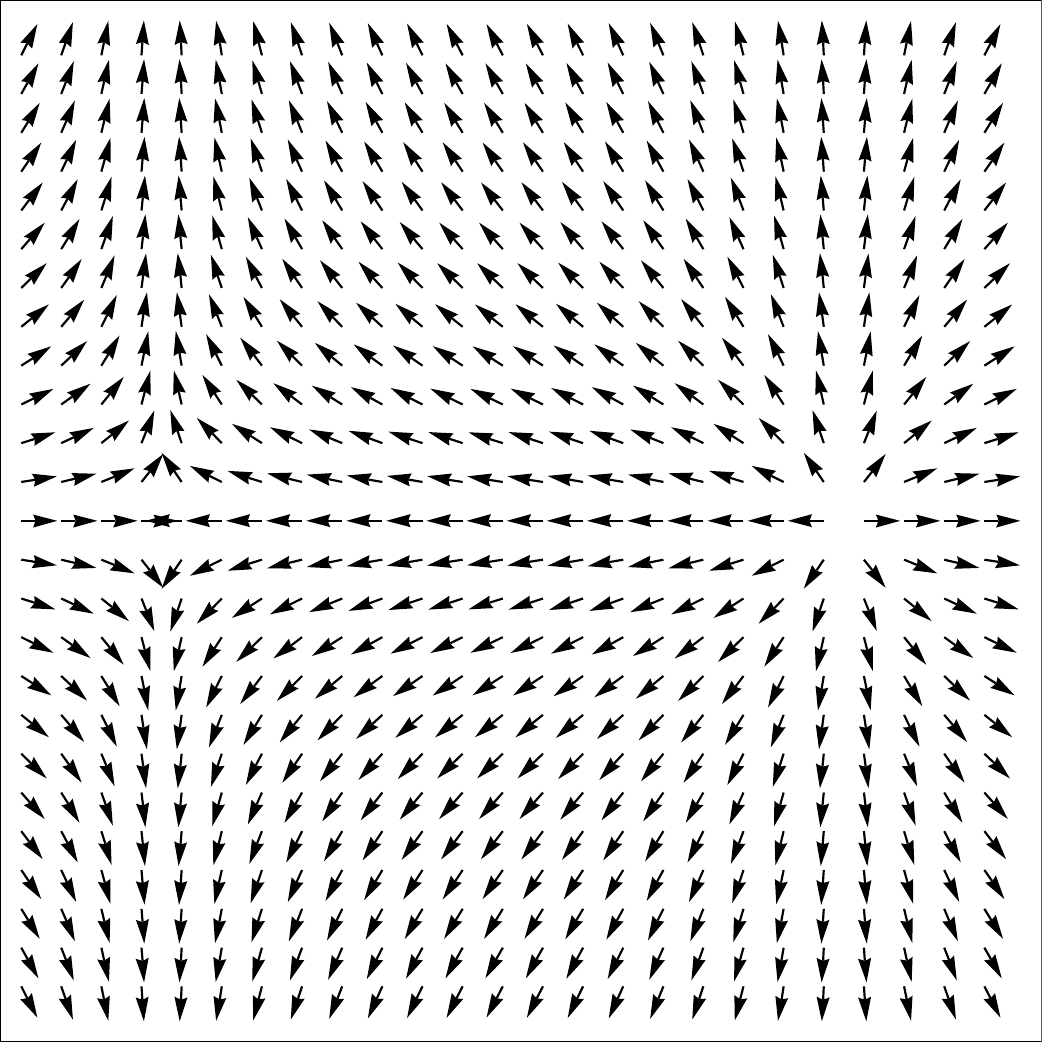}}
\subfigure[]{\includegraphics[height=4cm]{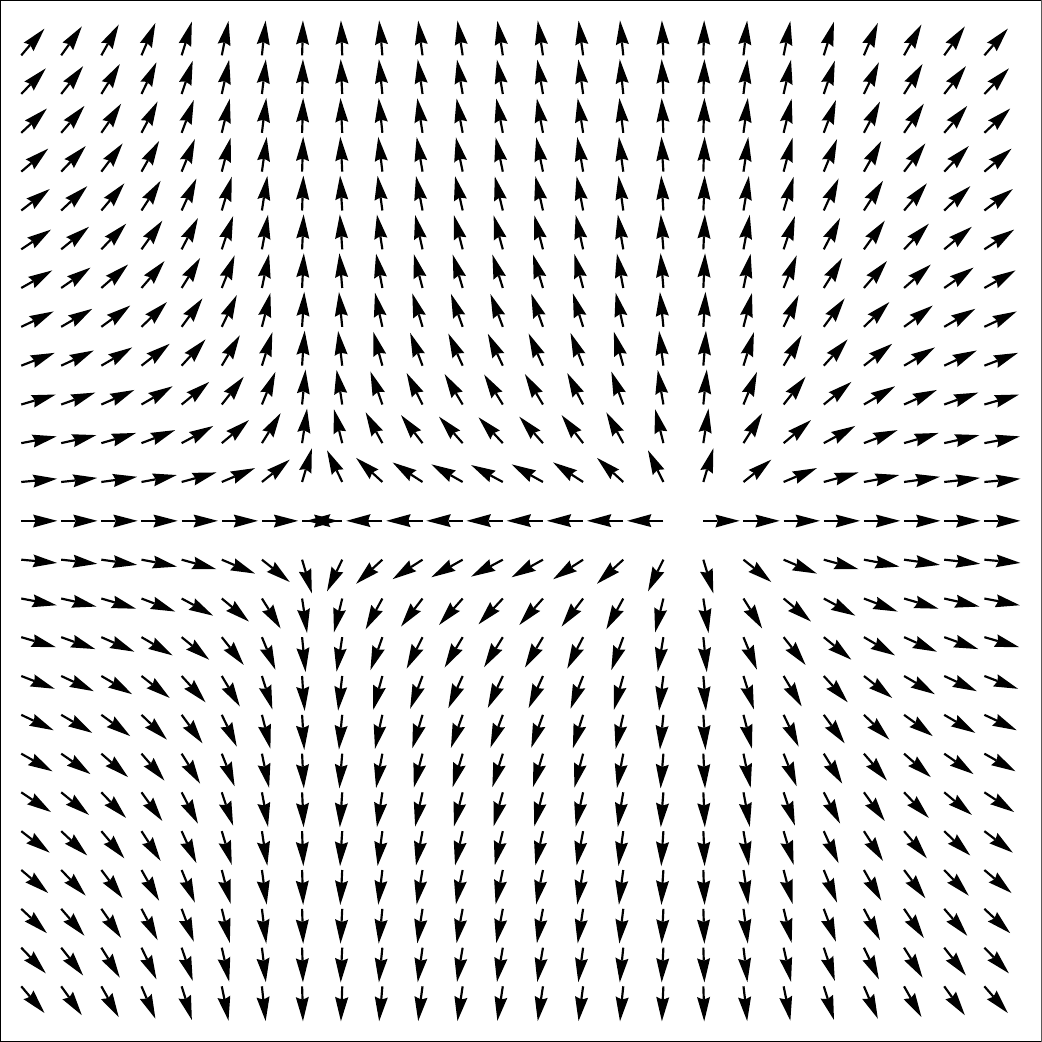}}
\subfigure[]{\includegraphics[height=4cm]{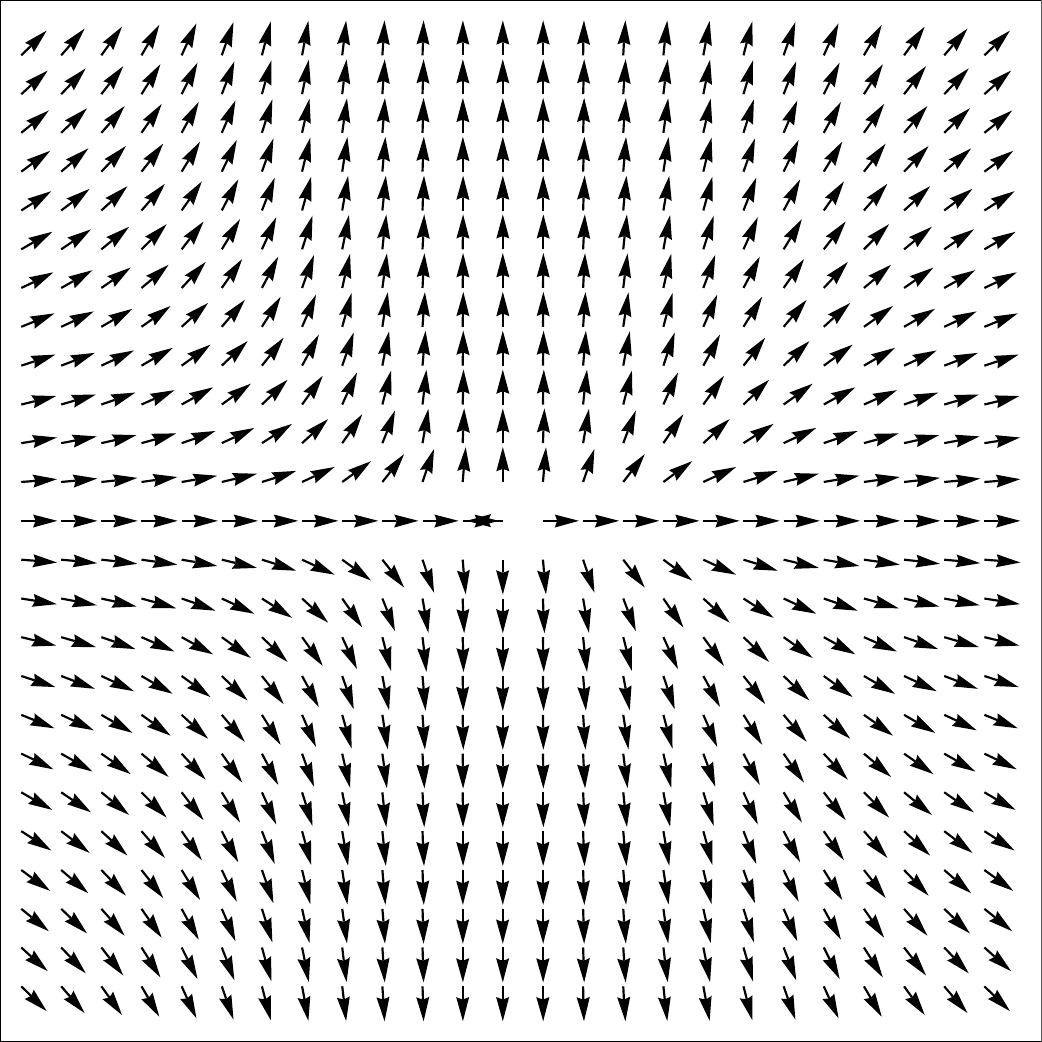}}
\subfigure[]{\includegraphics[height=4cm]{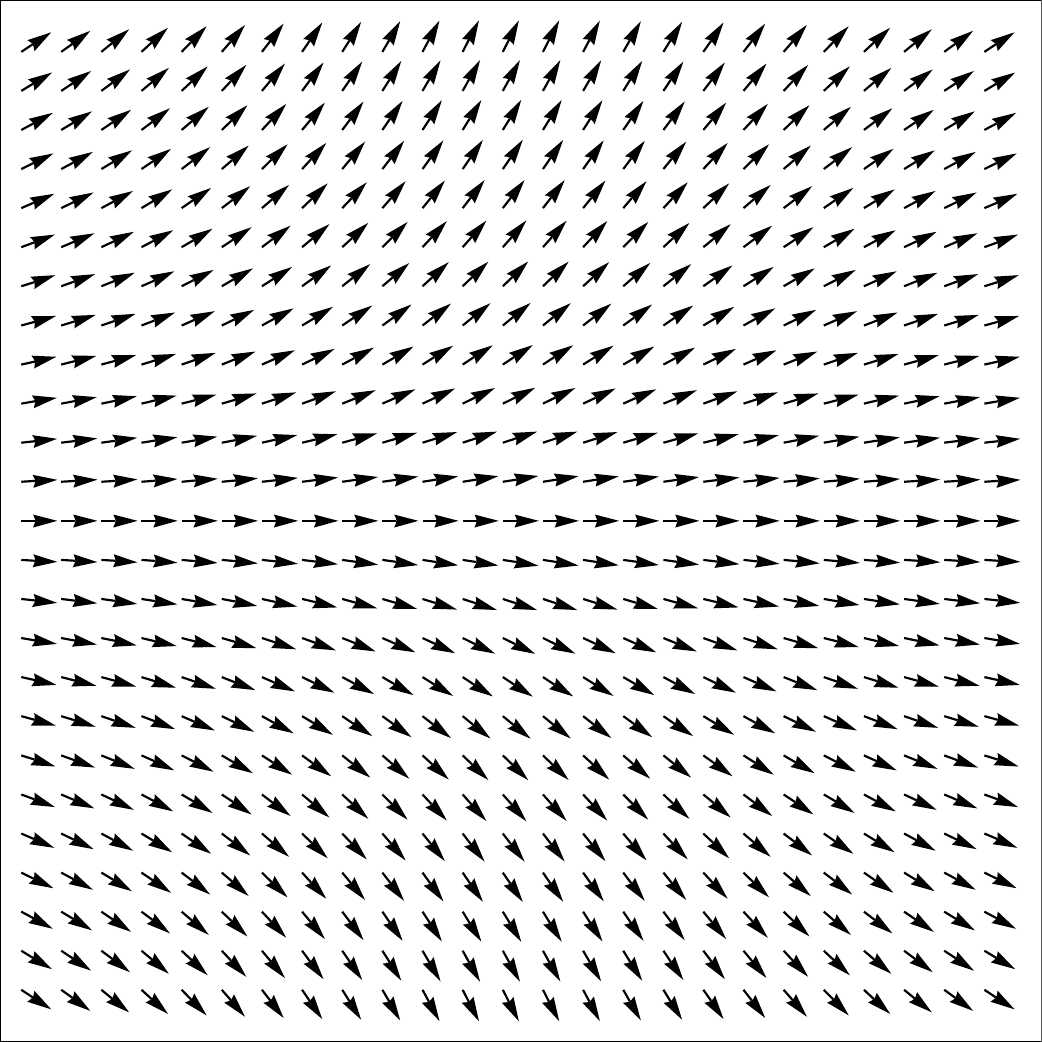}}
\caption{\label{vortexanni}Relative phase $\phi_{\vq}$ of the minimal Hamiltonian in reciprocal space $(q_x,q_y)$ [in units of $m^* c_y$] as a function of the driving parameter $\Delta$ [in units of $m^*c_y^2$]. The phase is given by $\phi_{\vq}=\textrm{Arg }[\Delta+\frac{q_x^2}{2m^*}+ic_yq_y]$. It shows the annihilation of two quantized vortices $+1 + (-1) \to 0$ across the merging transition. (a) $\Delta=-1$; (b)$\Delta=-0.3$; (c): $\Delta=0$ and (d) $\Delta=1$. From Montambaux et al. \cite{Montambaux2009b}.}
\end{center}
\end{figure}

The Hamiltonian ({\ref{newH}) has a remarkable structure and describes properly the vicinity of the topological transition, as shown in fig.~\ref{fig.bicones} for the spectrum given above and in fig.~\ref{vortexanni} for the relative phase $\phi\equiv \textrm{Arg }[\Delta+\frac{q_x^2}{2m^*}+ic_yq_y]$. When $ m^* \Delta$ is negative (we  choose $m^* >0$ without loss of generality), the spectrum exhibits two Dirac cones and a saddle point in $\D_0$ (at half distance between the two Dirac  points). Increasing $\Delta$ from negative to positive values, the saddle point evolves into the hybrid point at the transition ($\Delta=0$) before a gap $2 \Delta >0$ opens.

We stress that this Hamiltonian has the minimal structure needed to describe the physics of Dirac points, even far from the transition, since it captures quite simply the coupling between the two valleys associated with  the two Dirac points. In particular, we can relate the coupling between valleys to a double well potential problem. This is quite unusual, as effective low energy Hamiltonians describing the two valleys are typically $4\times 4$ matrices rather than $2\times 2$. Here the difference comes from expanding around the unique merging point $\vec{D}_0$ rather than around the two Dirac points $\pm \vec{D}$. The minimum Hamiltonian is not general or complete in the sense that we have removed unessential terms such as $q_y^2$ and $q_x q_y$ in the $\sigma_x$ term. Also, we have not considered terms on the diagonal $\sigma_0$ such as $q_x^2$, $q_y^2$ and $q_x q_y$ that exist and would account for the tilt of the cones. Therefore we name ({\ref{newH}) the minimal Hamiltonian\footnote{Even if in Refs.~\cite{Montambaux2009a,Montambaux2009b} we called it a ``universal Hamiltonian'', in hindsight, we find that ``minimal Hamiltonian'' is more appropriate.} as it is the simplest one (with only three independent and real parameters: $m^*>0$, $c_y>0$ and $\Delta$) that correctly captures the essentials -- such as the motion of the Dirac points, the annihilation of the quantized vortices, the anisotropy in the velocity, the semi-Dirac spectrum at the transition, etc. -- of the merging transition.

To give a concrete example, in the case of the $t-t'$ model, the three parameters of the minimal Hamiltonian can be taken as $\Delta=t'-2t$, $c_y=\frac{3t'a}{2}$ and $m^*=\frac{8}{3(2t+t')a^2}$ \cite{Montambaux2009b}. This Hamiltonian can be used, for example, to study how Landau levels evolve across the merging. When deep in the gapless phase ($-\Delta\gg t$), there are essentially two separate valleys giving each a set of relativistic Landau levels $\vep_n \propto \pm \sqrt{Bn}$. Then approaching the merging transition ($0<-\Delta\ll t$), the degeneracy between LLs is progressively lifted by the valley coupling induced by the magnetic field. This is known as magnetic breakdown and can be studied using semiclassical methods, which give the following gap opening for the $n=0$ LL $\Delta \vep_{n=0} \propto \exp(-\# \frac{|\Delta|}{(m^* c_y^2  \omega_c^2)^{1/3}})$ where $\omega_c\equiv eB/m^*$ \cite{Montambaux2009b} in agreement with a direct evaluation on the $t-t'$ model \cite{Esaki2009}. At the merging ($\Delta=0$), the LLs are unusual and scale as $\vep_n\propto \pm [B(n+1/2)]^{2/3}$ \cite{Dietl2008}, while deep in the gapped phase ($\Delta\gg t$), familiar LLs $\vep_n-\Delta \propto \pm B(n+1/2)$ are recovered. Once, $\Delta\geq 0$, the Landau index jumps $n\to n+1/2$ as a result of the annihilation of quantized vortices.

An interesting extension is to consider a situation in which the Dirac fermions are massive before the transition (such as occurs in boron nitride). A merging transition (annihilation of vortices) can also occur even if Dirac points, in the sense of contact points, are absent. This was recently studied in the context of the organic salts $\alpha$-(BEDT-TTF)$_2$I$_3$ \cite{Kobayashi2011}. The minimal model in this case depends on four parameters and is simply $H_{min}(\vk)+M\sigma_z$ where the last term breaks the inversion symmetry and endows the Dirac fermions with a Semenoff mass as in boron nitride.  

Another recent development in the study of deformed graphene (or strain engineering) is to consider inhomogeneous (rather than homogeneous) strain, which leads to the appearance of pseudo-gauge fields for the Dirac fermions. For example, a particular deformation that correspond to a uniform pseudo-magnetic field -- that does not break time-reversal symmetry -- has been proposed for the massless Dirac fermions of graphene \cite{GKG}, see also Ref.~\cite{Manes2007}.

Below, we study in detail a specific application of the minimal Hamiltonian (\ref{newH}) to the study of Landau-Zener-St\"uckelberg interferometry in an artificial graphene-like crystal made of cold atoms.
	
\subsection{Merging transition in artificial graphene}	
Very recently, the merging transition was observed in three instances of ``artificial graphene''. One consists of ultracold fermionic atoms loaded in a honeycomb-like optical lattice \cite{Tarruell2012}. The second is even more exotic as it is made of microwave photons hopping via evanescent waves in a deformed honeycomb lattice of dielectric resonators \cite{Mortessagne2013}. The third is a photonic honeycomb lattice \cite{Segev2012}. In the following we concentrate on the cold atoms realization, which we studied in detail in Refs.~\cite{Lim2012,Fuchs2012}.

\begin{figure}[htb]
\begin{center}
\includegraphics[height=5cm]{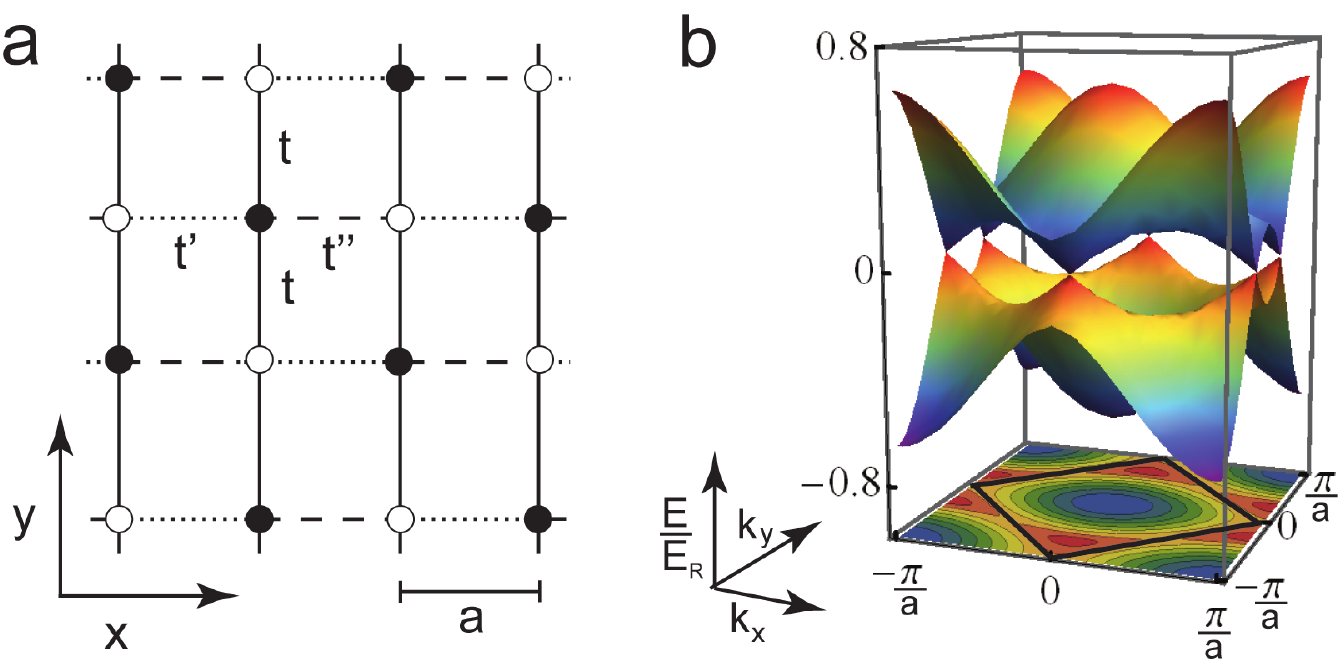}
\caption{\label{fig:brickwall}(a) Anisotropic square lattice model indicating the different hopping amplitudes $t$, $t$, $t'$, $t''$ and the two inequivalent sites. The nearest neighbor distance is $a$. (b) Band structure in the gapless phase featuring two inequivalent Dirac points ($t=t'=0.2E_R$ and $t''=0.05E_R$ where $E_R\equiv \frac{\pi^2\hbar^2}{2ma^2}$ is the so-called recoil energy with $m$ the atomic mass.). The first Brillouin zone is indicated by a black square. The portion of reciprocal space contained between $|k_x|,|k_y|\leq \frac{\pi}{a}$ and the first BZ corresponds to the second BZ. From Lim et al. \cite{Lim2012}.}
\end{center}
\end{figure}
In the Z\"urich experiment \cite{Tarruell2012}, fermionic atoms $^{40}$K in a single hyperfine state are confined in a harmonic trap. Because of a strong anisotropy in the external trapping, they realize a two-dimensional Fermi gas with a pancake shape\footnote{This was actually not the case in the experiment. The gas was only quasi -- rather than truly -- two-dimensional. However, this simplification makes the discussion easier. We will come back to this approximation later when discussing St\"uckelberg interferences.}. These atoms are cooled to quantum degeneracy such that the temperature $T$ is roughly 10\% of the Fermi energy. In addition there is a two-dimensional optical lattice realized by laser fields that can be tuned to produce different kinds of lattices such as triangular, checkerboard or honeycomb-like. The nearest neighbor distance is $a=\frac{\lambda}{2}$ in terms of the laser wavelength $\lambda$ and a convenient energy scale is given by the so-called recoil energy $E_R\equiv \frac{h^2}{2m\lambda^2}$, where $m$ is the atomic mass. An effective description -- keeping only the lowest two bands -- is provided by an anisotropic tight-binding model for atoms moving in a square lattice, with four nearest neighbor hopping amplitudes $t$, $t$, $t'$ and $t''$, see fig.~\ref{fig:brickwall} \cite{Lim2012}. When $t'\neq t''$, this model has two sites per unit cell and, when $t'+t''<2t$, it features two Dirac points. It is actually quite similar to the effective two-sites model describing the organic salt $\alpha$-(BEDT-TTF)$_2$I$_3$ \cite{Katayama2006,Montambaux2009a}. In particular, when $t''=0$, and $t'=t$ it is a brick-wall lattice, which has the same connectivity as the honeycomb lattice albeit with a rectangular rather than a triangular geometry. Therefore fermionic atoms moving in such a honeycomb-like lattice realize a kind of artificial graphene. The main advantage is that this graphene is highly tunable. Indeed, varying the parameters of the optical lattice changes the value of the hopping amplitudes and modifies the band structure in a controlled way. It is therefore possible to easily reach the situation where the two Dirac points merge. In particular, $t'+t''<2t$ corresponds to the gapless Dirac phase, $t'+t''=2t$ to the merging transition and $t'+t''>2t$ to the gapped phase.

\begin{figure}[htb]
\begin{center}
\subfigure[]{\includegraphics[height=5cm]{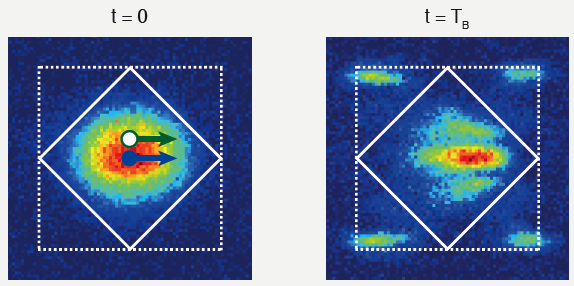}}
\subfigure[]{\includegraphics[height=5cm]{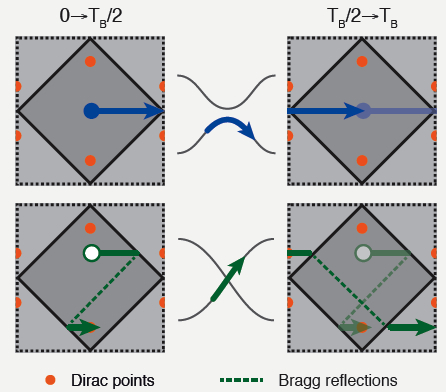}}
\caption{\label{fig:blochoscillation}Atomic density in reciprocal space $(k_x,k_y)$. The first/second BZ is indicated by a full/dashed line. The  dashed square goes from $-\frac{\pi}{a}$ to $\frac{\pi}{a}$. Single Bloch oscillation of the atomic cloud in the $x$ direction (``single cone case''). (a): initially ($t=0$), the Fermi sea is at equilibrium in the lower band (first BZ). After a single Bloch oscillation ($t=T_B$ which is the Bloch period), the Fermi sea has lost some of its atoms that are found in the second BZ corresponding to the upper band. (b): Two particular trajectories in reciprocal space. The blue atom moves far from the Dirac points and stays in the lower band, while the green atom encounters a Dirac point at which it jumps to the upper band. The position of Dirac points is indicated. From Tarruell et al. \cite{Tarruell2012}.}
\end{center}
\end{figure}
The atomic gas is essentially ideal as the atom-atom interaction are short ranged and the Pauli principle, which applies to this single species of fermions (equivalent to spin polarized electrons), forbids collisions. Such an isolated system is not connected to reservoirs or contacts and can therefore not be probed by transport-like experiments, as is common in solid state physics. In addition, because of the external harmonic trapping, the density of atoms is inhomogeneous and so is the Fermi level. Therefore it seems hard to have the Fermi level close to the Dirac points and to probe Dirac fermions and their motion. A way out of these two issues -- inhomogeneity of the Fermi level and absence of contacts -- was found by the Z\"urich group \cite{Tarruell2012}. They loaded a small number of atoms in their system, such that the Fermi level lies in the valence band, far below half-filling. In order to probe the presence/absence of Dirac points in their band structure, starting from the equilibrium Fermi sea, they applied a constant and homogeneous force $F$ to the atoms so that the latter perform Bloch oscillations in this ultra-clean system. The force is only applied for the expected duration of a single Bloch oscillation (with period $T_B=\frac{h}{Fa}$) and the atoms are then detected using a band mapping technique, which allows one to know whether an atom was found in the lower (valence) or in the upper (conduction) band. The basic idea is that an atom initially in the lower band is finally found in the upper band only if its Bloch trajectory passed close to a contact point between the two bands. In such a case, the atom can tunnel to the upper band via a Landau-Zener process. After a single Bloch oscillation a picture of the atomic cloud in reciprocal space reveals a Fermi sea with missing atoms in the lowest band (seen in the first BZ) and a few atoms in the upper band (seen in the second BZ), see fig.~\ref{fig:blochoscillation}. The result of a single experiment is summarized in the transferred fraction of atoms -- in other words, the number of atoms that were found in the upper band after a single Bloch oscillation divided by the total number of atoms in the gas -- for a given set of optical lattice parameters. Then the experiment is repeated many times in order to obtain the transferred fraction for each set of lattice parameters. The experimental results of \cite{Tarruell2012} are plotted in fig.~\ref{fig:tarruell}. Two perpendicular directions of motion were considered. In the first (``single cone'', motion along $x$) case, an atom encounters at most one Dirac point during a single Bloch oscillation (see fig.~\ref{fig:tarruell}(a)). In the second (``double cone'', motion along $y$) case, it may encounter the two inequivalent Dirac points during a single Bloch oscillation (see fig.~\ref{fig:tarruell}(b)). The figure shows that the transferred fraction is always zero in the gapped phase and can only be finite in the gapless phase. The threshold roughly corresponds to the expected merging transition but there is a striking difference between the single and the double cone cases. In particular the line of maximum transferred fraction (in red in both figures) is always inside the gapless phase, especially in the double cone case, where it is far from the merging transition. We explain these features below.
\begin{figure}[htb]
\begin{center}
\includegraphics[height=6cm]{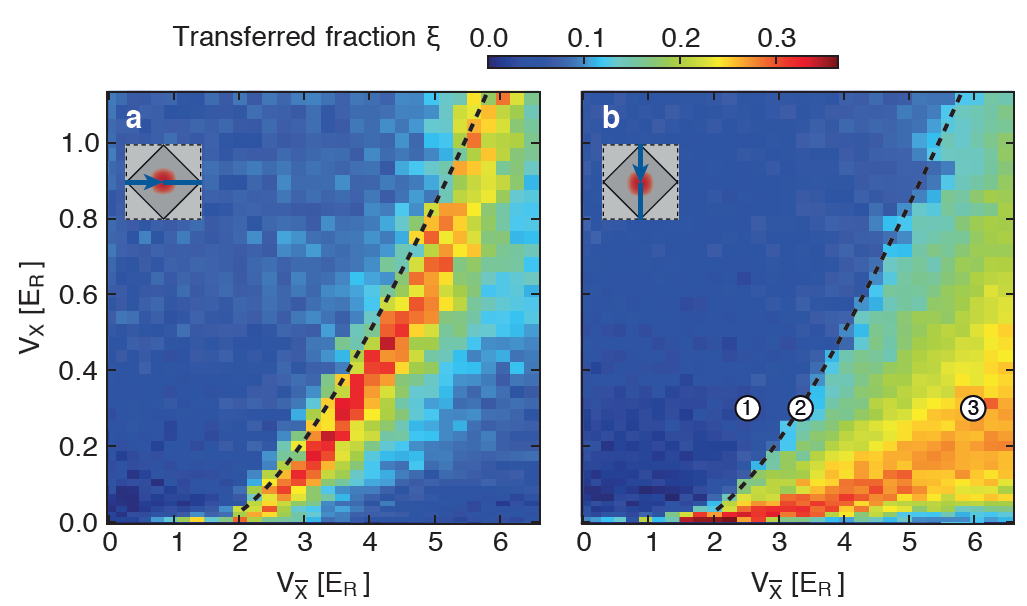}
\caption{\label{fig:tarruell} Color plot of the transferred fraction $\xi$ as a function of two optical lattice parameters $V_{\bar{X}}$ and $V_X$. Qualitatively $V_{\bar{X}}\sim t'-t'' =c_x$ and $V_X\sim 2t-t'-t''=-\Delta$ in terms of the hopping amplitudes $t$, $t'$, $t''$ and the parameters of the minimal Hamiltonian (velocity $c_x$ and driving parameter or merging gap $\Delta$).  The expected position of the merging transition ($\Delta=0$) is indicated by a dashed line. (a): Single cone case (motion along $x$). (b): Double cone case (motion along $y$). The numbered labels refer to the gapped phase (1), the merging transition (2) and the gapless Dirac phase (3). From Tarruell et al. \cite{Tarruell2012}.}
\end{center}
\end{figure}

From now on, we use units such that $\hbar \equiv 1$, $a=\frac{\lambda}{2}\equiv 1$ and $E_R=\frac{\pi^2 \hbar^2}{2ma^2}\equiv 1$. Compared to the previous section on the minimal Hamiltonian, and to confuse the reader, we exchange the $x$ and $y$ axis (this in order to stick to the notations in the experimental paper \cite{Tarruell2012}). Momenta in the full BZ are called $\vk$ while those in the vicinity of the merging point $\frac{\vec{G}}{2}=(0,\pi)$ are called $\vq=\vk-\frac{\vec{G}}{2}$. We start by writing the minimal Hamiltonian used to describe the Z\"urich experiment:
\be
H=\left(\begin{array}{cc}0&\Delta+\frac{q_y^2}{2m^*}-ic_xq_x\\\Delta+\frac{q_y^2}{2m^*}+ic_xq_x&0 \end{array}\right)
\ee
The parameters of the minimal Hamiltonian can be related to that of the tight-binding model: the merging gap $\Delta=t'+t''-2t$, the velocity $c_x=t'-t''$ and the effective mass $m^*=\frac{2}{2t+t'+t''}$ when $\Delta\leq 0$ and $m^*=\frac{1}{2t}$ when $\Delta\geq0$. The dispersion relation in the vicinity of the merging point $(0,\pi)$ is $\vep=\pm \sqrt{c_x^2q_x^2+(\frac{q_y^2}{2m^*}+\Delta)^2}$. The Dirac points only exist if $\Delta<0$ and are located at $q_x=0$ ($k_x=0$) and $q_y=\pm \sqrt{-2m^* \Delta}\equiv \pm q_D$ ($k_y=\pi \pm \sqrt{-2m^* \Delta}$).

First, consider the single cone case, in which the applied force $\vec{F}=F\vec{e}_x$ is in the $x$ direction. The semiclassical equation of motion for an atom with initial momentum $\vk(0)$ is $\vk(t)=\vk(0) + \vec{F}t$. As the Dirac points are always at $k_x=0$, an atom encounters at most a single Dirac cone when moving in the $x$ direction. The probability for an atom to tunnel to the upper band during a single Bloch oscillation is approximatively given by the Landau-Zener (LZ) probability\footnote{The Landau-Zener formula $P_Z=\exp(-\frac{\pi}{c F}(\Delta_g/2)^2)$ gives the probability for a quantum particle to tunnel at an avoided linear band crossing. The particle moves under the influence of a constant force $F$ and the avoided crossing is characterized by a minimal gap $\Delta_g$ and a slope of velocity $c$. The corresponding $2\times 2$ time-dependent Hamiltonian is $H(t)=cFt\sigma_z + \frac{\Delta_g}{2}\sigma_x$.}: 
\be
P_Z^x=\exp(-\frac{\pi}{c_xF}(\frac{q_y^2}{2m^*}+\Delta)^2)\, .
\ee
It only depends on the initial atomic position $q_y$ in reciprocal space (but not on $q_x$), which sets the minimal gap $2(\frac{q_y^2}{2m^*}+\Delta)$ between the two bands that is seen by the atom during its Bloch oscillation. If there is no gap (i.e. $q_y$ is such that $\frac{q_y^2}{2m^*}+\Delta=0$), the atom hits exactly a Dirac point in which case he moves to the upper band with certainty $P_z^x=1$. After averaging over the initial atomic density (i.e. over a distribution for $q_y$), we find the transferred fraction $\xi_x\equiv \langle P_Z^x \rangle$ as a function of $\Delta$, $m^*$ and $c_x$ (we assume that the force $F$ and the Fermi energy $\vep_F$ are fixed). In order to compare our results to the experiment, we calculated the ab initio band structure from the optical lattice potential and fitted the lowest two bands to that of the minimal Hamiltonian in the vicinity of the Dirac points \cite{Lim2012}. This gives a map between the minimal Hamiltonian parameters and the laser intensities $V_{\bar{X}}$ and $V_X$. The calculated transferred fraction $\xi_x$ is plotted as a function of $V_{\bar{X}}$ and $V_X$ in fig.~\ref{fig:lim}(a). As a rule of thumb $V_{\bar{X}}\sim t'-t'' = c_x$ and $V_X\sim 2t-t'-t''=-\Delta$. The agreement with the experiment is very good (compare with fig.~\ref{fig:tarruell}(a)). For example, we can explain why the red line of maximum transferred fraction is not exactly at $\Delta=0$ (merging transition) but slightly inside the gapped phase ($\Delta<0$). Indeed, an approximate calculation of the average gives $\xi_x\approx \exp\left(-\frac{\pi}{c_xF}\langle(\frac{q_y^2}{2m^*}+\Delta)^2\rangle\right)$, which shows that the maximum occurs at $\Delta \approx -\frac{\langle q_y^2\rangle}{2m^*}<0$ with $\langle q_y^2\rangle=\frac{k_{Fy}^2}{6}\propto \vep_F$.

Second, consider the double cone case, in which the applied force $\vec{F}=F\vec{e}_y$ is along $y$. As the Dirac points are always located at $k_x=0$, an atom may encounter two different Dirac cones during its trajectory $\vk(t)=\vk(0) + \vec{F}t$. There are therefore two ways an atom can end in the upper band: the atom may jump at the first Dirac cone or at the second. First assuming that the two tunneling events are incoherent, when $\Delta<0$, the interband transition probability is given by
\be
P_t^y=2P_Z^y(1-P_Z^y) \textrm{ where } P_Z^y=\exp(-\frac{\pi}{c_yF}(c_xq_x)^2) \textrm{ is the LZ probability.}
\ee   
Here $c_y\equiv \sqrt{\frac{-2\Delta}{m^*}}$ is the velocity in the $y$ direction and $2 c_x q_x$ is the minimal gap between the two bands that is seen by an atom during its Bloch oscillation. Averaging over the initial atomic density (i.e. over $q_x$), the transferred fraction is $\xi_y\equiv \langle P_t^y\rangle$, which is plotted in fig.~\ref{fig:lim}(b) as a function of the laser intensities. The agreement with the experimental result, see fig.~\ref{fig:tarruell}(b), is again very good. In particular, the red line of maximum transferred fraction is indeed found deep inside the gapless phase. In a simple Gaussian approximation $\xi_y\approx 2\langle P_Z^y\rangle(1-\langle P_Z^y\rangle)$ with $\langle P_Z^y\rangle\approx \exp(-\frac{\pi}{c_yF}c_x^2\langle q_x^2\rangle)$, which shows that the maximum transferred fraction is obtained when the LZ probability to jump at a single Dirac cone is $50\%$. This gives an equation for the red line of maximum probability $\frac{c_x^2}{c_y}=\frac{F \ln 2}{\pi \langle q_x^2\rangle}=$ cst with $\langle q_x^2\rangle=\frac{k_{Fx}^2}{6}\propto \vep_F$. In the inset of fig.~\ref{fig:lim}(b), lines of constant $\frac{c_x^2}{c_y}$ are plotted and indeed correspond to lines of constant transferred fraction. Compare fig.~\ref{fig:lim}(b) and (c) for two different values of the force $F$.
\begin{figure}[htb]
\begin{center}
\includegraphics[height=9cm]{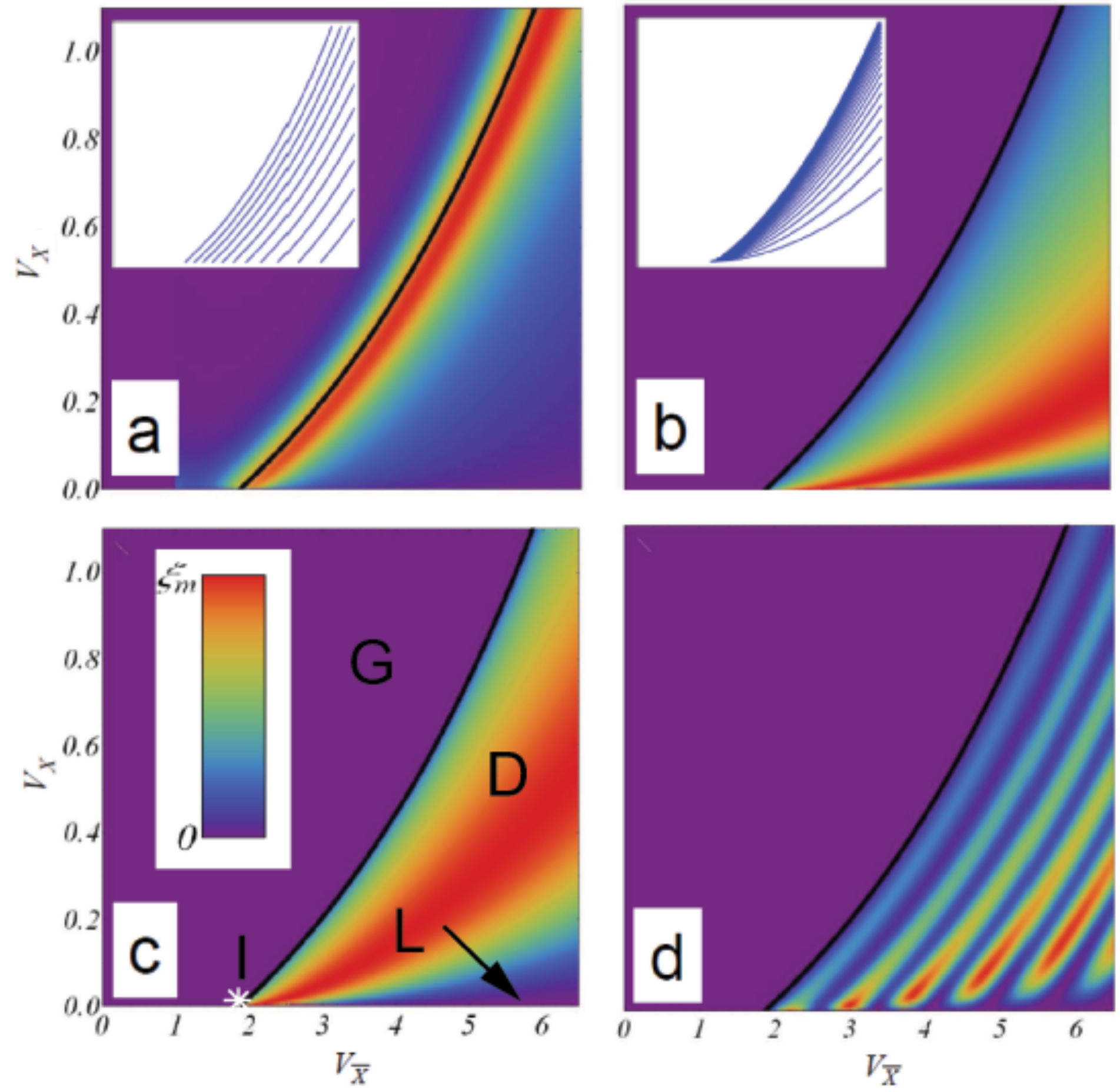}
\caption{\label{fig:lim} a) Transferred fraction $\xi_x$ (single cone case) as a function of the optical lattice parameters $V_{\bar{X}}\sim -\Delta$ and $V_X\sim c_x$ (here $k_{Fy}=\pi/2a, F=0.02E_R/a$). Inset: lines of constant $\Delta$. b) Transferred fraction $\xi_y$ (double cone case, with $k_{Fx}=2/a, F=0.02E_R/a$). Inset: lines of constant $c_x^2/c_y$. c) Same as (b) with a greater force $F=0.1E_R/a$.  d) Same parameters as (b) taking coherence into account and leading to St\"uckelberg oscillations. The color code for (a)-(d) is such that the maximum transferred fraction is $\xi_m=0.5,0.3,0.3,$ and $0.6$, respectively. The black line in (a)-(d) indicates the merging transition $\Delta=0$. In (c), G = gapped phase ($\Delta>0$), D= gapless Dirac phase ($\Delta<0$), I = isotropic square lattice ($t''=t'=t$), L = square lattice ($t'=t''\neq t$). From Lim et al. \cite{Lim2012}.}
\end{center}
\end{figure}

Actually, the two different paths to go from the lower to the upper band should interfere. Therefore we should add amplitudes instead of probabilities. Such a setup is known as a St\"uckelberg interferometer. The two linear avoided band crossings act as beam splitters and the fact that the two paths occur at different energies means that there is a phase difference of dynamical origin. When $\Delta<0$, the interband transition probability is \cite{LZSReview}
 \be
P_t^y=4P_Z^y(1-P_Z^y)\sin^2(\frac{\varphi_{dy}}{2}+\varphi_{St}) \to 2P_Z^y(1-P_Z^y) \textrm{ in the incoherent limit}
\label{eq:stuck}
\ee
where $\varphi_{dy}=F^{-1}\int_{-q_D}^{q_D}dq_y\, 2\sqrt{c_x^2q_x^2+(\Delta+\frac{q_y^2}{2m^*})^2}$ is the dynamical phase acquired between the two tunneling events and $\varphi_{St}=\frac{\pi}{4}+\delta(\ln \delta -1)+\textrm{Arg }\Gamma(1-i\delta)$ is the so-called Stokes phase (which is a phase delay for the reflected path), where $\delta\equiv \frac{c_x^2q_x^2}{2c_yF}$. The corresponding transferred fraction $\xi_y=\langle P_t^y\rangle$ is plotted in fig.~\ref{fig:lim}(d) and shows fringes, which are lines of constant $\Delta$ (see the inset in fig.~\ref{fig:lim}(a)). These fringes were not observed in the Z\"urich experiment. The decoherence time is expected to be much longer than the duration of a Bloch oscillation and can not be responsible for the absence of interferences. However, the experiment is actually performed on a 3D and not on a strictly 2D gas (see a preceding footnote). In the presence of the 2D optical lattice, the system is therefore best pictured as a honeycomb-like lattice (in the $xy$ plane) of tubes (in the $z$ direction). The motion is almost free along the $z$ direction, apart from the global harmonic trapping. We attribute the disappearance of fringes to the averaging over the third spatial direction (along the tubes). Indeed, interferences are lines of constant $\Delta$ with a fringe spacing $\sim 0.04 E_R$. For the experimentally given trapping frequency in the $z$ direction, we estimate that $\Delta$ varies along $z$ by $\sim 0.03 E_R$. This should be enough to wash out the interferences \cite{Lim2012,Fuchs2012}. An alternative explanation for the disappearance of the oscillations was recently proposed in \cite{Uehlinger2013}. It is based on the spatial inhomogeneity of the applied force in 2D (due to the presence of the harmonic trap in addition to the applied homogeneous force $\vec{F}$), which also leads to averaging and washing-out of the probability fringes.      

\begin{figure}[htb]
\begin{center}
\includegraphics[height=6cm]{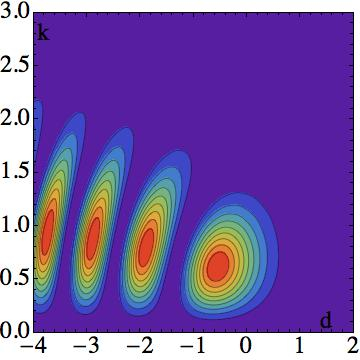}
\caption{\label{fig:pra2012}Contour plot of the interband transition probability $P_t^y$ in the double-cone case as a function of the merging gap $d\equiv \Delta$ and the perpendicular gap $k\equiv \hbar c_xq_x$ [in units of $\frac{(\hbar F)^{2/3}}{(2m^*)^{1/3}}$]. Interferences are clearly visible: they are essentially driven by $d$ but they also occur in the $k$ direction. The merging transition is at $d=0$, the gapless Dirac phase corresponds to $d<0$ and the gapped phase to $d>0$. Modulo the precise mapping between $(\Delta,c_xq_x)$ and $(V_{\bar{X}},V_X)$, which is qualitatively given by $(\Delta,c_xq_x)\sim (-V_{\bar{X}},V_X)$, this figure corresponds to fig.~\ref{fig:lim}(d). From Fuchs et al. \cite{Fuchs2012}.}
\end{center}
\end{figure}
Up to now, in the double cone case, we only considered the gapless phase ($\Delta<0$) and assumed that the global inter-band transition could be treated as a succession (coherent or not) of two LZ tunneling events. This is actually an approximation, valid when the Dirac points are sufficiently far apart ($\Delta \ll 0$). In a recent work \cite{Fuchs2012}, we treated the full inter-band transition problem starting from the minimal Hamiltonian and not linearizing the dispersion relation in the vicinity of the Dirac cones. This allowed us to explore the complete phase diagram as a function of the merging gap $\Delta$ and the transverse gap $c_x q_x$. In particular, we could explore the gapped phase ($\Delta>0$) and the close vicinity of the merging transition ($\Delta \approx 0$), which is not possible within the St\"uckelberg approach eq.~\ref{eq:stuck}. The probability $P_t^y$ for a single atom to tunnel from the lower to the upper band is shown in fig.~\ref{fig:pra2012} as a function of $\Delta\sim -V_{\bar{X}}$ and $c_xq_x\sim V_X$. This figure was obtained from a numerical solution of the Schr\"odinger equation. Analytically, we could solve the problem approximatively using a combination of adiabatic perturbation theory and the St\"uckelberg approach. We do not dwell into these calculations here and refer the interested reader to \cite{Fuchs2012}.

\section{Conclusion: stability of Dirac fermions and of Dirac points}
\label{conclusion2}
It is often thought that graphene's interest lies essentially in its peculiar energy spectrum featuring two linear contact points (Dirac points). But as important, if not more, is the fact that Dirac fermions correspond to quantized vortices (of minimal charge) in the relative phase $\phi$ between the two sublattices. This is an information, which is not contained in the energy spectrum alone, but in the Hamiltonian or in the eigenvectors\footnote{Hence, there is more to the Hamiltonian than its spectrum. In particular the topological properties are contained in the phase $\phi$. In the simplest tight binding model, the Hamiltonian is $H(\vk)=\left(\begin{array}{cc}0&f(\vk)^*\\f(\vk)&0\end{array}\right)=|f(\vk)|\left(\begin{array}{cc}0&e^{-i\phi_{\vk}}\\e^{i\phi_{\vk}}&0\end{array}\right)$. Its eigenvalues $\vep_\alpha=\alpha |f(\vk)|$ depend on the modulus of $f(\vk)$, while the corresponding eigenvectors $\frac{1}{\sqrt{2}}\left(\begin{array}{c}1\\\alpha e^{i\phi}\end{array} \right)$ depend on the phase of $f(\vk)$. We have seen that Landau levels are affected by $\phi$ and do not only depend on the zero field spectrum. This affects measurable quantities such as the orbital susceptibility \cite{Raoux2013}. Another example is provided by the angle resolved photo-emission spectroscopy (ARPES) of graphene, which shows extinctions of signal due to the chirality factor, i.e. to the phase $\phi$, where energy bands would be expected \cite{Mucha2008}.}. It is a consequence of coupling between two bands. Indeed hopping in graphene is predominantly between the $A$ and the $B$ sublattices (and not say between $A$ and $A$ or $B$ and $B$). Therefore having two (or more) orbitals per unit cell is not enough to have coupled bands. What is needed is hopping preferentially between different orbitals (or different sites). In the simplest case, this is related to the chiral (sublattice) symmetry resulting from the bipartite nature of the lattice.
	
Dirac fermions can also exist in the absence of contact points between the bands. This is the case of boron nitride for example, which exhibits massive Dirac fermions at low energy. The pair of quantized vortices is also present in that case and many of the interesting properties of graphene remain. It is therefore important to distinguish between Dirac fermions -- corresponding to quantized vortices in the relative phase of the wavefunction on two sublattices --, which can be massive or massless, and Dirac points -- corresponding to linear band contacts -- and which, by definition, are massless.

Dirac fermions appear and disappear in pairs of opposite vorticity (fermion doubling, see the general introduction). If they exist and are well separated in reciprocal space, any small perturbation that aims at destroying them has to bring them together in order to make them annihilate. Dirac fermions are therefore robust to small perturbations. Here, we have studied a specific type of perturbation, namely lattice deformation, and seen that a large strain is indeed required for merging. The stability of Dirac fermions is related to that of quantized vortices. The further apart in reciprocal space, the harder it is to couple the two valleys, and the more stable the Dirac fermions are.
 
However, contact (Dirac) points are not as stable as the vortices. In the introduction (see section \ref{mergingintro}), we mentioned that small symmetry breaking perturbations are able to open a gap (cf. inversion symmetry breaking leading to a Semenoff mass \cite{Semenoff} or time-reversal symmetry breaking leading to a Haldane mass \cite{Haldane}). In the following, we exclude such space-time inversion symmetry breaking terms and discuss the robustness of Dirac points with respect to other perturbations. Apart from lattice deformations, other possible perturbations include interactions between electrons, disorder and the magnetic field. 

We first discuss interactions. Power counting shows that repulsive short range interactions have a naive scaling dimension $1-d$ and are therefore irrelevant in $d=2$ spatial dimensions \cite{SunFradkin}. Detailed calculation on the honeycomb lattice with nearest and next-nearest neighbor repulsion confirm that a critical strength for interactions is needed to gap the system \cite{Herbut2006}. In contrast to irrelevant short range repulsion, long-range Coulomb interactions are superficially marginal according to power counting. However, perturbative RG calculations show that they are actually marginally irrelevant \cite{Kotov2012}. Therefore, massless Dirac fermions are immune to weak repulsive interactions, whether long or short ranged. Another way to see that interactions are not so crucial for massless Dirac fermions comes from the fact that the density of states vanishes linearly at $\vep_F=0$. This inhibits instabilities, including the BCS instability when the interaction between electrons is attractive \cite{BCSDirac}. An interesting example to compare with is that of a quadratic band crossing such as that occurring in bilayer graphene. This case corresponds to vortices of charge $\pm 2$. These are not the minimally charged vortices and hence they are unstable with respect to fractionalization in smaller vortices $2\to 1+1$. In this case there is no threshold for the loss of the quadratic band crossing and arbitrary small perturbations may open a gap \cite{SunFradkin}. Here the density of states is finite even when the Fermi surface is point-like. Instabilities of undoped bilayer graphene have recently been extensively studied, see e.g. \cite{NL,MacDo}.

What can be said in general of the robustness of Dirac points with respect to disorder? To discuss this question we consider either one of two quantities related to the existence of Dirac points and that have been computed in the literature: the zero-energy density of states $\rho(0)$ in the presence of disorder or the (minimal) conductivity $\sigma$. In clean graphene, the density of states vanishes linearly and, surprisingly, the minimal conductivity remains finite $\sigma=\frac{1}{\pi}\frac{e^2}{h}$ (per valley and per spin) at zero doping \cite{Shon1998,Katsnelson2006,Tworzydlo2006}. Actually, Dirac points are quite immune to disorder: the self-consistent Born approximation (SCBA, and other more refined techniques) show that the density of states either remains zero or is exponentially small at zero doping depending on the type of disorder \cite{Shon1998,Ostrovsky2006}. The immunity of Dirac points is related to both (i) the absence of backscattering within a valley and (ii) to the fact that valleys are not coupled by impurities of range larger than the lattice spacing. For uncoupled valleys, even interference effects -- beyond the SCBA -- lead to anti-localization (rather than localization) so that the conductivity \textit{increases} with respect to its clean value \cite{NM2007,Bardarson,NomuraRyu}. Typically $\sigma\sim \frac{e^2}{h}$ (per spin and valley) in disordered graphene at the neutrality point. However, with short range impurities inducing coupling between valleys, quantum interferences eventually lead to strong localization $\sigma \ll \frac{e^2}{h}$ as in a standard 2D electron gas \cite{AleinerEfetov}. In a recent work, the effect of disorder on Dirac points was studied as a function of the distance to the merging transition using the minimal Hamiltonian \cite{Carpentier2013}. When valleys are far apart, results are quite similar to that for uncoupled massless Dirac fermions. Upon approaching the merging transition, Dirac points become less stable and the effect of disorder is stronger, as seen in a finite -- and no longer exponentially small -- density of states. Formulated differently, disorder is a relevant perturbation for semi-Dirac fermions (at the merging), while it is marginal for massless Dirac fermions (far from the merging).
	
The presence of a magnetic field also changes the above considerations for the stability of Dirac points. It reintroduces a finite density of states at zero energy, due to the macroscopic degeneracy of the $n=0$ LL, and therefore the possibility of instabilities (this is known as the magnetic catalysis of an instability). In addition to a finite density of states, the other required ingredient is some coupling between the valleys. As we have seen in the section on the interaction-induced integer quantum Hall effect, in the presence of interactions (whether electron-electron, short/long range or electron-lattice), the valley degeneracy of the $n=0$ Landau level is lifted. As another example of magnetic field induced instability of Dirac points, we mention that a magnetic field opens an exponentially small gap at zero energy in graphene due to ``magnetic breakdown'' \cite{Montambaux2009b,Esaki2009}. This effect is due to quantum tunneling between cyclotron orbits and crucially depends on the distance in reciprocal space between the two valleys. It is therefore much more pronounced close to merging, when graphene is strongly deformed and the two valleys are close by. Finally, a last example is that of a vortex superlattice on top of graphene's honeycomb lattice, which is also able to open a gap at zero energy if its periodicity\footnote{The reciprocal vectors of the superlattice should connect the $K$ and $K'$ points.} is such that it couples the two valleys \cite{Kamfor2011}.

Conclusion on the stability of Dirac fermions and of Dirac points. Stability should be discussed with respect to (1) the disappearance of Dirac fermions (due to merging) and (2) the opening of a gap for Dirac fermions (without merging; massive Dirac fermions). Concerning (1), Dirac fermions are topologically protected as they are quantized vortices of minimal charge. These can only disappear via merging which is not easy as, usually, Dirac fermions are far apart in reciprocal space. Concerning (2), as we have seen, the existence of contact points is usually due to a discrete symmetry (such as chiral or space-time inversion). If the perturbations respect this symmetry, then there is symmetry protected topological stability of the Dirac points. If it does not, instabilities are nevertheless inhibited by the linearly vanishing density of states and the difficulty to couple valleys which are far apart in reciprocal space.

		
		

	\chapter{Conclusion and perspectives}
	
	{\it  ``Not well digested, but brilliantly shat.'' (attributed to Einstein)}
	
	\vspace{0.5cm}
	
	Ending this manuscript, we first draw a brief conclusion\footnote{We have given longer separate conclusions to the two parts of this manuscript, see section \ref{conclusion1} and section \ref{conclusion2}.} and then present a few perspectives. Rather than listing our achievements, we think it is more useful to extract the essence of what we have learned on two-dimensional Dirac fermions during these years. Concerning massless Dirac fermions in a perpendicular magnetic field (first part of the manuscript), the most interesting aspect is arguably the existence of a zero-energy Landau level related to the parity anomaly (see section \ref{conclusion1}). In the second part of this thesis, we saw that more important than the linear spectrum or contact points is the existence of quantized vortices in the relative phase of the electron wavefunction on the two sublattices in graphene. These vortices are responsible for the stability of Dirac fermions (see section \ref{conclusion2}). They are quantized vortices of minimal non-zero charge and appear in pairs of opposite charge or vorticity (fermion doubling). Their appearance/disappearance is only via a merging transition, which is not easily reached. The existence and stability of Dirac points -- i.e. linear contact points between two bands -- is not the same thing as that of Dirac fermions -- i.e. quantized vortices.
	
\begin{figure}[htb]
\begin{center}
\subfigure[]{\includegraphics[height=5cm]{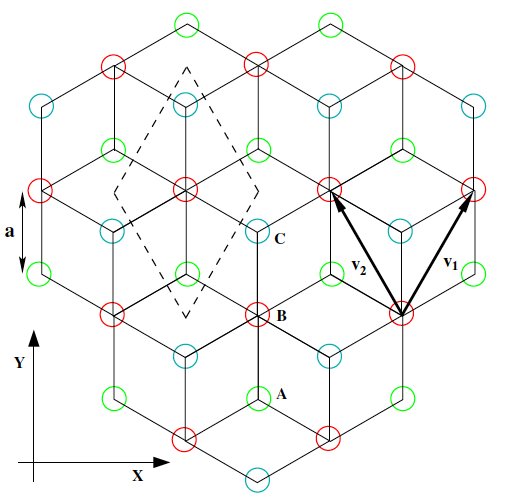}}
\subfigure[]{\includegraphics[height=5cm]{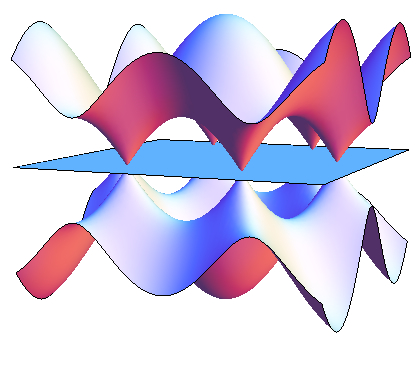}}
\subfigure[]{\includegraphics[height=4.5cm]{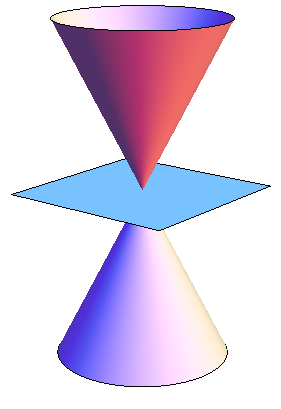}}
\caption{\label{fig:dice}(a) Dice (or T$_3$) lattice: triangular Bravais lattice with three atoms ($A$, $B$ and $C$) per unit cell. From M. Morigi \cite{MorigiStage2012}. (b) Corresponding tight-binding band structure featuring three bands, one of which is flat. (c) Zoom at low energy: pseudo-spin 1 Weyl fermions.}
\end{center}
\end{figure}
In the future, we plan to study massless Dirac fermions in a more general sense by looking at the dice (or T$_3$) lattice. This is a tight-binding model with a triangular Bravais lattice and three atoms per unit cell, see fig.~\ref{fig:dice}(a). Two atoms ($A$ and $C$) are threefold connected, while the third ($C$) is sixfold connected. Hence the lattice is bipartite, the tight-binding Hamiltonian has a chiral/sublattice symmetry and the corresponding energy spectrum therefore has a particle-hole symmetry. As there is an odd number of bands, there is necessarily a zero energy flat band, see fig.~\ref{fig:dice}(b). This effect was found by Sutherland, which he named topological localization \cite{Sutherland1986}. Overall, the band structure of the dice lattice is very similar to that of graphene, except for the additional flat band. At low energy around half-filling, the effective description is in terms of pseudo-spin 1 Weyl fermions, with dispersion $\vep_\alpha(\vp)=\alpha v_F p$, where the band index $\alpha=\pm 1$ or $0$, see fig.~\ref{fig:dice}(c). The effective Hamiltonian for a single cone is $H=v_F (p_xS_x+p_yS_y)$ where $S_x$ and $S_y$ are $3\times 3$ pseudo-spin 1 matrices, see e.g. \cite{Bercioux}.  

Our main idea is to introduce a parameter dependent model (called $\beta$-T$_3$, where $\beta$ is a dimensionless parameter) that interpolates continuously between the honeycomb ($\beta=0$) and the dice ($\beta=1$) lattices. The model has 3 atoms per unit cell with the same geometry as the dice lattice and always has a flat band. However at $\beta=0$, one atom out of three is completely decoupled from the others and we essentially recover a honeycomb lattice. We are interested in two specific properties related to this $\beta$-T$_3$ model. 

First, we would like to study localization effects and flat bands. We already mentioned the zero energy flat band found by Sutherland. There is another possibility of obtaining localization and flat bands in the dice lattice. Indeed in a perpendicular magnetic field, at half a quantum of flux per plaquette (rhombus), there is an interference effect known as Aharonov-Bohm cages \cite{Vidal1998} that creates a discrete spectrum made of two flat bands at finite energies in addition to the zero energy flat band. We wish to study the robustness of these localization effects with respect to several perturbations such as small additional magnetic fields, the $\beta$ parameter, staggered on-site potentials, disorder, etc. This project was recently started during the master internship of Matteo Morigi. For example, we have studied in detail the Hofstadter spectrum of the dice lattice at remarkable fluxes (such as zero, one-third and one-half). We have also seen that, at zero flux, the Berry phase associated to the Dirac cone (fig.~\ref{fig:dice}(c)) is topological (it does not depend on the precise path around the contact point)  but not quantized as its value continuously evolves between $\pi$ and $0$ with $\beta$ \cite{MorigiStage2012}.

Second, we wish to investigate the orbital magnetism of massless Dirac-like electrons. Indeed, undoped graphene is predicted to have a very strong diamagnetism at zero temperature, whereas doping should make the system paramagnetic \cite{StauberGomez}. This subject was started during the master internships of Maurice Tia and of Arnaud Raoux. We found that, as a function of $\beta$ varying between 0 and 1, the orbital susceptibility changes from diamagnetic to paramagnetic at a critical value of $\beta$ \cite{Raoux2013}. We also computed numerically the orbital susceptibility for the honeycomb tight-binding model and found that it changes sign several times as a function of doping. We plan to study several other tight-binding models with the tools we developed to show the importance of band coupling effects on the orbital magnetism.

Another research direction, unrelated to the dice lattice, is to use the St\"uckelberg interferometer, uncovered in cold atoms' experiments on the merging transition of artificial graphene \cite{Tarruell2012,Lim2012}, to detect changes in the sign of the mass of Dirac fermions and hence to access topological quantities such as Chern numbers. This is on-going work with Lih-King Lim and G. Montambaux. Related ideas have recently appeared in a one-dimensional context \cite{DemlerBloch}.


\chapter{Publications}
Below is a list of my publications related to the material presented in this habilitation thesis. There are web links to the articles.
 
 \vspace{0.5cm}
 
\noindent \textbf{Chapter 2: Massless Dirac fermions in a strong magnetic field}

\noindent \underline{Landau levels and quantum Hall effects}
     
\cite{Plochocka} P. Plochocka, C. Faugeras, M. Orlita, M.L. Sadowski, G. Martinez, M. Potemski, M.O. Goerbig, J.N. Fuchs, C. Berger and W.A. de Heer, ``High-Energy Limit of Massless Dirac Fermions in Multilayer Graphene using Magneto-Optical Transmission Spectroscopy'', \href{http://prl.aps.org/abstract/PRL/v100/i8/e087401}{Phys. Rev. Lett. \textbf{100}, 087401 (2008)}. 

\cite{Goerbig2008} M.O. Goerbig, J.N. Fuchs, G. Montambaux and F. Pi\'echon, ``Tilted anisotropic Dirac cones in quinoid-type graphene and alpha-(BEDT-TTF)$_2$I$_3$'', \href{http://prb.aps.org/abstract/PRB/v78/i4/e045415}{Phys. Rev. B \textbf{78}, 045415 (2008)}.

\cite{Goerbig2009} M.O. Goerbig, J.N. Fuchs, G. Montambaux and F. Pi\'echon, ``Electric-field-induced lifting of the valley degeneracy in alpha-(BEDT-TTF)$_2$I$_3$ Dirac-like Landau levels'', \href{http://epljournal.edpsciences.org/index.php?option=com_article&access=standard&Itemid=129&url=/articles/epl/abs/2009/05/epl11656/epl11656.html}{Europhys. Lett. \textbf{85}, 57005 (2009)}.

     
\cite{FL} J.N. Fuchs and P. Lederer, ``Spontaneous parity breaking of graphene in the quantum Hall regime'', \href{http://prl.aps.org/abstract/PRL/v98/i1/e016803}{Phys. Rev. Lett. \textbf{98}, 016803 (2007)}.
     
\noindent \underline{Particle-hole excitations}     
     
\cite{Roldan2009} R. Rold\'an, J.N. Fuchs and M.O. Goerbig, ``Collective Excitations of Doped Graphene in a Strong Magnetic Field'', \href{http://prb.aps.org/abstract/PRB/v80/i8/e085408}{Phys. Rev. B \textbf{80}, 085408 (2009)}.

\cite{Roldan2010a} R. Rold\'an, M.O. Goerbig and J.N. Fuchs, ``The magnetic field particle-hole excitation spectrum in doped graphene and in a standard two-dimensional electron gas'', \href{http://iopscience.iop.org/0268-1242/25/3/034005}{Semicond. Sci. Technol. \textbf{25}, 034005 (2010)}.

\cite{Roldan2010b} R. Rold\'an, J.N. Fuchs and M.O. Goerbig, `` Spin-flip excitations, spin waves, and magneto-excitons in graphene Landau levels at integer filling factors'', \href{http://prb.aps.org/abstract/PRB/v82/i20/e205418}{Phys. Rev. B \textbf{82}, 205418 (2010)}.

\cite{Roldan2011} R. Rold\'an, M.O. Goerbig and J.N. Fuchs, ``Theory of Bernstein modes in graphene'',  \href{http://prb.aps.org/abstract/PRB/v83/i20/e205406}{Phys. Rev. B \textbf{83}, 205406 (2011)}.

\noindent \underline{Magneto-phonon resonance}

\cite{Goerbig2007} M.O. Goerbig, J.N. Fuchs, K. Kechedzhi and V.I. Fal'ko, ``Filling-Factor-Dependent Magnetophonon Resonance in Graphene'', \href{http://prl.aps.org/abstract/PRL/v99/i8/e087402}{Phys. Rev. Lett. \textbf{99}, 087402 (2007)}.

\noindent \underline{Magneto-transport}

\cite{Monteverde2010} M. Monteverde, C. Ojeda-Aristizabal, R. Weil, K. Bennaceur, M. Ferrier, S. Gueron, C. Glattli, H. Bouchiat, J.N. Fuchs and D. Maslov, ``Transport and elastic scattering times as probes of the nature of impurity scattering in single and bilayer graphene'', \href{http://prl.aps.org/abstract/PRL/v104/i12/e126801}{Phys. Rev. Lett \textbf{104}, 126801 (2010)}.

\vspace{0.5cm}

\noindent \textbf{Chapter 3: Topological properties and stability of Dirac fermions}

\noindent \underline{Berry phases}

\cite{Fuchs2010} J.N. Fuchs, F. Pi\'echon, M.O. Goerbig and G. Montambaux, ``Topological Berry phase and semiclassical quantization of cyclotron orbits for two dimensional electrons in coupled band models'', \href{http://epjb.epj.org/index.php?option=com_article&access=doi&doi=10.1140/epjb/e2010-00259-2&Itemid=129}{Eur. Phys. J. B \textbf{77}, 351 (2010)}.

\noindent \underline{Merging transition}

\cite{Montambaux2009a} G. Montambaux, F. Pi\'echon, J.N. Fuchs and M.O. Goerbig, ``Merging of Dirac points in a two-dimensional crystal'', \href{http://prb.aps.org/abstract/PRB/v80/i15/e153412}{Phys. Rev. B \textbf{80}, 153412 (2009)}.
     
\cite{Montambaux2009b} G. Montambaux, F. Pi\'echon, J.N. Fuchs and M.O. Goerbig, ``A universal Hamiltonian for the motion and the merging of Dirac cones in a two-dimensional crystal'', \href{http://epjb.epj.org/index.php?option=com_article&access=doi&doi=10.1140/epjb/e2009-00383-0&Itemid=129}{Eur. Phys. J. B \textbf{72}, 509 (2009)}.     
     
\cite{Lim2012} L.K. Lim, J.N. Fuchs and G. Montambaux, ``Bloch-Zener oscillations across a merging transition of Dirac points'', \href{http://prl.aps.org/abstract/PRL/v108/i17/e175303}{Phys. Rev. Lett. \textbf{108}, 175303 (2012)}.     
     
\cite{Fuchs2012} J.N. Fuchs, L.K. Lim and G. Montambaux, ``Inter-band tunneling near the merging transition of Dirac cones'', \href{http://pra.aps.org/abstract/PRA/v86/i6/e063613}{Phys. Rev. A \textbf{86}, 063613 (2012)}.



	
	
	

	
	
	

\end{document}